\documentclass[aps,pre,twocolumn,superscriptaddress,showpacs,floatfix,longbibliography]{revtex4-1}

\makeatletter
\def\@bibdataout@aps{%
\immediate\write\@bibdataout{%
@CONTROL{%
apsrev41Control%
\longbibliography@sw{%
    ,author="08",editor="1",pages="1",title="0",year="1"%
    }{%
    ,author="08",editor="1",pages="1",title="",year="1"%
    }%
  }%
}%
\if@filesw \immediate \write \@auxout {\string \citation {apsrev41Control}}\fi
}
\makeatother

\usepackage{graphicx}
\usepackage{dcolumn}
\usepackage{bm}

\usepackage{amssymb}
\usepackage{amsmath}
\usepackage{graphicx}
\usepackage{xcolor}

\usepackage[colorlinks,citecolor=blue,linkcolor=red,urlcolor=blue]{hyperref}

\begin{document}
\title{Finite-time scaling with two characteristic time scales: Driven critical dynamics with emergent symmetry}
\author{Yu-Rong Shu}
\affiliation{School of Physics and Materials Science, Guangzhou University, Guangzhou 510006, China}

\author{Li-Ying Yang}
\affiliation{School of Physics and Materials Science, Guangzhou University, Guangzhou 510006, China}

\author{Shuai Yin}
\email{yinsh6@mail.sysu.edu.cn}
\affiliation{School of Physics, Sun Yat-Sen University, Guangzhou 510275, China}
\affiliation{Guangdong Provincial Key Laboratory of Magnetoelectric Physics and Devices, Sun Yat-Sen University, Guangzhou 510275, China}

\date{\today}

\begin{abstract}
Critical points with emergent symmetry exhibit intriguing scaling properties induced by two divergent length scales, attracting extensive investigations recently. We study the driven critical dynamics in a three-dimensional $q$-state clock model, in which the ordered phase breaks the $Z_q$ discrete symmetry, while an emergent $U(1)$ symmetry appears at the critical point. By increasing the temperature at a finite velocity $v$ to traverse the critical point from the ordered phase, we uncover rich dynamic scaling properties beyond the celebrated Kibble-Zurek mechanism. Our findings reveal the existence of two finite-time scaling (FTS) regions, characterized by two driving-induced time scales $\zeta_d\propto v^{-z/r}$ and $\zeta_d'\propto v^{-z/r'}$, respectively. Here $z$ is the dynamic exponent, $r$ is the usual critical exponent of $v$, and $r'$ represents an additional critical exponent of $v$ associated with the dangerously irrelevant scaling variable. While the square of the order parameter $M^2$ obeys the usual FTS form, the angular order parameter $\phi_q$ shows remarkably distinct scaling behaviors controlled by both FTS regions. For small $v$, $\phi_q$ is dominated by the time scale $\zeta_d$, whereas for large $v$, $\phi_q$ is governed by the second time scale $\zeta_d'$. We verify the universality of these scaling properties in models with both isotropic and anisotropic couplings. Our theoretical insights provide a promising foundation for further experimental investigations in the hexagonal RMnO$_3$ (R=rare earth) materials.
\end{abstract}

\maketitle

\section{Introduction}
Nonequilibrium critical dynamics represents a crucial facet of critical phenomena, drawing continuous interest through both theoretical and experimental studies~\cite{Hohenberg1977rmp,Folk2006jpa,Tauber2014book}. In a second-order phase transition, when the critical point $T_c$ is traversed with a finite cooling velocity $v$ starting from the disordered phase ($T>T_c$), the celebrated Kibble-Zurek mechanism (KZM) predicts that the evolution of the system stops being adiabatic owing to critical slowing down in the vicinity of the critical point and enters an impulse stage~\cite{Kibble1976,Kibble2007,Zurek1985,ZUREK1996177}. In the ordered side ($T<T_c$), disparate local choices of the ordered domain lead to the formation of topological defects~\cite{Kibble1976,Kibble2007,Zurek1985,ZUREK1996177}. Furthermore, the KZM asserts that the density of the resulting topological defects satisfies a scaling relation depending on the cooling rate~\cite{Kibble1976,Kibble2007,Zurek1985,ZUREK1996177}.

Along this line, significant progress has been made in two directions.
One involves the theoretical extension of the KZM, including generalizing the KZM to quantum phase transitions~\cite{Zoller2005prl,Dziarmaga2005prl,Polkovnikov2005prbr,Damski2007prl,Grandi2011prb,Deng2025prl}, accounting for inhomogeneous effects~\cite{Zurek2009prl,delcampo2010prl,delCampo2013}, exploring nonlinear driving scenarios~\cite{Sen2008prl,chandran2012prb}, combining the KZM with other dynamic mechanisms~\cite{Biroli2010prb,Huangyy2016prb,Jeong2019pre,Tang2025}, among other developments. 
Within this context, a finite-time scaling (FTS) theory was proposed~\cite{Gong2010njp}. The FTS theory introduces a notion of driving-induced time scale $\zeta_d\sim v^{-z/r}$, in which $r=z+1/\nu$ with $z$ and $\nu$ being the the dynamic critical exponent and the correlation length critical exponent, respectively, and shows that $\zeta_d$ can characterize the critical dynamics near the critical point. This framework provides a comprehensive understanding of the universal dynamic scaling behaviors throughout the entire critical region~\cite{Zhongf2005prb,Zhong2006pre,Fan2007pre,Huangxz2010pre,Yin2014prb,Huang2014prb,feng2016prb,Fan2009pre}. Moreover, the FTS theory demonstrates that, apart from the topological defects, other macroscopic quantities---such as the order parameter, correlation functions, entanglement entropy---also exhibit scaling behaviors dependent on the driving rate $v$~\cite{Zhongf2005prb,Zhong2006pre,Fan2007pre,Huangxz2010pre,Yin2014prb,Huang2014prb,feng2016prb,Fan2009pre,huqj2015prb,Cao2018prb}. In addition, the FTS theory can also accommodate other types of driving, including heating dynamics starting from the ordered initial state~\cite{Zhongf2005prb,Zhong2006pre,Fan2007pre,Huangxz2010pre,Yin2014prb,Huang2014prb,feng2016prb,Fan2009pre,huqj2015prb,Cao2018prb}, driven dynamics for changing the symmetry-breaking field~\cite{Gong2010njp}, temperature-driven dynamics in approaching quantum phase transitions~\cite{Yin2014prb}, and others. 
Recently, the FTS has been generalized to cover dynamical phase transition~\cite{Liyh2019pre,Yinarxiv2024} and has integrated with relaxation critical dynamics, extending the KZM to beyond adiabaticity~\cite{Zhong2006pre,Huangyy2016prb,Yin2016prb,Zeng2024arx_1,Zeng2024arx_2}. Full scaling forms similar to the FTS have also been discussed from other settings~\cite{Deng2008,chandran2012prb,Grandi2011prb,Huse2012prl,liuprb2014,Zurek2016prb}.

The other direction involves the experimental examination of the KZM. A variety of experiments have been designed aiming to confirm the scaling predictions of the KZM, including studies on superfluid $^3$He and $^4$He~\cite{Hendry1994nature,Dodd1998prl,Bunkov1996nature,Xuwen1996nature}, superconductivity rings~\cite{Carmi2000prl,Monaco2002prl,Monacoprb2009}, Bose-Einstein condensates~\cite{Brand2013prl,Labeyrie2016prl,Navon2015science}, Berezinskii-Kosterlitz-Thouless transition in two-dimensional ($2$D) Bose gases~\cite{Shinichi2023science}, liquid crystals~\cite{Chuang1991sci,Ducci1999prl}, quantum computational devices~\cite{Keesling2019,Ebadi2021,King2022}. Among these experimental systems, rare-earth multiferroics~\cite{Chae2012prl,Griffin2012prx,Lin2014natphy,Meier2017prx,Skjaervo2019prx,Meier2020prb,Zhang2021prb,OWSandvik2023nl,Kang2023jap,Baghizadeh2019jpcc,Juraschek2020prl} have provided particularly compelling evidence. Experiments show that cooling RMnO$_3$ (R=rare earth) materials through their critical points yields vortices located at the focal points of domains induced by local symmetry breaking. It was found that the dependence of the vortex density on the cooling rate conforms to the KZM prediction of the three-dimensional ($3$D) $U(1)$ universality class, offering the first possible experimental setting compatible with the KZM beyond the mean-field theory~\cite{Chae2012prl,Skjaervo2019prx,Griffin2012prx,Lin2014natphy,Meier2017prx}.

\begin{figure}[tbp]
\centering
  \includegraphics[width=\linewidth,clip]{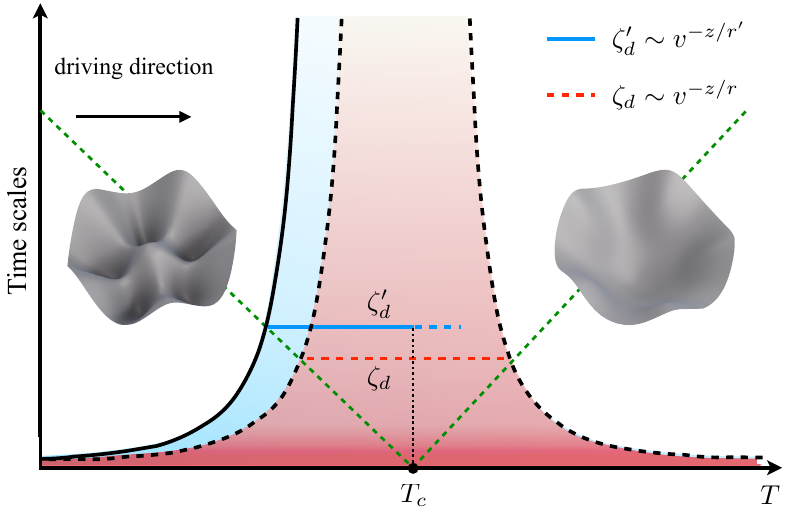}
  \vskip-3mm
  \caption{Illustration of heating critical dynamics with emergent symmetry in $5$-state clock model. The free-energy landscapes at low and high temperatures are shown on the left and right sides, respectively. At low temperature, the puddles represent the five possible directions, while at high temperature, the free-energy landscape has a single minimum.
  Near the critical point, the usual correlation time $\zeta$ diverges as $\zeta\propto |g|^{-\nu z}$ (dashed black curve). An additional critical time scale related to the DISV develops on the ordered side, scaling as $\zeta'\propto |g|^{-\nu'z}$ (solid black curve). $\zeta'$ characterizes the time scale beyond which the discrete symmetry breaking fixed point dominates the dynamic critical behaviors. Under external driving with a rate $v$, starting from the ordered phase (indicated by the arrow), two FTS regions emerge. The intersections between the time distance $|t|$ (green dashed line) and curves of $\zeta$ and $\zeta'$ determine the boundaries of FTS Region I and II, respectively. In the FTS Region I, the typical time scale is $\zeta_d\propto v^{-z/r}$, whereas in the FTS Region II, the typical time scale is $\zeta_d'\propto v^{-z/r'}$. Note that although in equilibrium, DISV only works in the ordered phase, FTS Region II can extend even in the side of $T>T_c$ (dashed blue line). 
  }
  \label{fig:quench}
\end{figure}

Here we point out that special attention should be directed towards the critical dynamics in rare-earth multiferroics, as the $U(1)$ symmetry detected at the critical point is an emergent symmetry, which is not originally possessed by the system.
Symmetry analyses reveal that the microscopic effective Hamiltonian characterizing this phase transition is the clock model with discrete $Z_q$ symmetry~\cite{Chae2012prl,Griffin2012prx,Lin2014natphy,Meier2017prx,Skjaervo2019prx,Meier2020prb,Zhang2021prb,OWSandvik2023nl,Kang2023jap,Baghizadeh2019jpcc,Juraschek2020prl}.

For the general $Z_q$ clock model, it was shown that similar critical properties appear when $q\geq 5$. The critical point separating the ferromagnetic and paramagnetic phases exhibits emergent $U(1)$ symmetry~\cite{Oshikawa2000prb}, accompanied by a dangerously irrelevant scaling variable (DISV), which is irrelevant at the critical point but becomes relevant in the ordered phase, in response to the discrete symmetry breaking. Additionally, a second divergent length scale $\xi'$ emerges with its critical exponent $\nu'$, characterizing the crossover from $U(1)$ to $Z_q$ symmetry breaking in the ordered phase, alongside the conventional correlation length $\xi$~\cite{Oshikawa2000prb,Lou2007prl,Okubo2015prb,Leonard2015prl,Pujari2015prb,Ding2016prb,Hasenbusch2011prb,Shao2020prl,Patil2021prb,Li2017nc,Jian2017prb,Jian2018prb,Senthil2004sci,Senthil2004prb,Nahum2015prl,Wang2017prx,Takahashi2020prr,Ma2019prl,Torres2018prb,Classen2017prb,Nelson1976prb,Miyashita1997jpsj,Shao2016Sci}. 
The relationship between these two length scales is encapsulated by the scaling law $\nu'=\nu(1+|y_q|/p)$, wherein $y_q$ is the scaling dimension of the DISV and $p=2$ (or $p=3$) for isotropic (or anisotropic) couplings~\cite{Oshikawa2000prb,Lou2007prl,Okubo2015prb,Leonard2015prl,Pujari2015prb,Ding2016prb,Hasenbusch2011prb,Shao2020prl,Patil2021prb}.
The interplay between $\xi$ and $\xi'$~\cite{Oshikawa2000prb,Lou2007prl,Okubo2015prb,Leonard2015prl,Pujari2015prb,Ding2016prb,Hasenbusch2011prb,Shao2020prl,Patil2021prb,Li2017nc,Jian2017prb,Jian2018prb,Senthil2004sci,Senthil2004prb,Nahum2015prl,Wang2017prx,Takahashi2020prr,Ma2019prl,Torres2018prb,Classen2017prb,Nelson1976prb,Amit1982annph,Miyashita1997jpsj,Shao2016Sci} gives rise to exotic equilibrium scaling behaviors,
highlighting the richness of phase transitions in systems with emergent symmetries.

Given the unique universal equilibrium critical properties and the novel experimental results in RMnO$_3$, it is immensely desired to systematically investigate the nonequilibrium properties in critical point with emergent symmetry, and particularly, to establish a general theoretical framework to describe the driven critical dynamics in the presence of DISV. 
In prior work, we studied the driven dynamics of the $Z_6$-clock model with isotropic ferromagnetic couplings and found that the angular order parameter exhibits a remarkable piecewise scaling behaviors for different driving rates~\cite{Shu2023kz}. However, the universality of these dynamic phenomena remains unclear.

To further uncover the nonequilibrium critical dynamics with emergent symmetry, in this paper, we investigate the driven dynamics in the $Z_5$ clock model with both isotropic and anisotropic couplings. In order to capture the effects induced by the two typical length scales, we consider the heating dynamics starting from the ordered phase, since the DISV only resides in the ordered side and heating with a finite velocity can bring universal features contributed by two length scales into the critical region.

For a given driving velocity $v$, we identify two FTS regions as illustrated in Fig.~\ref{fig:quench}. FTS Region I is characterized by the usual driving-induced time scale $\zeta_d\sim v^{-z/r}$, while FTS Region II is characterized by an additional time scale $\zeta_d'\sim v^{-z/r'}$ with $r'=z+1/\nu'$.

We find that the square of the order parameter $M^2$ conforms to the usual FTS form and is only affected by the time scale $\zeta_d$. In contrast, the angular order parameter $\phi_q$ displays distinct scaling properties dictated by both FTS regions. For small $v$, $\phi_q$ is governed by $\zeta_d$, while for large $v$, $\phi_q$ is dominated by $\zeta_d'$. Additionally, we show that for small $v$, the distance to the critical point $g$ is rescaled as $gL^{1/\nu}$; while for large $v$, $g$ is rescaled as $gL^{1/\nu'}$. These results are consistent with those found in $q=6$ case~\cite{Shu2023kz}, thereby confirming the universality of piecewise FTS form in driven dynamics with emergent symmetry. Our findings bring new vitality into experimental investigations in hexagonal RMnO$_3$ materials~\cite{Chae2012prl,Griffin2012prx,Lin2014natphy,Meier2017prx,Skjaervo2019prx,Meier2020prb,Zhang2021prb,OWSandvik2023nl,Kang2023jap,Baghizadeh2019jpcc,Juraschek2020prl}, and potentially exert influence on the nonequilibrium dynamics of supersolid quantum phase transitions in materials such as Na$_2$BaCo(PO$_4$)$_2$~\cite{Xiang2024nat,Chi2024,Gao2024}.

The rest of paper is organized as follows. In Sec.~\ref{model}, we introduce the clock model studied and the numerical method employed.
In Sec.~\ref{generalt}, we develop a general FTS theory to describe the driven dynamics in the presence of the DISV. In Sec.~\ref{orderpara}, we show that the dynamic scaling of $M^2$ satisfies the usual FTS form. In Sec.~\ref{aorderpara}, we explore the driven dynamics of the angular order parameter $\phi_q$ and find that its scaling behaviors satisfy the piecewise FTS form governed by two characteristic time scales. A summary and discussion is given in Sec.~\ref{sum}.

\section{\label{model}Model and Method}
The Hamiltonian of the ``hard'' $Z_q$-clock model in $3$D cubic lattice is~\cite{Oshikawa2000prb,Lou2007prl,Okubo2015prb,Leonard2015prl,Pujari2015prb,Ding2016prb,Hasenbusch2011prb,Shao2020prl,Patil2021prb}
\begin{equation}
\label{eq:hamiltonian}
H=-J_{\parallel}\sum_{\langle ij\rangle_{\parallel}}\cos(\theta_i-\theta_j)-J_{\perp}\sum_{\langle ij\rangle_{\perp}}\cos(\theta_i-\theta_j),
\end{equation}
in which $\theta=2n\pi/q$ with $n\in \{0,\cdots,q-1\}$, $\parallel$ ($\perp$) denotes the dimension parallel (perpendicular) to the plane supporting $\theta$ angle, and the summation is taken among nearest pairs denoted by $\langle\rangle$. We parameterize $J_{\parallel}$ and $J_{\perp}$ using a single parameter $\lambda$ as 
\begin{equation}
\label{JJ}
J_{\parallel}=1-\lambda,~~~J_{\perp}=1+\lambda,
\end{equation}
with $\lambda\in[0,1)$ measuring the strength of anisotropy. For $\lambda=0$, the isotropic case is recovered. Model~(\ref{eq:hamiltonian}) has a ``soft'' version wherein the angles vary continuously within $[0,2\pi)$, with an additional discrete symmetric term $-h\sum_i \cos(q \theta_i)$~\cite{Shao2020prl}. 
Here we only consider the ``hard'' version.

For $q=2$ and $3$, model~(\ref{eq:hamiltonian}) reduces to the Ising and $3$-state Potts models, respectively. For $q=4$, it is equivalent to two copies of the Ising model. In $2$D, for $q\geq5$, there is an intermediate $U(1)$ symmetric phase spanning a finite phase region between the low-temperature ordered phase with the spontaneously broken $Z_q$ symmetry and the high-temperature disordered phase~\cite{Nelson1983book,xiang2020pre}.

In contrast, in $3$D, model~(\ref{eq:hamiltonian}) for $q\geq 5$ undergoes a direct continuous phase transition from the ordered phase to the disordered phase at $g\equiv T-T_c=0$, as illustrated in Fig.~\ref{fig:quench}. The critical point exhibits an emergent $U(1)$ symmetry and the phase transition belongs to the $3$D XY universality class~\cite{Oshikawa2000prb,Lou2007prl,Okubo2015prb,Leonard2015prl,Pujari2015prb,Ding2016prb,Hasenbusch2011prb,Shao2020prl,Patil2021prb}. However, when $g<0$ the ordered phase breaks $Z_q$ symmetry. To account for the discrete symmetry breaking from the emergent symmetry, the DISV with negative scaling dimension $y_q$ is identified, which corresponds to the field term $h$ for the soft $Z_q$ clock model but is implicitly included in the hard version~\cite{Shao2020prl,Patil2021prb}. While the DISV is irrelevant at the critical point, it becomes relevant in the ordered phase, corresponding to a discrete symmetry breaking fixed point of the renormalization group flow~\cite{Okubo2015prb,Leonard2015prl,Shao2020prl,Patil2021prb}.

The presence of DISV introduces a new typical length scale $\xi'$, characterized by a new critical exponent $\nu'$, in addition to the usual correlation length $\xi$ with critical exponent $\nu$~\cite{Ueno1991prb,Chubukov1994prb,Oshikawa2000prb,Leonard2015prl,Banerjee2018prl}. Beyond $\xi'$, the fixed point corresponding to discrete symmetry breaking becomes dominant. It was shown that $\nu'$ increases with $q$ and satisfies the relation $\nu'=\nu(1+|y_q|/p)$, where $p=2$ for $\lambda=0$ and $p=3$ for $\lambda\neq 0$ ~\cite{Lou2007prl,Okubo2015prb,Leonard2015prl,Pujari2015prb,Ding2016prb,Hasenbusch2011prb,Shao2020prl,Patil2021prb,Ueno1991prb,Chubukov1994prb,Banerjee2018prl}.

The interplay between $\xi$ and $\xi'$ contributes intriguing scaling behaviors. For the squared order parameter $M^2$, defined as $M^2=M_x^2+M_y^2$, it was shown that $\xi$ dominates its scaling behavior near the critical point~\cite{Lou2007prl}. In contrast, the angular order parameter $\phi_q$ defined as $\phi_q=\langle\cos(q \Theta)\rangle$ with $\Theta\equiv \arccos(M_x/M)$ (where $\langle\rangle$ stands for statistical average), exhibits remarkably distinct scaling properties. Specifically, $\phi_q$ captures information about the transverse fluctuations with scale $\xi'$~\cite{Lou2007prl,Okubo2015prb,Pujari2015prb,Shao2020prl,Patil2021prb}.
Recent numerical studies combined with the finite-size scaling analysis reveal that for small $g$, $\phi_q$ is controlled by $\nu$ and $\xi$; while for large $g$, it is controlled by $\nu'$ and $\xi'$~\cite{Shao2020prl}.

Here we study the driven dynamics across the critical point from an ordered initial state, as shown in Fig.~\ref{fig:quench}. The nonequilibrium evolution is simulated using Monte Carlo method with standard Metropolis dynamics~\cite{binderbook}. 
To implement the heating dynamics, we first generate an equilibrium state with the same set of parameters ($q$ and $\lambda$) at a temperature $T_0<T_c$, and then gradually increase the temperature with a given velocity as $T=T_0+vt$ until the target temperautre is reached. The total simulation time is determined by the initial temperature, target temperature and driving velocity, with the time unit defined as a full Monte Carlo sweep through the lattice.
It has been established that the Metropolis Monte Carlo dynamics falls within the Model A universality class~\cite{Hohenberg1977rmp,Folk2006jpa,Tauber2014book}, and is straightforward to implement in experiments~\cite{Chae2012prl,Skjaervo2019prx,Griffin2012prx,Lin2014natphy,Meier2017prx}.

\section{\label{generalt}General scaling theory}

In usual critical points, FTS demonstrates that during linear heating $g=vt$ from ordered phase, the external driving rate introduces a typical time scale $\zeta_d\sim v^{-z/r}$~\cite{Huang2014prb,feng2016prb}. Near the critical point, $\zeta_d$ dominates the critical dynamics. For a quantity $P$ with a scaling dimension $\kappa$, the driven dynamics is described by the scaling form form~\cite{Gong2010njp,Zhongf2005prb,Zhong2006pre,Fan2007pre,Huangxz2010pre,Yin2014prb,Huang2014prb,feng2016prb,Fan2009pre}
\begin{equation}
\label{generalp}
P(g,L,v)=L^{-\kappa}f_P(gL^{1/\nu},vL^{r}),
\end{equation}
in which $L$ is the system size and $f_P$ is a scaling function.

For critical points with emergent symmetry, the critical behaviors are characterized by both $\xi$ and $\xi'$. Naturally, there are two corresponding equilibrium time scales $\zeta\propto \xi^z$ and $\zeta'\propto \xi'^z$~\cite{Hohenberg1977rmp,Folk2006jpa,Tauber2014book}. The former represents the usual correlation time scale, while the latter is the time scale beyond which the discrete fixed point controls the dynamics. Similarly, when the system is driven by changing $g$ with velocity $v$, the external driving induces two time scales $\zeta_d\sim v^{-z/r}$ and $\zeta_d'\sim v^{-z/r'}$ with $r'=z+1/\nu'$. Each of them can contribute a FTS region. As shown in Fig.~\ref{fig:quench}, we denote the scaling region controlled by $\zeta_d$ as FTS Region I while the scaling region controlled by $\zeta_d'$ as FTS Region II. Since $\nu'>\nu$~\cite{Lou2007prl,Okubo2015prb,Leonard2015prl,Pujari2015prb,Ding2016prb,Hasenbusch2011prb,Shao2020prl,Patil2021prb,Ueno1991prb,Chubukov1994prb,Banerjee2018prl}, and consequently $r'>r$, the time scale $\zeta_d'$ is larger than $\zeta_d$, as illustrated in Fig.~\ref{fig:quench}.

In the heating process, affected by the interplay between these two FTS regions, the dynamics of $P$ should satisfy the general scaling form:
\begin{equation}
\label{generalp1}
P(g,L,v)=L^{-\kappa}f_{P1}(gL^{1/\nu},gL^{1/\nu'};vL^{r},vL^{r'}),
\end{equation}
where $f_{P1}$ is another scaling function. In the absence of DISV, Eq.~(\ref{generalp1}) retreats to the usual FTS form Eq.~(\ref{generalp}). Without external driving, Eq.~(\ref{generalp1}) recovers the equilibrium finite-size scaling in the presence of the DISV~\cite{Shao2020prl}.

\section{\label{orderpara}Dynamic scaling of the order parameter $M^2$}

\begin{figure}[tbp]
\centering
  \includegraphics[width=\linewidth,clip]{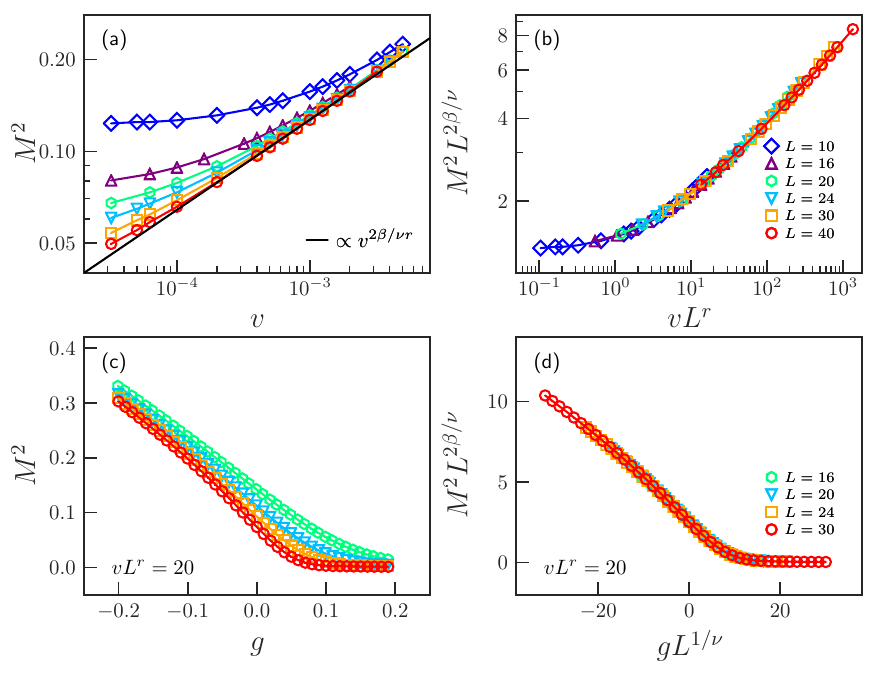}
  \vskip-3mm
  \caption{Heating critical dynamics of $M^2$ for isotropic coupling. Upper row: Results for the average squared magnetization after linear driving to $T_c=2.20502$ from the ordered phase at $T_0=0.20502$. Curves of $M^2$ as a function of $v$ for different system sizes before (a) and after (b) rescaling. The solid line in (a) indicates the expected power-law behavior, $M^2\propto v^{2\beta/\nu r}$, with $3$D XY exponents. Lower row: For fixed $vL^r=20$ (for instance, $v=5.84\times 10^{-3}$ for $L=16$), curves of $M^2$ versus the distance to the critical point $g=T-T_c$ before (c) and after (d) rescaling, verifying the dynamic scaling behavior near the critical point.
  }
  \label{fig:op}
\end{figure}

\begin{figure}[tbp]
\centering
  \includegraphics[width=\linewidth,clip]{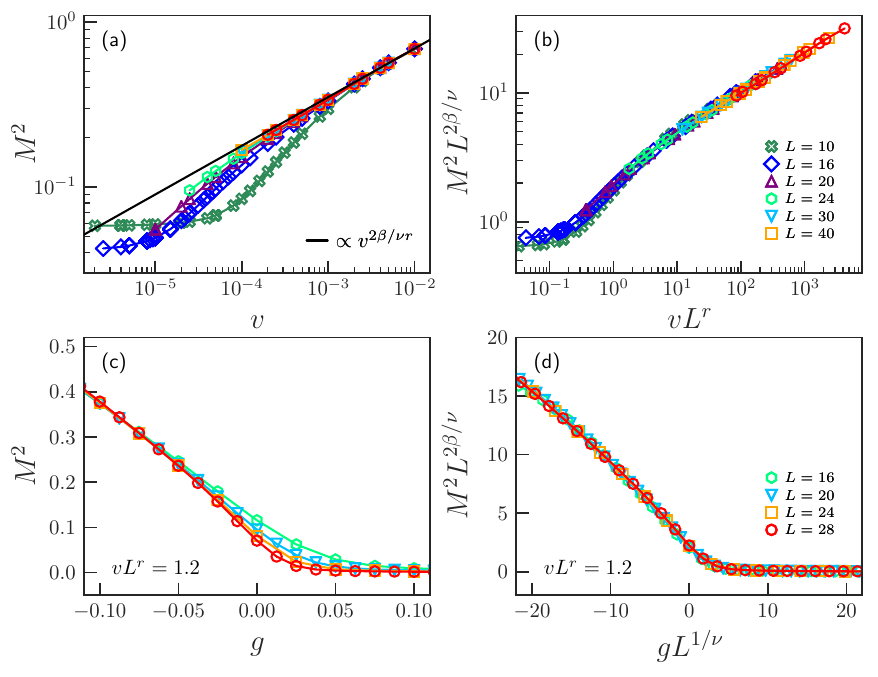}
  \vskip-3mm
  \caption{Heating critical dynamics of $M^2$ for anisotropic coupling with $\lambda=0.9$. Upper row: Results for the average squared magnetization after linear driving to $T_c=0.9184$ from the ordered phase at $T_0=0.4184$. Curves of $M^2$ as a function of $v$ for different system sizes before (a) and after (b) rescaling. The solid line in (a) indicates the expected power-law behavior, $M^2\propto v^{2\beta/\nu r}$, with $3$D XY exponents. Lower row: For fixed $vL^r=1.2$ (for instance, $v=7.06\times 10^{-5}$ when $L=16$), curves of $M^2$ versus the distance to the critical point $g$ before (c) and after (d) rescaling.
  }
  \label{fig:op1}
\end{figure}

In this section, we explore the driven dynamics of the square of the order parameter $M^2$. We focus on the case of $q=5$ in Eq.~(\ref{eq:hamiltonian}). We first investigate the scaling of $M^2$ for the isotropic case with $\lambda=0$. At the critical point $T_c=2.20502$~\cite{Shao2020prl}, Fig.~\ref{fig:op} (a) shows that for large $v$, the finite-size effects become negligible and $M^2\propto v^{2\beta/\nu r}$, where $\beta=0.3486(1)$, $\nu=0.6717(1)$~\cite{Campostrini2006prb,Campostrini2001prb,Chester2020jhep}, and $r=3.5134$ using $z=2.0246(10)$~\cite{Adzhemyan2022pa} as input. These critical exponents belong to the $3$D XY universality class. For small $v$, $M^2$ approaches the equilibrium value as $M^2\propto L^{-2\beta/\nu}$. Combining these two limit cases gives the scaling form of $M^2$ at $g=0$~\cite{Gong2010njp,Zhongf2005prb,Zhong2006pre,Fan2007pre,Huangxz2010pre,Yin2014prb,Huang2014prb,feng2016prb,Fan2009pre}:
\begin{equation}
\label{op1}
M^2(L,v)=L^{-2\beta/\nu}f_{M}(vL^{r}),
\end{equation}
where $f_{M}$ is the scaling function. For small $v$, $f_M$ tends to a constant, yielding $M^2(L,v)\propto L^{-2\beta/\nu}$; while for large $v$, $f_M(vL^{r})$ develops a power law as $f_M(vL^{r})\sim (vL^r)^{2\beta/\nu r}$, leading to $M^2\propto v^{2\beta/\nu r}$, consistent with the results shown in Fig.~\ref{fig:op} (a). Rescaling $M^2$ and $v$ as $M^2 L^{2\beta/\nu}$ and $vL^r$, respectively, we find that the rescaled curves collapse well, as shown in Fig.~\ref{fig:op} (b), confirming Eq.~(\ref{op1}).

We also study the dynamic scaling of $M^2$ during the driven process as shown in Fig.~\ref{fig:op} (c). For an arbitrary fixed $vL^r$, the rescaled curves of $M^2 L^{2\beta/\nu}$ versus $gL^{1/\nu}$ for different system sizes collapse well as shown in Fig.~\ref{fig:op} (d). This demonstrates that the full scaling form of $M^2$ is~\cite{Gong2010njp,Zhongf2005prb,Zhong2006pre,Fan2007pre,Huangxz2010pre,Yin2014prb,Huang2014prb,feng2016prb,Fan2009pre}
\begin{equation}
\label{op2}
M^2(g,L,v)=L^{-2\beta/\nu}f_{M1}(gL^{1/\nu},vL^{r}).
\end{equation}

For the anisotropic case of $\lambda=0.9$, similar behaviors of $M^2$ are observed. Figures~\ref{fig:op1} (a)-(b) confirm that the dynamics of $M^2$ at $g=0$ obeys Eq.~(\ref{op1}) and exhibits similar asymptotic behaviors to the $\lambda=0$ case for both large and small $v$. Moreover, Figs.~\ref{fig:op1} (c)-(d) demonstrate that the behaviors of $M^2$ in the driven process is also described by Eq.~(\ref{op2}). These results are also consistent with those for the $q=6$ case~\cite{Shu2023kz}. 

From these results, we conclude that the emergent symmetry can manifest in the driven process, since the driven dynamics of $M^2$ in the heating process is fully controlled by the FTS Region I with critical exponents of the $3$D XY universality class, whereas the FTS Region II with the length scale $\xi'$ and time scale $\zeta_d'$, does not contribute. A possible explanation is that $M^2$, describing the amplitude of the order parameter, is insensitive to the angular fluctuations, which are remarkably different for the Nambu-Goldstone fixed point of the $3$D XY model and the discrete symmetry breaking fixed point of the $Z_q$ clock model.

\section{\label{aorderpara}Dynamic scaling of the angular order parameter $\phi_q$}

To reveal the effects induced by two FTS regions, it is essential to consider the quantities that are sensitive to transverse fluctuations. In this section, we examine the dynamics of the angular order parameter $\phi_q$. Recent studies indicate that when $|g|$ is small, $\phi_q\propto L^{-|y_q|}$ where $y_q$ is the scaling dimension of the DISV, and the exponent for $g$ is $\nu$; while for large $|g|$, $\phi_q$ becomes dimensionless and the exponent for $g$ changes to $\nu'=\nu(1+|y_q|/p)$~\cite{Shao2020prl}. For the isotropic case, $p=2$~\cite{Shao2020prl}, whereas for the anisotropic case, $p=3$~\cite{Patil2021prb}. In the following, we extend these scaling analyses to the nonequilibrium case in Sec.~\ref{scalingana} and verify the scaling theory for the isotropic and anisotropic cases in Sec.~\ref{numerical1} and Sec.~\ref{numerical2}, respectively.

\begin{figure*}[htbp]
\centering
  \includegraphics[width=\linewidth,clip]{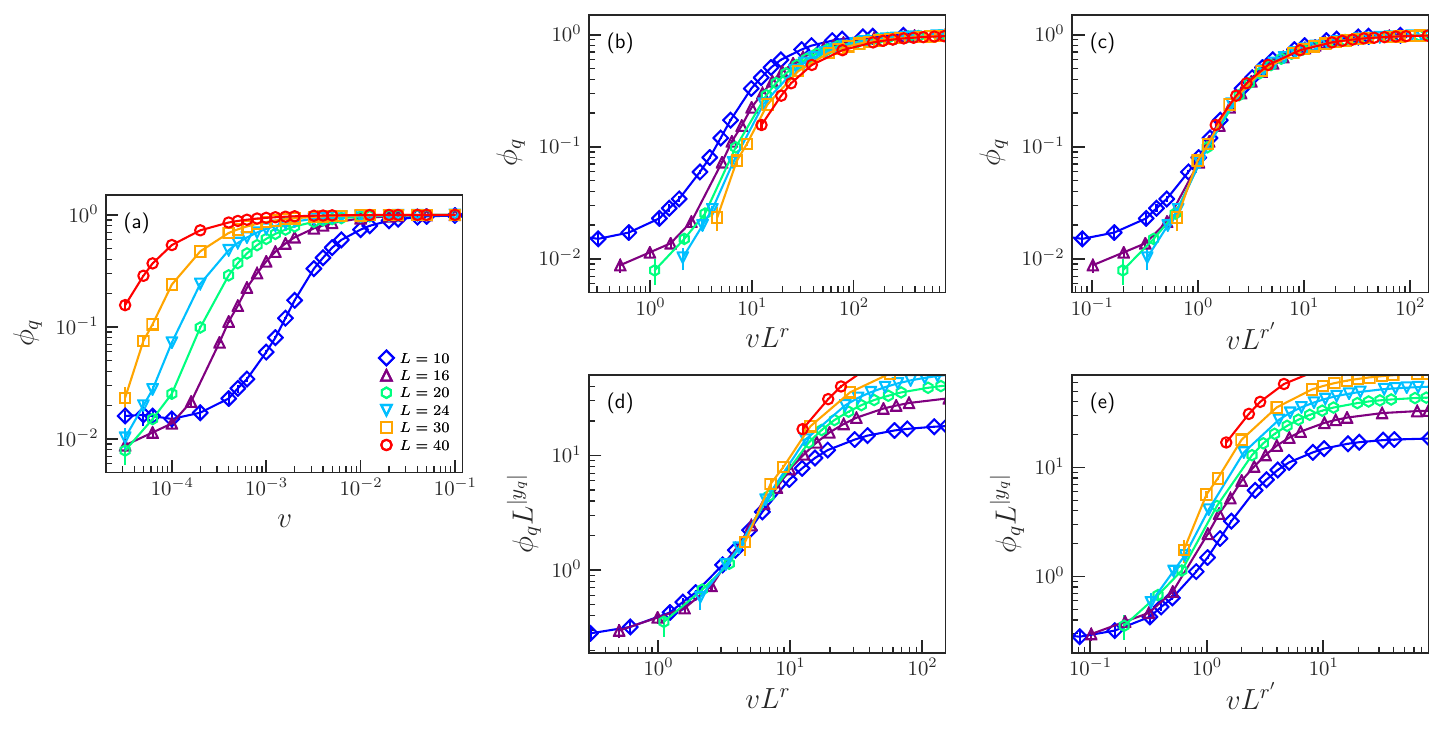}
  \vskip-3mm
  \caption{Dynamic scaling of the angular order parameter $\phi_q$ at the critical point $T_c=2.20502$ of the isotropic clock model. (a) Dependence of $\phi_q$ on $v$ for system of size $L = 10$ to $40$, starting from the ordered state at $T_0=0.20502$. (b) Rescaling $v$ as $vL^r$ does not lead to data collapse for $\phi_q<1$. (c) Rescaling $v$ as $vL^{r'}$ results in good data collapse when $\phi_q$ is close to $1$. (d) Closer to equilibrium, $\phi_q$ should have a scaling dimension of $y_q$ and rescaling with $L^{|y_q|}$ indeed leads to data collapse for small values of $\phi_qL^{|y_q|}$ vs $vL^{r}$. (e) In contrast, there is no data collapse for the curves of $\phi_q L^{|y_q|}$ vs $vL^{r'}$ in nonequilibrium region.
  }
  \label{fig:phi1}
\end{figure*}

\subsection{\label{scalingana}Scaling analyses}

We first focus on the scaling of $\phi_q$ at the critical point. According to Eq.~(\ref{generalp1}), in the presence of the DISV, the scaling form of $\phi_q$ should follow~\cite{Shu2023kz}
\begin{equation}
\label{sphi1}
\phi_q(L,v)=L^{-|y_q|}f_{\phi1}(vL^r,vL^{r'}),
\end{equation}
in which, for small $v$, $\phi_q$ should approach its equilibrium value and obey the scaling relation of $\phi_q\propto L^{-|y_q|}$. Consequently, $f_{\phi1}$ should be analytic in both $vL^r$ and $vL^{r'}$. Since $r>r'$, the term $vL^r$ dominates for small $v$, leading to
\begin{equation}
\label{sphi2}
\phi_q(L,v)\approx L^{-|y_q|}f_{\phi2}(vL^r),
\end{equation}
which suggests that, for small $v$, $\phi_q$ obeys the dynamic scaling controlled by the FTS Region I.

As $v$ increases, the value of $\phi_q$ also grows. As learned from the equilibrium, for large $\phi_q$, $\xi'$ and $\nu'$ start to play a role in affecting the scaling behaviors~\cite{Shao2020prl}. Therefore, in the driven dynamics, for large $v$, the term $vL^{r'}$ must also be taken into account. In this case, $f_{\phi1}$ in Eq.~(\ref{sphi1}) develops a power-law dependence on $vL^r$, resulting in~\cite{Shu2023kz}
\begin{equation}
\label{sphi3}
\phi_q(L,v)\approx L^{-|y_q|}(vL^r)^af_{\phi3}(vL^{r'}),
\end{equation}
where $a$ is a crossover exponent to be determined later.

As the increase of $v$ continues, $\phi_q$ saturates at $1$, becoming completely controlled by the fixed point of discrete symmetry breaking. At this stage, the equilibrium scaling indicates that $\nu'$ dominates~\cite{Shao2020prl}, yielding the dynamic scaling form
\begin{equation}
\label{sphi4}
\phi_q(L,v)\approx f_{\phi4}(vL^{r'}),
\end{equation}
in which $\phi_q$ becomes dimensionless and $f_{\phi4}(vL^{r'})\rightarrow 1$ for large $v$, indicating that $\phi_q$ obeys the dynamic scaling in the FTS Region II for fast driving.

Notably, Eq.~(\ref{sphi3}) should analytically crossover into Eq.~(\ref{sphi4}). This condition imposes the requirement that $f_{\phi3}(vL^{r'})$ satisfies
\begin{equation}
\label{sphi5}
f_{\phi3}(vL^{r'})=(vL^{r'})^b f_{\phi5}(vL^{r'}),
\end{equation}
in which $b=-a$ since the coefficient of the scaling function $f_{\phi4}$ is constant and independent of $v$, and $f_{\phi5}$ is another scaling function, such that $f_{\phi5}=cf_{\phi4}$ ( where $c$ is a constant) for large $v$. To eliminate the dependence on $L$ in the coefficient preceding $f_{\phi3}$ of Eq.~(\ref{sphi3}), the condition $-|y_q|+ar-ar'=0$ must hold, leading to~\cite{Shu2023kz}
\begin{equation}
\label{sphi6}
a=\frac{|y_q|}{r-r'}=\frac{|y_q|}{1/\nu-1/\nu'}.
\end{equation}

Physically, these scaling analyses can be interpreted as follows: When an external driving with velocity $v$ is applied to a system starting from an ordered initial state, the evolution is initially adiabatic. When the driving time scale $\zeta'_d\sim v^{-z/r'}$ is shorter than $\zeta'\sim \xi'^z\sim |g|^{-\nu' z}$, the system first enters the FTS region II, where $\zeta'_d$ dominates. As the driving continues, when $\zeta\sim \xi^z\sim |g|^{-\nu z}$ exceeds $\zeta_d\sim v^{-z/r}$, the system enters the FTS Region I governed by $\zeta_d$.

However, for the dynamic crossover from Region II to Region I, due to the critical slowing down,
the memory effects should be considered.
When driven into Region I from Region II, the system retains memories of the dynamic features possessed by Region II but these memories decay slowly, while $\zeta_d$ gradually takes control of the evolution. 
For small $v$, the driving time is sufficient for $\zeta_d$ to fully establish its dominance, ensuring that the critical behavior at the critical point is dictated by $\zeta_d$. Conversely, when driving with a large $v$, the system can reach the critical point before $\zeta_d$ completely governs the dynamics. In such cases, both $\zeta_d$ and $\zeta_d'$ contribute to shaping the critical behaviors. In the extreme scenario of very large $v$, the memory of Region II can penetrate into Region I near the critical point, and therefore the scaling behaviors are governed primarily by $\zeta_d'$.

\begin{figure}[!htbp]
\centering
  \includegraphics[width=\linewidth,clip]{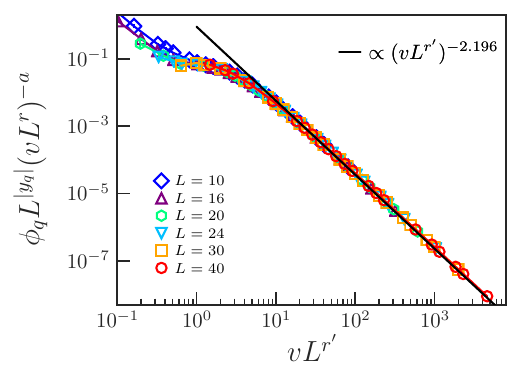}
  \vskip-3mm
  \caption{Crossover scaling property of $\phi_q$ from small-$v$ to large-$v$ ranges for $\lambda=0$. After rescaling $v$ and $\phi_q$ as $vL^{r'}$ and $\phi_qL^{|y_q|}(vL^r)^{-a}$ with the predicted exponent $a=2.196$ and with the other exponents set to their known values for the $q=5$ clock model, the data for different system sizes and velocities show good collapse when $v$ is not small. Deviations appear in the small-$v$ region. The solid line indicates a power law of $(vL^{r'})^{-a}$ with the predicted exponent $a=2.196$.
  }
  \label{fig:phi2}
\end{figure}

Besides $g=0$, these scaling analyses also apply to the driven process near the critical point. For small $v$, the dynamics near the critical point is controlled by the FTS Region I, described by
\begin{equation}
\label{sphi7}
\phi_q(g,L,v)=L^{-|y_q|}f_{\phi5}(gL^{1/\nu},vL^{r}).
\end{equation}
In contrast, for large $v$, the dynamic scaling near the critical point is governed by the FTS Region II with the corresponding scaling form being
\begin{equation}
\label{sphi8}
\phi_q(g,L,v)=f_{\phi_6}(gL^{1/\nu'},vL^{r'}).
\end{equation}
Note that in Eqs.~(\ref{sphi7}) and (\ref{sphi8}), the dimensions of $g$ and $v$ should be consistent subjecting to the constraint $g=vt$.

\subsection{\label{numerical1} Iostropic case $\lambda=0$}

In this section, we numerically verify the above scaling analyses of the dynamics of $\phi_q$ for the $5$-state clock model with isotropic coupling. For $q=5$ and $\lambda=0$, the critical exponents are $\nu'=1.0982$~\cite{Shao2020prl} and $r'=2.9352$.

We first examine the dependence of $\phi_q$ on the driving rate $v$ for various $L$ at $g=0$, as shown in Fig.~\ref{fig:phi1} (a). It is evident that $\phi_q$ increases as $v$ grows and $\phi_q\rightarrow 1$ for large $v$. From Fig.~\ref{fig:phi1} (b), one finds that the curves do not achieve a good collapse when $v$ is rescaled as $vL^{r}$; while Fig.~\ref{fig:phi1} (c) demonstrates a successful collapse in the large-$v$ region when $v$ is rescaled as $vL^{r'}$. These results confirm the scaling form of Eq.~(\ref{sphi4}), verifying that for large $v$, the scaling behavior of $\phi_q$ at $g=0$ is controlled by FTS Region II.  However, for small $v$, clear deviations from the curves in Fig.~\ref{fig:phi1} (c) are observed, indicating that Eq.~(\ref{sphi4}) is not valid in this velocity range.
For small $v$ when rescaling $v$ and $\phi_q$ as $vL^{r}$ and $\phi_q L^{|y_q|}$, respectively, the curves collapse in the small-$v$ range and becoming almost flat as $v\rightarrow 0$ as shown in Fig.~\ref{fig:phi1} (d). As a comparison, Fig.~\ref{fig:phi1}(e) shows that the rescaled curves of $\phi_q L^{|y_q|}$ versus $vL^{r'}$ cannot collapse. This indicates that in this velocity range, $v$ should be rescaled as $vL^{r}$ rather than $vL^{r'}$ and $\phi_q$ approaches its equilibrium value and acquires a scaling dimension of ${y_q}$. These findings validate Eq.~(\ref{sphi2}), indicating that the scaling behavior of $\phi_q$ at $g=0$ is controlled by the FTS Region I for small $v$.

\begin{figure}[!htbp]
\centering
  \includegraphics[width=\linewidth,clip]{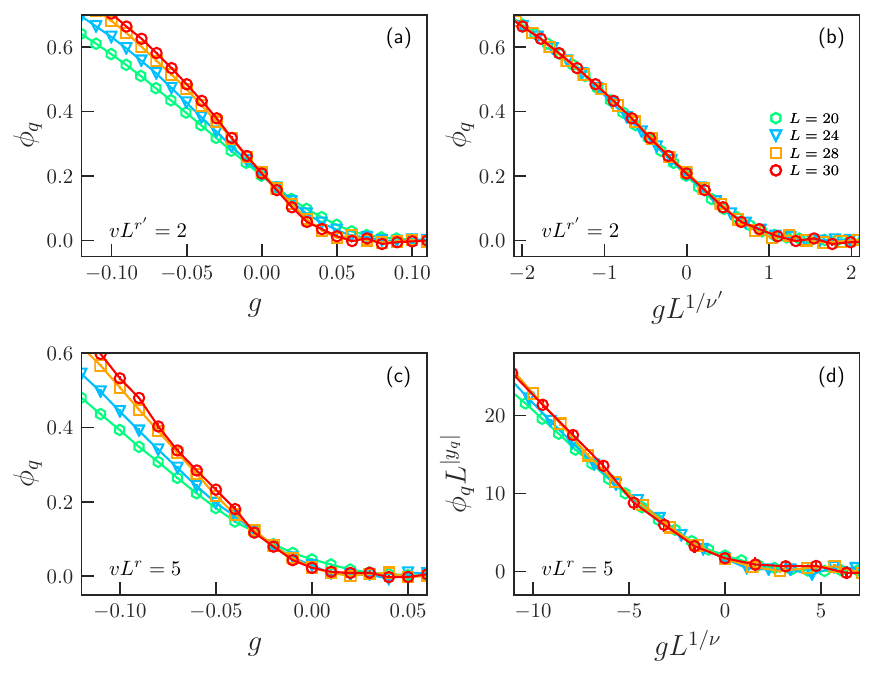}
  \vskip-3mm
  \caption{Dynamic scaling of the angular order parameter $\phi_q$ during the heating process from the ordered state for the $\lambda=0$ case. (a) Dependence of $\phi_q$ on $g$ for system of size $L = 20$ to $30$, starting from the ordered state at $T_0=0.20502$, with fixed $vL^{r'}=2$ (for instance, $v=3.04\times 10^{-4}$ for $L=20$) in the large-$v$ region. (b) Rescaling $v$ as $gL^{1/\nu'}$ leads to good data collapse, confirming that near the critical point the large-$v$ region is governed by the FTS Region II. (c) Dependence of $\phi_q$ on $g$ for the same set of system size and starting state in (a), with $vL^{r}=5$ (for instance, $v=1.34\times 10^{-4}$ when $L=20$) in the small-$v$ region. Near equilibrium, $\phi_q$ should have a scaling dimension of $y_q$ and rescaling with $L^{|y_q|}$ indeed leads to data collapse for small values of $\phi_qL^{|y_q|}$ vs $gL^{1/\nu}$, verifying that the small-$v$ region is controlled by the FTS Region I near the critical point.
  }
  \label{fig:phi3}
\end{figure}

We further explore the crossover behavior between scaling regions for small and large $v$. As discussed in Sec.~\ref{scalingana}, Eq.~(\ref{sphi3}) serves as a bridge connecting the two regions described by Eq.~(\ref{sphi2}) and Eq.~(\ref{sphi4}). 
As shown in Fig.~\ref{fig:phi2}, when rescaling $\phi_q$ and $v$ as $\phi_qL^{|y_q|}(vL^r)^{-a}$ and $vL^{r'}$, one finds good collapse for large $v$, while deviations emerge for small $v$, verifying Eq.~(\ref{sphi3}) and defining its range of applicability.
For very large $v$, the rescaled curves satisfy a power-law with an exponent of $-2.196$, which matches $-a$ of $q=5$, further validating Eq.~(\ref{sphi5}). The presence of a plateau in the small-$v$ region suggests that the influence of $\zeta_d'$ fades out as $v$ decreases. At very small $v$, the curves turn upward. This is not a signal of the revival of $\zeta_d'$, but instead indicates that this stage is beyond the applicable range of Eq.~(\ref{sphi3}). Alternatively, the scaling behavior here is described by Eq.~(\ref{sphi2}).

Previous investigations~\cite{Shu2023kz} have provided scaling results on the $q=6$ clock model. Although the value of $\nu'$ and $r'$ for $q=6$ are different from the current $q=5$ case, consistent results have been obtained, confirming that Eqs.~(\ref{sphi2})-(\ref{sphi7}) provide a universal framework describing of the driven critical dynamics of $\phi_q$.

Next we examine the off-critical effects in the driven process. For large $v$, by fixing $vL^{r'}$ to an arbitrary constant, we consider the dependence of $\phi_q$ on $g$. When rescaling $g$ as $gL^{1/\nu'}$, we observe a good collapse of the rescaled curves in Figs.~\ref{fig:phi3} (a)-(b), verifying Eq.~(\ref{sphi8}). In addition, Fig.~\ref{fig:phi3} (a) show that curves of $\phi_q$ versus $g$ for different $L$ cross at $g=0$, confirming that for large $v$, $\phi_q$ is dimensionless. Moreover, Figs.~\ref{fig:phi3} (a)-(b) also show that Eq.~(\ref{sphi8}) even applies in the region for $T>T_c$, although in this region the DISV does not play a role in equilibrium. These results demonstrate that the external driving can bring the influence of DISV to much broader critical region. Thus, for large $v$, the FTS Region II governs the critical region around the critical point.

\begin{figure*}[!htbp]
\centering
  \includegraphics[width=\linewidth,clip]{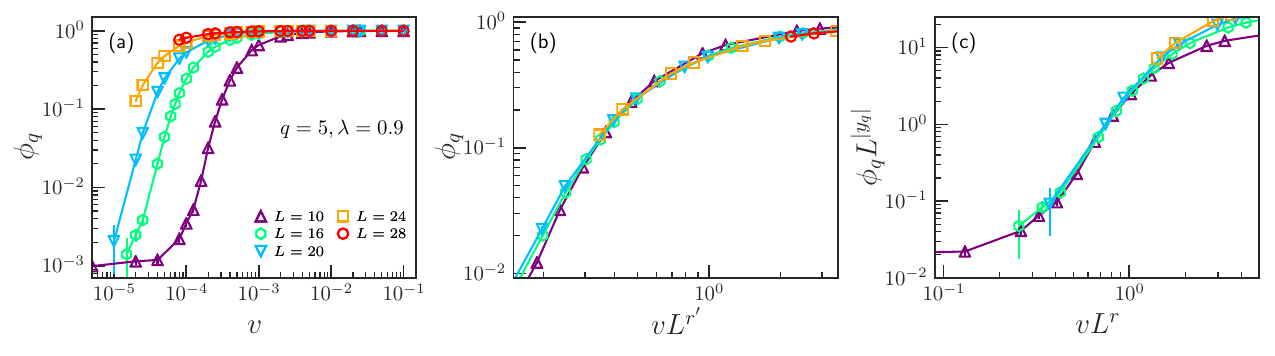}
  \vskip-3mm
  \caption{Dynamic scaling of the angular order parameter $\phi_q$ at the critical point $T_c=0.9184$ in heating dynamics from the ordered state for $\lambda=0.9$. (a) Dependence of $\phi_q$ on $v$ for system of size $L = 10$ to $28$, starting from the ordered state at $T_0=0.4184$. (b) Rescaling $v$ as $vL^{r'}$ leads to good data collapse when $\phi_q$ is close to $1$. (c) Near equilibrium, $\phi_q$ acquires a scaling dimension of $y_q$ and rescaling with $L^{|y_q|}$ indeed leads to data collapse for small values of $\phi_qL^{|y_q|}$ vs $vL^{r}$. 
  }
  \label{fig:phi4}
\end{figure*}

Similarly, for small $v$, the rescaled curves of $\phi_qL^{|y_q|}$ and $gL^{1/\nu}$ with a fixed $vL^{r}$ also collapse well, as shown in Figs.~\ref{fig:phi3} (c)-(d), confirming Eq.~(\ref{sphi7}). In addition, Fig.~\ref{fig:phi3} (c) shows that $\phi_q$ does not cross at $g=0$ for different $L$, demonstrating that the scaling dimension of $\phi_q$ is not zero in this region. Therefore, Figs.~\ref{fig:phi3} (c)-(d) confirm that for small $v$, the typical time scale switch to $\zeta_d$ and the FTS Region I becomes dominant near the critical point.

\begin{figure}[tbp]
\centering
  \includegraphics[width=\linewidth,clip]{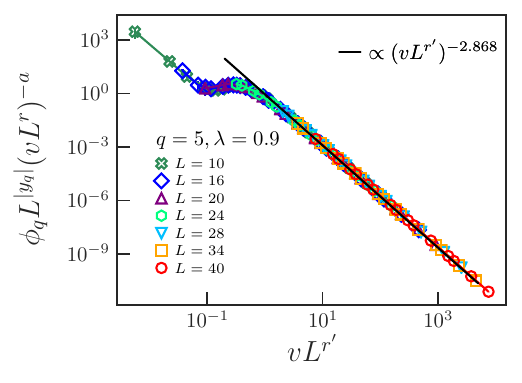}
  \vskip-3mm
  \caption{Crossover scaling property of $\phi_q$ from small-$v$ to large-$v$ ranges for $\lambda=0.9$. After rescaling $v$ and $\phi_q$ as $vL^{r'}$ and $\phi_qL^{|y_q|}(vL^r)^{-a}$ using the predicted exponent $a=2.868$ and with the other exponents set to their known values for the $q=5$ clock model, the $\phi_q$ data for different system sizes and velocities collapse well when $v$ is not small. Deviations appear for small-$v$ range. The solid line indicates a power law of $(vL^{r'})^{-a}$ with the predicted exponent $a=2.868$.
  }
  \label{fig:phi5}
\end{figure}

\begin{figure}[tbp]
\centering
  \includegraphics[width=\linewidth,clip]{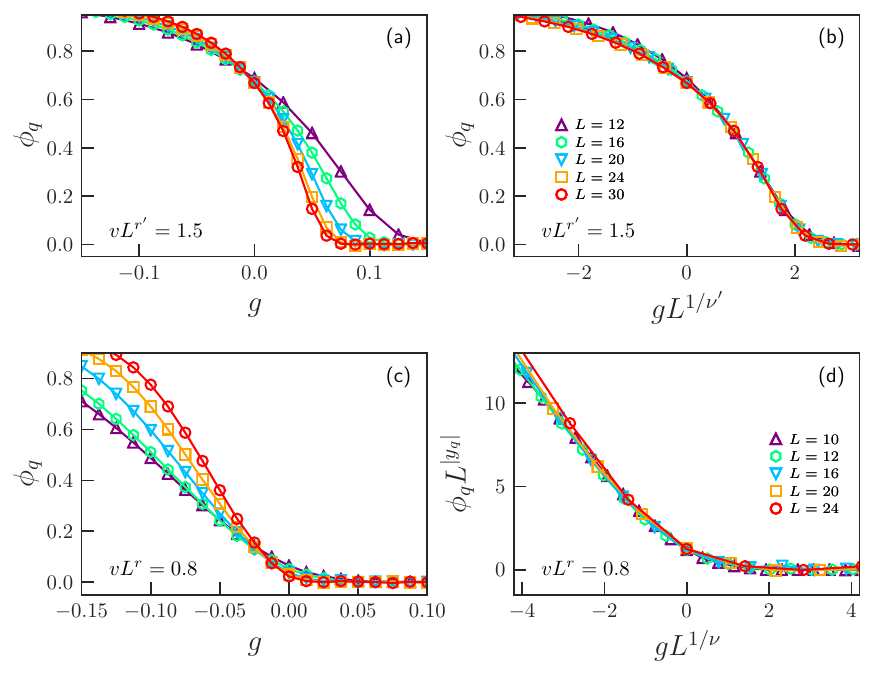}
  \vskip-3mm    
    \caption{Dynamic scaling of the angular order parameter $\phi_q$ during the heating process from the ordered state for the $\lambda=0.9$ case. (a) Dependence of $\phi_q$ on $g$ for system of size $L=12$ to $30$, starting from the ordered state at $T_0=0.4186$, with fixed $vL^{r'}=1.5$ (for instance, $v=7.29\times 10^{-4}$ when $L=12$) in the large-$v$ range. (b) Rescaling $v$ as $gL^{1/\nu'}$ leads to good data collapse, confirming that near the critical point the large-$v$ range is governed by the FTS Region II. (c) Dependence of $\phi_q$ on $g$ for system size from $L=10$ to $24$, with the same starting state in (a), with $vL^{r}=0.8$ (for instance, $v=1.29\times 10^{-4}$ when $L=12$) in the small-$v$ range. (d) Near equilibrium, $\phi_q$ should have a scaling dimension of $y_q$ and rescaling with $L^{|y_q|}$ indeed leads to data collapse for small values of $\phi_qL^{|y_q|}$ vs $gL^{1/\nu}$, verifying that the small-$v$ range is controlled by the FTS Region I near the critical point.
  }
  \label{fig:phi6}
\end{figure}

\subsection{\label{numerical2} Anisotropic case $\lambda\neq 0$}

Next, we show results of the anisotropic case to verify the universality of the scaling analyses discussed in Sec.~\ref{scalingana}. For $\lambda\neq0$, it was found that $p=3$ and $\nu'=0.9560$~\cite{Patil2021prb}, leading to $r'=3.0706$ with $z=2.2046$~\cite{Adzhemyan2022pa}. To make the anisotropy effect more pronounced, here we adopt a strong anisotropy parameter $\lambda=0.9$.

We first examine the dependence of $\phi_q$ on the velocity at the critical point $T_c=0.9184(2)$ starting from an equilibrium state at $T_0=0.4184$ (for the determination of $T_c$, see Appendix A). As shown in Fig.~\ref{fig:phi4} (a), similar to the isotropic case, $\phi_q$ grows as $v$ increases and approaches $1$ for large $v$. When $v$ is rescaled as $vL^{r'}$, the rescaled curves collapse well in the large-$v$ range, confirming that the behavior of $\phi_q$ at $g=0$ is controlled by the FTS Region II with the scaling form of Eq.~(\ref{sphi4}), as seen in Fig.~\ref{fig:phi4} (b). However, deviations appear in the small-$v$ range, suggesting that Eq.~(\ref{sphi4}) is not applicable in this velocity range. Instead, as shown in Fig.~\ref{fig:phi4} (c), the rescaled curves of $\phi_q L^{|y_q|}$ versus $vL^{r}$ collapse in the range with small $v$, demonstrating that the scaling behavior of $\phi_q$ at $g=0$ is controlled by the FTS Region I with Eq.~(\ref{sphi2}) for small $v$.

The crossover scaling form Eq.~(\ref{sphi3}) connecting FTS Region I and Region II is verified in Fig.~\ref{fig:phi5}. Here, $a=2.868$ is determined by substituting the corresponding values of $\nu$ and $\nu'$ into Eq.~(\ref{sphi6}). After rescaling $\phi_q$ and $v$ as $\phi_qL^{|y_q|}(vL^r)^{-a}$ and $vL^{r'}$, the rescaled curves collapse well for large $v$, as shown in Fig.~\ref{fig:phi5}. In a large range of velocity, the rescaled curves satisfy a power law with an exponent of $-2.868$, which is consistent with $-a$ for $q=5$ and $\lambda\neq 0$, thereby confirming Eq.~(\ref{sphi5}). Similar to the isotropic case, a plateau appears in the small-$v$ region, suggesting that the influence of $r'$ decays as $v$ decreases. For much smaller $v$, the curves tend to turn upward, entering the governing range of Eq.~(\ref{sphi2}). However, due to the moderately small velocity, the deviations similar to those in Fig.~\ref{fig:phi2} are not yet observed.

Off-critical effects in the driven process are also analyzed.
For large $v$, by fixing $vL^{r'}$ to an arbitrary constant, we investigate the dependence of $\phi_q$ on $g$. The rescaled curves for $\phi_q$ versus $gL^{1/\nu'}$ collapse well, as shown in Figs.~\ref{fig:phi6} (a)-(b), validating Eq.~(\ref{sphi8}). Moreover, Fig.~\ref{fig:phi6} (a) show that curves of $\phi_q$ versus $g$ for different $L$ almost cross at $g=0$, confirming that for large $v$, $\phi_q$ is a dimensionless variable. In contrast, for small $v$, Fig.~\ref{fig:phi6} (c) shows that $\phi_q$ at $g=0$ changes with $L$, demonstrating that the scaling dimension of $\phi_q$ is not zero in this region. Moreover, successful collapse is achieved for the rescaled curve of $\phi_qL^{|y_q|}$ and $gL^{1/\nu}$ with a fixed $vL^{r}$, as shown in Figs.~\ref{fig:phi6} (d), confirming Eq.~(\ref{sphi7}).

Given the these consistent results, we therefore establish the universality of the scaling analyses describing the driven dynamics of $\phi_q$ in Sec.~\ref{scalingana} for the models with both the isotropic and anisotropic cases.

\section{\label{sum}Summary and Discussion}

In this paper, we have investigated the heating critical dynamics of the $Z_q$ clock model with emergent $U(1)$ symmetry at the critical point. The $q=5$ models with both isotropic and anisotropic couplings are taken as examples. Two FTS regions have been identified, and our analyses have revealed that the square of the order parameter $M^2$ follows the usual FTS form, controlled by the typical time scale $\zeta_d$ in the FTS Region I. In contrast, the angular order parameter $\phi_q$  exhibits remarkable different scaling properties influenced by both FTS regions. For small $v$, $\phi_q$ is controlled by the time scale $\zeta_d$. But for large $v$, $\phi_q$ is controlled by a different time scale $\zeta_d'$ associated to the DISV. Combining with the results for $q=6$ reported previously, we have established a universal scaling theory for the driven critical dynamics in the critical points with emergent symmetry.

In the context of relaxation critical dynamics, a new dynamic exponent $z'$, slightly larger than the usual $z$~\cite{Shu2024prb}, has been found to govern the relaxation dynamics of $\phi_q$ in the long-time stage. Therefore, it cannot be excluded that there may also exist other time scales related to this new dynamics exponent $z'$. However, our study did not observe evidence of such time scales, likely because that the  corresponding scaling regions for these potential time scales are too narrow to be clearly resolved.

Our findings provide new ingredients to recent experiments on the nonequilibrium critical dynamics in the hexagonal RMnO$_3$ materials~\cite{Chae2012prl,Griffin2012prx,Lin2014natphy,Meier2017prx,Skjaervo2019prx,Meier2020prb,Zhang2021prb,OWSandvik2023nl,Kang2023jap,Baghizadeh2019jpcc,Juraschek2020prl}. 
Dynamic scaling behaviors affected by the DISV can be detected in the heating process of these systems. Although here we only focus on the heating dynamics, our results may bring new elements to the cooling dynamics. In particular, when considering the impact of phase ordering on the KZM, especially for large driving rate~\cite{Biroli2010prb}, it is expected that the DISV can play a significant role. This might offer another route to explain the scaling anomalies observed in experiments.

Furthermore, our results also shed light on the dynamics of quantum criticality with emergent symmetry. For instance, it is intuitive to extend our results to the phase transitions beyond the usual Landau paradigm, such as fermion-induced quantum criticality and deconfined quantum criticality, which both exhibit emergent symmetry at the critical point~\cite{Li2017nc,Jian2017prb,Jian2018prb,Torres2018prb,Classen2017prb,Senthil2004sci,Senthil2004prb,Nahum2015prl,Wang2017prx,Takahashi2020prr,Ma2019prl}. Spin-supersolid quantum phase transitions in materials such as Na$_2$BaCo(PO$_4$)$_2$ have been observed and the quantum phase transition therein also displays emergent $U(1)$ symmetry~\cite{Xiang2024nat}. Besides, the driven critical dynamics can be readily realized in fast-developing quantum annealers. Recently, the critical dynamics of quantum $Z_q$ clock model have been studied~\cite{Ali2024} and our findings may also be detected in these systems.

\section*{Acknowledgements}
The authors would like to thank Chengxiang Ding for helpful discussions. S.Y. is supported by the National Natural Science Foundation of China (Grants No. 12222515 and No. 12075324) and the Science and Technology Projects in Guangdong Province (Grants No. 2021QN02X561). Y.R.S. acknowledges support from the National Natural Science Foundation of China (Grant No. 12104109), the Science and Technology Projects in Guangzhou (Grant No. 2024A04J2092).

\appendix
\setcounter{equation}{0}
\renewcommand\theequation{A\arabic{equation}}
\setcounter{figure}{0}
\renewcommand\thefigure{A\arabic{figure}}

\section{Determination of $T_c$ for the anisotropic $Z_5$ model}

\begin{figure}[htbp]
\centering
  \includegraphics[width=\linewidth,clip]{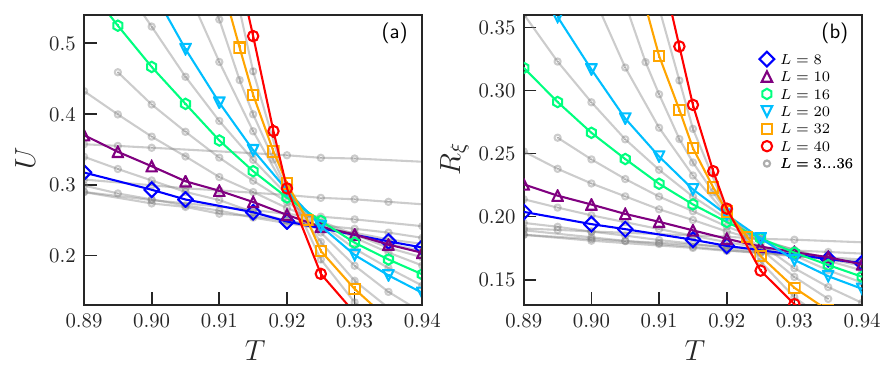}
  \vskip-3mm
  \caption{(a) Dependence of the Binder cumulant $U$ and the correlation length ratio $R_{\xi}$ on $T$ for different system sizes. The behaviors of $U$ and $R_{\xi}$ are similar. (b) The size-dependent critical temperature $T_c(L)$ is extracted using the crossing point in the results of size $L$ and $2L$.
  }
  \label{fig:aniur}
\end{figure}

\begin{figure}[htbp]
\centering
  \includegraphics[width=0.85\linewidth,clip]{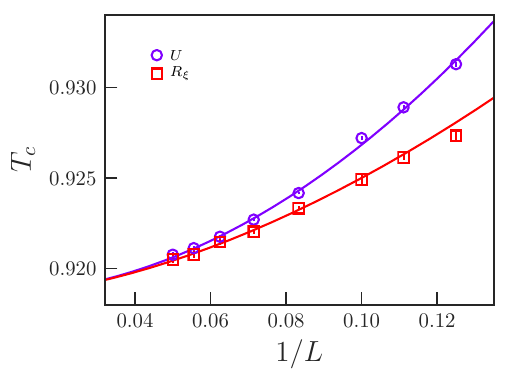}
  \vskip-3mm
  \caption{Extrapolation of the size-dependent critical temperature $T_c(L)$. The curves are fits to the form $T_c(L)=T_c+aL^{-\omega}$. Extrapolating to $L\rightarrow\infty$, both quantities yield consistent results of $T_c=0.9186(5)$ and $T_c=0.9184(2)$, for $U$ and $R_{\xi}$, respectively.
  }
  \label{fig:anitc}
\end{figure}

In order to extract the critical point for the $Z_5$ clock model with $\lambda=0.9$, we perform Monte Carlo simulations to obtain equilibrium results of the Binder cumulant, defined as
\begin{equation}
U=2-\frac{\langle M^4\rangle}{\langle M^2\rangle^2},
\label{eq:um}   
\end{equation}
which is dimensionless and follows 
\begin{equation}
U(g,L,v)=f_{U}(gL^{1/\nu},vL^{r})
\label{eq:umf}   
\end{equation}
Another quantity studied is the correlation length ratio $R_{\xi}=\xi/L$. The correlation length is given by
\begin{equation}
\xi=\frac{1}{q_1}\sqrt{\frac{S(0)}{S(q_1)}-1},
\label{eq:xi}   
\end{equation}
in which $S(q_1)$ is the structure factor at momentum $q_1$ defined as $S(q_1)=\langle M(q_1)M(-q_1) \rangle$, and $q_1=(2\pi/L,0,0)$ is one of the wave-vectors that are closest to the ordering momentum.  $R_\xi$ is also dimensionless and has a similar scaling form with Eq.~(\ref{eq:umf}).

Both $U$ and $R_{\xi}$ are dimensionless and invariant under renormalization group transformation in the critical limit. In equilibrium, the term $vL^{r}$ vanishes, so that $U$ for different system sizes cross at $g=0$ and can be employed in determining the critical point (so does $R_{\xi}$).
The dependence of $U$ and $R_{\xi}$ on different temperature is shown in Fig.~\ref{fig:aniur}.
Following the standard curve-crossing process, we extract the size-dependent critical temperature $T_c(L)$, as shown in Fig.~\ref{fig:anitc}. Extrapolating the power-law to the thermodynamic limit, the results obtained using $U$ and $R_{\xi}$ yield consistent result, giving $T_c=0.9186(5)$ and $T_c=0.9184(2)$, respectively. The errorbars are estimated by carrying out a large number of fits with Gaussian noise added to the data and compute the standard deviation of the distributions of the fitting parameters. For the calculations in the main text, we use the value $T_c=0.9184(2)$ since the errorbar is smaller.

\bibliography{refzq}

\begin{thebibliography}{109}%
\makeatletter
\providecommand \@ifxundefined [1]{%
 \@ifx{#1\undefined}
}%
\providecommand \@ifnum [1]{%
 \ifnum #1\expandafter \@firstoftwo
 \else \expandafter \@secondoftwo
 \fi
}%
\providecommand \@ifx [1]{%
 \ifx #1\expandafter \@firstoftwo
 \else \expandafter \@secondoftwo
 \fi
}%
\providecommand \natexlab [1]{#1}%
\providecommand \enquote  [1]{``#1''}%
\providecommand \bibnamefont  [1]{#1}%
\providecommand \bibfnamefont [1]{#1}%
\providecommand \citenamefont [1]{#1}%
\providecommand \href@noop [0]{\@secondoftwo}%
\providecommand \href [0]{\begingroup \@sanitize@url \@href}%
\providecommand \@href[1]{\@@startlink{#1}\@@href}%
\providecommand \@@href[1]{\endgroup#1\@@endlink}%
\providecommand \@sanitize@url [0]{\catcode `\\12\catcode `\$12\catcode
  `\&12\catcode `\#12\catcode `\^12\catcode `\_12\catcode `\%12\relax}%
\providecommand \@@startlink[1]{}%
\providecommand \@@endlink[0]{}%
\providecommand \url  [0]{\begingroup\@sanitize@url \@url }%
\providecommand \@url [1]{\endgroup\@href {#1}{\urlprefix }}%
\providecommand \urlprefix  [0]{URL }%
\providecommand \Eprint [0]{\href }%
\providecommand \doibase [0]{http://dx.doi.org/}%
\providecommand \selectlanguage [0]{\@gobble}%
\providecommand \bibinfo  [0]{\@secondoftwo}%
\providecommand \bibfield  [0]{\@secondoftwo}%
\providecommand \translation [1]{[#1]}%
\providecommand \BibitemOpen [0]{}%
\providecommand \bibitemStop [0]{}%
\providecommand \bibitemNoStop [0]{.\EOS\space}%
\providecommand \EOS [0]{\spacefactor3000\relax}%
\providecommand \BibitemShut  [1]{\csname bibitem#1\endcsname}%
\let\auto@bib@innerbib\@empty
\bibitem [{\citenamefont {Hohenberg}\ and\ \citenamefont
  {Halperin}(1977)}]{Hohenberg1977rmp}%
  \BibitemOpen
  \bibfield  {author} {\bibinfo {author} {\bibfnamefont {P.~C.}\ \bibnamefont
  {Hohenberg}}\ and\ \bibinfo {author} {\bibfnamefont {B.~I.}\ \bibnamefont
  {Halperin}},\ }\bibfield  {title} {\enquote {\bibinfo {title} {{Theory of
  dynamic critical phenomena}},}\ }\href {\doibase 10.1103/RevModPhys.49.435}
  {\bibfield  {journal} {\bibinfo  {journal} {Rev. Mod. Phys.}\ }\textbf
  {\bibinfo {volume} {49}},\ \bibinfo {pages} {435--479} (\bibinfo {year}
  {1977})}\BibitemShut {NoStop}%
\bibitem [{\citenamefont {Folk}\ and\ \citenamefont
  {Moser}(2006)}]{Folk2006jpa}%
  \BibitemOpen
  \bibfield  {author} {\bibinfo {author} {\bibfnamefont {R.}~\bibnamefont
  {Folk}}\ and\ \bibinfo {author} {\bibfnamefont {G.}~\bibnamefont {Moser}},\
  }\bibfield  {title} {\enquote {\bibinfo {title} {{Critical dynamics: a
  field-theoretical approach}},}\ }\href {\doibase 10.1088/0305-4470/39/24/R01}
  {\bibfield  {journal} {\bibinfo  {journal} {Journal of Physics A:
  Mathematical and General}\ }\textbf {\bibinfo {volume} {39}},\ \bibinfo
  {pages} {R207} (\bibinfo {year} {2006})}\BibitemShut {NoStop}%
\bibitem [{\citenamefont {T\"auber}(2014)}]{Tauber2014book}%
  \BibitemOpen
  \bibfield  {author} {\bibinfo {author} {\bibfnamefont {U.~C.}\ \bibnamefont
  {T\"auber}},\ }\href {\doibase 10.1017/CBO9781139046213} {\emph {\bibinfo
  {title} {{Critical Dynamics: A Field Theory Approach to Equilibrium and
  Non-Equilibrium Scaling Behavior}}}}\ (\bibinfo  {publisher} {Cambridge
  University Press},\ \bibinfo {year} {2014})\BibitemShut {NoStop}%
\bibitem [{\citenamefont {Kibble}(1976)}]{Kibble1976}%
  \BibitemOpen
  \bibfield  {author} {\bibinfo {author} {\bibfnamefont {T.~W.~B.}\
  \bibnamefont {Kibble}},\ }\bibfield  {title} {\enquote {\bibinfo {title}
  {Topology of cosmic domains and strings},}\ }\href {\doibase
  10.1088/0305-4470/9/8/029} {\bibfield  {journal} {\bibinfo  {journal}
  {Journal of Physics A: Mathematical and General}\ }\textbf {\bibinfo {volume}
  {9}},\ \bibinfo {pages} {1387} (\bibinfo {year} {1976})}\BibitemShut
  {NoStop}%
\bibitem [{\citenamefont {Kibble}(2007)}]{Kibble2007}%
  \BibitemOpen
  \bibfield  {author} {\bibinfo {author} {\bibfnamefont {T.}~\bibnamefont
  {Kibble}},\ }\bibfield  {title} {\enquote {\bibinfo {title}
  {{Phase-transition dynamics in the lab and the universe}},}\ }\href {\doibase
  10.1063/1.2784684} {\bibfield  {journal} {\bibinfo  {journal} {Physics
  Today}\ }\textbf {\bibinfo {volume} {60}},\ \bibinfo {pages} {47--52}
  (\bibinfo {year} {2007})}\BibitemShut {NoStop}%
\bibitem [{\citenamefont {Zurek}(1985)}]{Zurek1985}%
  \BibitemOpen
  \bibfield  {author} {\bibinfo {author} {\bibfnamefont {W.~H.}\ \bibnamefont
  {Zurek}},\ }\bibfield  {title} {\enquote {\bibinfo {title} {{Cosmological
  experiments in superfluid helium?}}}\ }\href {\doibase 10.1038/317505a0}
  {\bibfield  {journal} {\bibinfo  {journal} {Nature}\ }\textbf {\bibinfo
  {volume} {317}},\ \bibinfo {pages} {505--508} (\bibinfo {year}
  {1985})}\BibitemShut {NoStop}%
\bibitem [{\citenamefont {Zurek}(1996)}]{ZUREK1996177}%
  \BibitemOpen
  \bibfield  {author} {\bibinfo {author} {\bibfnamefont {W.}~\bibnamefont
  {Zurek}},\ }\bibfield  {title} {\enquote {\bibinfo {title} {Cosmological
  experiments in condensed matter systems},}\ }\href {\doibase
  https://doi.org/10.1016/S0370-1573(96)00009-9} {\bibfield  {journal}
  {\bibinfo  {journal} {Physics Reports}\ }\textbf {\bibinfo {volume} {276}},\
  \bibinfo {pages} {177--221} (\bibinfo {year} {1996})}\BibitemShut {NoStop}%
\bibitem [{\citenamefont {Zurek}\ \emph {et~al.}(2005)\citenamefont {Zurek},
  \citenamefont {Dorner},\ and\ \citenamefont {Zoller}}]{Zoller2005prl}%
  \BibitemOpen
  \bibfield  {author} {\bibinfo {author} {\bibfnamefont {W.~H.}\ \bibnamefont
  {Zurek}}, \bibinfo {author} {\bibfnamefont {U.}~\bibnamefont {Dorner}}, \
  and\ \bibinfo {author} {\bibfnamefont {P.}~\bibnamefont {Zoller}},\
  }\bibfield  {title} {\enquote {\bibinfo {title} {Dynamics of a quantum phase
  transition},}\ }\href {\doibase 10.1103/PhysRevLett.95.105701} {\bibfield
  {journal} {\bibinfo  {journal} {Phys. Rev. Lett.}\ }\textbf {\bibinfo
  {volume} {95}},\ \bibinfo {pages} {105701} (\bibinfo {year}
  {2005})}\BibitemShut {NoStop}%
\bibitem [{\citenamefont {Dziarmaga}(2005)}]{Dziarmaga2005prl}%
  \BibitemOpen
  \bibfield  {author} {\bibinfo {author} {\bibfnamefont {J.}~\bibnamefont
  {Dziarmaga}},\ }\bibfield  {title} {\enquote {\bibinfo {title} {Dynamics of a
  quantum phase transition: Exact solution of the quantum ising model},}\
  }\href {\doibase 10.1103/PhysRevLett.95.245701} {\bibfield  {journal}
  {\bibinfo  {journal} {Phys. Rev. Lett.}\ }\textbf {\bibinfo {volume} {95}},\
  \bibinfo {pages} {245701} (\bibinfo {year} {2005})}\BibitemShut {NoStop}%
\bibitem [{\citenamefont {Polkovnikov}(2005)}]{Polkovnikov2005prbr}%
  \BibitemOpen
  \bibfield  {author} {\bibinfo {author} {\bibfnamefont {A.}~\bibnamefont
  {Polkovnikov}},\ }\bibfield  {title} {\enquote {\bibinfo {title} {Universal
  adiabatic dynamics in the vicinity of a quantum critical point},}\ }\href
  {\doibase 10.1103/PhysRevB.72.161201} {\bibfield  {journal} {\bibinfo
  {journal} {Phys. Rev. B}\ }\textbf {\bibinfo {volume} {72}},\ \bibinfo
  {pages} {161201} (\bibinfo {year} {2005})}\BibitemShut {NoStop}%
\bibitem [{\citenamefont {Damski}\ and\ \citenamefont
  {Zurek}(2007)}]{Damski2007prl}%
  \BibitemOpen
  \bibfield  {author} {\bibinfo {author} {\bibfnamefont {B.}~\bibnamefont
  {Damski}}\ and\ \bibinfo {author} {\bibfnamefont {W.~H.}\ \bibnamefont
  {Zurek}},\ }\bibfield  {title} {\enquote {\bibinfo {title} {Dynamics of a
  quantum phase transition in a ferromagnetic bose-einstein condensate},}\
  }\href {\doibase 10.1103/PhysRevLett.99.130402} {\bibfield  {journal}
  {\bibinfo  {journal} {Phys. Rev. Lett.}\ }\textbf {\bibinfo {volume} {99}},\
  \bibinfo {pages} {130402} (\bibinfo {year} {2007})}\BibitemShut {NoStop}%
\bibitem [{\citenamefont {De~Grandi}\ \emph {et~al.}(2011)\citenamefont
  {De~Grandi}, \citenamefont {Polkovnikov},\ and\ \citenamefont
  {Sandvik}}]{Grandi2011prb}%
  \BibitemOpen
  \bibfield  {author} {\bibinfo {author} {\bibfnamefont {C.}~\bibnamefont
  {De~Grandi}}, \bibinfo {author} {\bibfnamefont {A.}~\bibnamefont
  {Polkovnikov}}, \ and\ \bibinfo {author} {\bibfnamefont {A.~W.}\ \bibnamefont
  {Sandvik}},\ }\bibfield  {title} {\enquote {\bibinfo {title} {Universal
  nonequilibrium quantum dynamics in imaginary time},}\ }\href {\doibase
  10.1103/PhysRevB.84.224303} {\bibfield  {journal} {\bibinfo  {journal} {Phys.
  Rev. B}\ }\textbf {\bibinfo {volume} {84}},\ \bibinfo {pages} {224303}
  (\bibinfo {year} {2011})}\BibitemShut {NoStop}%
\bibitem [{\citenamefont {Deng}\ \emph {et~al.}(2025)\citenamefont {Deng},
  \citenamefont {Sun},\ and\ \citenamefont {Li}}]{Deng2025prl}%
  \BibitemOpen
  \bibfield  {author} {\bibinfo {author} {\bibfnamefont {M.}~\bibnamefont
  {Deng}}, \bibinfo {author} {\bibfnamefont {Z.}~\bibnamefont {Sun}}, \ and\
  \bibinfo {author} {\bibfnamefont {F.}~\bibnamefont {Li}},\ }\bibfield
  {title} {\enquote {\bibinfo {title} {Defect production across higher-order
  phase transitions beyond kibble-zurek scaling},}\ }\href {\doibase
  10.1103/PhysRevLett.134.010409} {\bibfield  {journal} {\bibinfo  {journal}
  {Phys. Rev. Lett.}\ }\textbf {\bibinfo {volume} {134}},\ \bibinfo {pages}
  {010409} (\bibinfo {year} {2025})}\BibitemShut {NoStop}%
\bibitem [{\citenamefont {Zurek}(2009)}]{Zurek2009prl}%
  \BibitemOpen
  \bibfield  {author} {\bibinfo {author} {\bibfnamefont {W.~H.}\ \bibnamefont
  {Zurek}},\ }\bibfield  {title} {\enquote {\bibinfo {title} {Causality in
  condensates: Gray solitons as relics of bec formation},}\ }\href {\doibase
  10.1103/PhysRevLett.102.105702} {\bibfield  {journal} {\bibinfo  {journal}
  {Phys. Rev. Lett.}\ }\textbf {\bibinfo {volume} {102}},\ \bibinfo {pages}
  {105702} (\bibinfo {year} {2009})}\BibitemShut {NoStop}%
\bibitem [{\citenamefont {del Campo}\ \emph {et~al.}(2010)\citenamefont {del
  Campo}, \citenamefont {De~Chiara}, \citenamefont {Morigi}, \citenamefont
  {Plenio},\ and\ \citenamefont {Retzker}}]{delcampo2010prl}%
  \BibitemOpen
  \bibfield  {author} {\bibinfo {author} {\bibfnamefont {A.}~\bibnamefont {del
  Campo}}, \bibinfo {author} {\bibfnamefont {G.}~\bibnamefont {De~Chiara}},
  \bibinfo {author} {\bibfnamefont {G.}~\bibnamefont {Morigi}}, \bibinfo
  {author} {\bibfnamefont {M.~B.}\ \bibnamefont {Plenio}}, \ and\ \bibinfo
  {author} {\bibfnamefont {A.}~\bibnamefont {Retzker}},\ }\bibfield  {title}
  {\enquote {\bibinfo {title} {Structural defects in ion chains by quenching
  the external potential: The inhomogeneous kibble-zurek mechanism},}\ }\href
  {\doibase 10.1103/PhysRevLett.105.075701} {\bibfield  {journal} {\bibinfo
  {journal} {Phys. Rev. Lett.}\ }\textbf {\bibinfo {volume} {105}},\ \bibinfo
  {pages} {075701} (\bibinfo {year} {2010})}\BibitemShut {NoStop}%
\bibitem [{\citenamefont {del Campo}\ \emph {et~al.}(2013)\citenamefont {del
  Campo}, \citenamefont {Kibble},\ and\ \citenamefont {Zurek}}]{delCampo2013}%
  \BibitemOpen
  \bibfield  {author} {\bibinfo {author} {\bibfnamefont {A.}~\bibnamefont {del
  Campo}}, \bibinfo {author} {\bibfnamefont {T.~W.~B.}\ \bibnamefont {Kibble}},
  \ and\ \bibinfo {author} {\bibfnamefont {W.~H.}\ \bibnamefont {Zurek}},\
  }\bibfield  {title} {\enquote {\bibinfo {title} {Causality and
  non-equilibrium second-order phase transitions in inhomogeneous systems},}\
  }\href {\doibase 10.1088/0953-8984/25/40/404210} {\bibfield  {journal}
  {\bibinfo  {journal} {Journal of Physics: Condensed Matter}\ }\textbf
  {\bibinfo {volume} {25}},\ \bibinfo {pages} {404210} (\bibinfo {year}
  {2013})}\BibitemShut {NoStop}%
\bibitem [{\citenamefont {Sen}\ \emph {et~al.}(2008)\citenamefont {Sen},
  \citenamefont {Sengupta},\ and\ \citenamefont {Mondal}}]{Sen2008prl}%
  \BibitemOpen
  \bibfield  {author} {\bibinfo {author} {\bibfnamefont {D.}~\bibnamefont
  {Sen}}, \bibinfo {author} {\bibfnamefont {K.}~\bibnamefont {Sengupta}}, \
  and\ \bibinfo {author} {\bibfnamefont {S.}~\bibnamefont {Mondal}},\
  }\bibfield  {title} {\enquote {\bibinfo {title} {Defect production in
  nonlinear quench across a quantum critical point},}\ }\href {\doibase
  10.1103/PhysRevLett.101.016806} {\bibfield  {journal} {\bibinfo  {journal}
  {Phys. Rev. Lett.}\ }\textbf {\bibinfo {volume} {101}},\ \bibinfo {pages}
  {016806} (\bibinfo {year} {2008})}\BibitemShut {NoStop}%
\bibitem [{\citenamefont {Chandran}\ \emph {et~al.}(2012)\citenamefont
  {Chandran}, \citenamefont {Erez}, \citenamefont {Gubser},\ and\ \citenamefont
  {Sondhi}}]{chandran2012prb}%
  \BibitemOpen
  \bibfield  {author} {\bibinfo {author} {\bibfnamefont {A.}~\bibnamefont
  {Chandran}}, \bibinfo {author} {\bibfnamefont {A.}~\bibnamefont {Erez}},
  \bibinfo {author} {\bibfnamefont {S.~S.}\ \bibnamefont {Gubser}}, \ and\
  \bibinfo {author} {\bibfnamefont {S.~L.}\ \bibnamefont {Sondhi}},\ }\bibfield
   {title} {\enquote {\bibinfo {title} {Kibble-zurek problem: Universality and
  the scaling limit},}\ }\href {\doibase 10.1103/PhysRevB.86.064304} {\bibfield
   {journal} {\bibinfo  {journal} {Phys. Rev. B}\ }\textbf {\bibinfo {volume}
  {86}},\ \bibinfo {pages} {064304} (\bibinfo {year} {2012})}\BibitemShut
  {NoStop}%
\bibitem [{\citenamefont {Biroli}\ \emph {et~al.}(2010)\citenamefont {Biroli},
  \citenamefont {Cugliandolo},\ and\ \citenamefont {Sicilia}}]{Biroli2010prb}%
  \BibitemOpen
  \bibfield  {author} {\bibinfo {author} {\bibfnamefont {G.}~\bibnamefont
  {Biroli}}, \bibinfo {author} {\bibfnamefont {L.~F.}\ \bibnamefont
  {Cugliandolo}}, \ and\ \bibinfo {author} {\bibfnamefont {A.}~\bibnamefont
  {Sicilia}},\ }\bibfield  {title} {\enquote {\bibinfo {title} {Kibble-zurek
  mechanism and infinitely slow annealing through critical points},}\ }\href
  {\doibase 10.1103/PhysRevE.81.050101} {\bibfield  {journal} {\bibinfo
  {journal} {Phys. Rev. E}\ }\textbf {\bibinfo {volume} {81}},\ \bibinfo
  {pages} {050101} (\bibinfo {year} {2010})}\BibitemShut {NoStop}%
\bibitem [{\citenamefont {Huang}\ \emph {et~al.}(2016)\citenamefont {Huang},
  \citenamefont {Yin}, \citenamefont {Hu},\ and\ \citenamefont
  {Zhong}}]{Huangyy2016prb}%
  \BibitemOpen
  \bibfield  {author} {\bibinfo {author} {\bibfnamefont {Y.}~\bibnamefont
  {Huang}}, \bibinfo {author} {\bibfnamefont {S.}~\bibnamefont {Yin}}, \bibinfo
  {author} {\bibfnamefont {Q.}~\bibnamefont {Hu}}, \ and\ \bibinfo {author}
  {\bibfnamefont {F.}~\bibnamefont {Zhong}},\ }\bibfield  {title} {\enquote
  {\bibinfo {title} {Kibble-zurek mechanism beyond adiabaticity: Finite-time
  scaling with critical initial slip},}\ }\href {\doibase
  10.1103/PhysRevB.93.024103} {\bibfield  {journal} {\bibinfo  {journal} {Phys.
  Rev. B}\ }\textbf {\bibinfo {volume} {93}},\ \bibinfo {pages} {024103}
  (\bibinfo {year} {2016})}\BibitemShut {NoStop}%
\bibitem [{\citenamefont {Jeong}\ \emph {et~al.}(2019)\citenamefont {Jeong},
  \citenamefont {Kim},\ and\ \citenamefont {Lee}}]{Jeong2019pre}%
  \BibitemOpen
  \bibfield  {author} {\bibinfo {author} {\bibfnamefont {K.}~\bibnamefont
  {Jeong}}, \bibinfo {author} {\bibfnamefont {B.}~\bibnamefont {Kim}}, \ and\
  \bibinfo {author} {\bibfnamefont {S.~J.}\ \bibnamefont {Lee}},\ }\bibfield
  {title} {\enquote {\bibinfo {title} {Growth kinetics of the two-dimensional
  ising model with finite cooling rates},}\ }\href {\doibase
  10.1103/PhysRevE.99.022113} {\bibfield  {journal} {\bibinfo  {journal} {Phys.
  Rev. E}\ }\textbf {\bibinfo {volume} {99}},\ \bibinfo {pages} {022113}
  (\bibinfo {year} {2019})}\BibitemShut {NoStop}%
\bibitem [{\citenamefont {{Chen Tang and Wanzhou Zhang and Chengxiang
  Ding}}(2025)}]{Tang2025}%
  \BibitemOpen
  \bibfield  {author} {\bibinfo {author} {\bibnamefont {{Chen Tang and Wanzhou
  Zhang and Chengxiang Ding}}},\ }\href {https://arxiv.org/abs/2502.05598}
  {\enquote {\bibinfo {title} {{Dynamics of Baxter-Wu model}},}\ } (\bibinfo
  {year} {2025})\BibitemShut {NoStop}%
\bibitem [{\citenamefont {Gong}\ \emph {et~al.}(2010)\citenamefont {Gong},
  \citenamefont {Zhong}, \citenamefont {Huang},\ and\ \citenamefont
  {Fan}}]{Gong2010njp}%
  \BibitemOpen
  \bibfield  {author} {\bibinfo {author} {\bibfnamefont {S.}~\bibnamefont
  {Gong}}, \bibinfo {author} {\bibfnamefont {F.}~\bibnamefont {Zhong}},
  \bibinfo {author} {\bibfnamefont {X.}~\bibnamefont {Huang}}, \ and\ \bibinfo
  {author} {\bibfnamefont {S.}~\bibnamefont {Fan}},\ }\bibfield  {title}
  {\enquote {\bibinfo {title} {Finite-time scaling via linear driving},}\
  }\href {\doibase 10.1088/1367-2630/12/4/043036} {\bibfield  {journal}
  {\bibinfo  {journal} {New Journal of Physics}\ }\textbf {\bibinfo {volume}
  {12}},\ \bibinfo {pages} {043036} (\bibinfo {year} {2010})}\BibitemShut
  {NoStop}%
\bibitem [{\citenamefont {Zhong}\ and\ \citenamefont
  {Xu}(2005)}]{Zhongf2005prb}%
  \BibitemOpen
  \bibfield  {author} {\bibinfo {author} {\bibfnamefont {F.}~\bibnamefont
  {Zhong}}\ and\ \bibinfo {author} {\bibfnamefont {Z.}~\bibnamefont {Xu}},\
  }\bibfield  {title} {\enquote {\bibinfo {title} {Dynamic monte carlo
  renormalization group determination of critical exponents with linearly
  changing temperature},}\ }\href {\doibase 10.1103/PhysRevB.71.132402}
  {\bibfield  {journal} {\bibinfo  {journal} {Phys. Rev. B}\ }\textbf {\bibinfo
  {volume} {71}},\ \bibinfo {pages} {132402} (\bibinfo {year}
  {2005})}\BibitemShut {NoStop}%
\bibitem [{\citenamefont {Zhong}(2006)}]{Zhong2006pre}%
  \BibitemOpen
  \bibfield  {author} {\bibinfo {author} {\bibfnamefont {F.}~\bibnamefont
  {Zhong}},\ }\bibfield  {title} {\enquote {\bibinfo {title} {Probing
  criticality with linearly varying external fields: Renormalization group
  theory of nonequilibrium critical dynamics under driving},}\ }\href {\doibase
  10.1103/PhysRevE.73.047102} {\bibfield  {journal} {\bibinfo  {journal} {Phys.
  Rev. E}\ }\textbf {\bibinfo {volume} {73}},\ \bibinfo {pages} {047102}
  (\bibinfo {year} {2006})}\BibitemShut {NoStop}%
\bibitem [{\citenamefont {Fan}\ and\ \citenamefont {Zhong}(2007)}]{Fan2007pre}%
  \BibitemOpen
  \bibfield  {author} {\bibinfo {author} {\bibfnamefont {S.}~\bibnamefont
  {Fan}}\ and\ \bibinfo {author} {\bibfnamefont {F.}~\bibnamefont {Zhong}},\
  }\bibfield  {title} {\enquote {\bibinfo {title} {Determination of the dynamic
  and static critical exponents of the two-dimensional three-state potts model
  using linearly varying temperature},}\ }\href {\doibase
  10.1103/PhysRevE.76.041141} {\bibfield  {journal} {\bibinfo  {journal} {Phys.
  Rev. E}\ }\textbf {\bibinfo {volume} {76}},\ \bibinfo {pages} {041141}
  (\bibinfo {year} {2007})}\BibitemShut {NoStop}%
\bibitem [{\citenamefont {Huang}\ \emph {et~al.}(2010)\citenamefont {Huang},
  \citenamefont {Gong}, \citenamefont {Zhong},\ and\ \citenamefont
  {Fan}}]{Huangxz2010pre}%
  \BibitemOpen
  \bibfield  {author} {\bibinfo {author} {\bibfnamefont {X.}~\bibnamefont
  {Huang}}, \bibinfo {author} {\bibfnamefont {S.}~\bibnamefont {Gong}},
  \bibinfo {author} {\bibfnamefont {F.}~\bibnamefont {Zhong}}, \ and\ \bibinfo
  {author} {\bibfnamefont {S.}~\bibnamefont {Fan}},\ }\bibfield  {title}
  {\enquote {\bibinfo {title} {Finite-time scaling via linear driving:
  Application to the two-dimensional potts model},}\ }\href {\doibase
  10.1103/PhysRevE.81.041139} {\bibfield  {journal} {\bibinfo  {journal} {Phys.
  Rev. E}\ }\textbf {\bibinfo {volume} {81}},\ \bibinfo {pages} {041139}
  (\bibinfo {year} {2010})}\BibitemShut {NoStop}%
\bibitem [{\citenamefont {Yin}\ \emph {et~al.}(2014)\citenamefont {Yin},
  \citenamefont {Mai},\ and\ \citenamefont {Zhong}}]{Yin2014prb}%
  \BibitemOpen
  \bibfield  {author} {\bibinfo {author} {\bibfnamefont {S.}~\bibnamefont
  {Yin}}, \bibinfo {author} {\bibfnamefont {P.}~\bibnamefont {Mai}}, \ and\
  \bibinfo {author} {\bibfnamefont {F.}~\bibnamefont {Zhong}},\ }\bibfield
  {title} {\enquote {\bibinfo {title} {Nonequilibrium quantum criticality in
  open systems: The dissipation rate as an additional indispensable scaling
  variable},}\ }\href {\doibase 10.1103/PhysRevB.89.094108} {\bibfield
  {journal} {\bibinfo  {journal} {Phys. Rev. B}\ }\textbf {\bibinfo {volume}
  {89}},\ \bibinfo {pages} {094108} (\bibinfo {year} {2014})}\BibitemShut
  {NoStop}%
\bibitem [{\citenamefont {Huang}\ \emph {et~al.}(2014)\citenamefont {Huang},
  \citenamefont {Yin}, \citenamefont {Feng},\ and\ \citenamefont
  {Zhong}}]{Huang2014prb}%
  \BibitemOpen
  \bibfield  {author} {\bibinfo {author} {\bibfnamefont {Y.}~\bibnamefont
  {Huang}}, \bibinfo {author} {\bibfnamefont {S.}~\bibnamefont {Yin}}, \bibinfo
  {author} {\bibfnamefont {B.}~\bibnamefont {Feng}}, \ and\ \bibinfo {author}
  {\bibfnamefont {F.}~\bibnamefont {Zhong}},\ }\bibfield  {title} {\enquote
  {\bibinfo {title} {Kibble-zurek mechanism and finite-time scaling},}\ }\href
  {\doibase 10.1103/PhysRevB.90.134108} {\bibfield  {journal} {\bibinfo
  {journal} {Phys. Rev. B}\ }\textbf {\bibinfo {volume} {90}},\ \bibinfo
  {pages} {134108} (\bibinfo {year} {2014})}\BibitemShut {NoStop}%
\bibitem [{\citenamefont {Feng}\ \emph {et~al.}(2016)\citenamefont {Feng},
  \citenamefont {Yin},\ and\ \citenamefont {Zhong}}]{feng2016prb}%
  \BibitemOpen
  \bibfield  {author} {\bibinfo {author} {\bibfnamefont {B.}~\bibnamefont
  {Feng}}, \bibinfo {author} {\bibfnamefont {S.}~\bibnamefont {Yin}}, \ and\
  \bibinfo {author} {\bibfnamefont {F.}~\bibnamefont {Zhong}},\ }\bibfield
  {title} {\enquote {\bibinfo {title} {Theory of driven nonequilibrium critical
  phenomena},}\ }\href {\doibase 10.1103/PhysRevB.94.144103} {\bibfield
  {journal} {\bibinfo  {journal} {Phys. Rev. B}\ }\textbf {\bibinfo {volume}
  {94}},\ \bibinfo {pages} {144103} (\bibinfo {year} {2016})}\BibitemShut
  {NoStop}%
\bibitem [{\citenamefont {Fan}\ and\ \citenamefont {Zhong}(2009)}]{Fan2009pre}%
  \BibitemOpen
  \bibfield  {author} {\bibinfo {author} {\bibfnamefont {S.}~\bibnamefont
  {Fan}}\ and\ \bibinfo {author} {\bibfnamefont {F.}~\bibnamefont {Zhong}},\
  }\bibfield  {title} {\enquote {\bibinfo {title} {Critical dynamics of the
  two-dimensional random-bond potts model with nonequilibrium monte carlo
  simulations},}\ }\href {\doibase 10.1103/PhysRevE.79.011122} {\bibfield
  {journal} {\bibinfo  {journal} {Phys. Rev. E}\ }\textbf {\bibinfo {volume}
  {79}},\ \bibinfo {pages} {011122} (\bibinfo {year} {2009})}\BibitemShut
  {NoStop}%
\bibitem [{\citenamefont {Hu}\ \emph {et~al.}(2015)\citenamefont {Hu},
  \citenamefont {Yin},\ and\ \citenamefont {Zhong}}]{huqj2015prb}%
  \BibitemOpen
  \bibfield  {author} {\bibinfo {author} {\bibfnamefont {Q.}~\bibnamefont
  {Hu}}, \bibinfo {author} {\bibfnamefont {S.}~\bibnamefont {Yin}}, \ and\
  \bibinfo {author} {\bibfnamefont {F.}~\bibnamefont {Zhong}},\ }\bibfield
  {title} {\enquote {\bibinfo {title} {Scaling of the entanglement spectrum in
  driven critical dynamics},}\ }\href {\doibase 10.1103/PhysRevB.91.184109}
  {\bibfield  {journal} {\bibinfo  {journal} {Phys. Rev. B}\ }\textbf {\bibinfo
  {volume} {91}},\ \bibinfo {pages} {184109} (\bibinfo {year}
  {2015})}\BibitemShut {NoStop}%
\bibitem [{\citenamefont {Cao}\ \emph {et~al.}(2018)\citenamefont {Cao},
  \citenamefont {Hu},\ and\ \citenamefont {Zhong}}]{Cao2018prb}%
  \BibitemOpen
  \bibfield  {author} {\bibinfo {author} {\bibfnamefont {X.}~\bibnamefont
  {Cao}}, \bibinfo {author} {\bibfnamefont {Q.}~\bibnamefont {Hu}}, \ and\
  \bibinfo {author} {\bibfnamefont {F.}~\bibnamefont {Zhong}},\ }\bibfield
  {title} {\enquote {\bibinfo {title} {Scaling theory of entanglement entropy
  in confinements near quantum critical points},}\ }\href {\doibase
  10.1103/PhysRevB.98.245124} {\bibfield  {journal} {\bibinfo  {journal} {Phys.
  Rev. B}\ }\textbf {\bibinfo {volume} {98}},\ \bibinfo {pages} {245124}
  (\bibinfo {year} {2018})}\BibitemShut {NoStop}%
\bibitem [{\citenamefont {Li}\ \emph {et~al.}(2019)\citenamefont {Li},
  \citenamefont {Zeng},\ and\ \citenamefont {Zhong}}]{Liyh2019pre}%
  \BibitemOpen
  \bibfield  {author} {\bibinfo {author} {\bibfnamefont {Y.}~\bibnamefont
  {Li}}, \bibinfo {author} {\bibfnamefont {Z.}~\bibnamefont {Zeng}}, \ and\
  \bibinfo {author} {\bibfnamefont {F.}~\bibnamefont {Zhong}},\ }\bibfield
  {title} {\enquote {\bibinfo {title} {Driving driven lattice gases to identify
  their universality classes},}\ }\href {\doibase 10.1103/PhysRevE.100.020105}
  {\bibfield  {journal} {\bibinfo  {journal} {Phys. Rev. E}\ }\textbf {\bibinfo
  {volume} {100}},\ \bibinfo {pages} {020105} (\bibinfo {year}
  {2019})}\BibitemShut {NoStop}%
\bibitem [{\citenamefont {Wang}\ \emph {et~al.}(2024)\citenamefont {Wang},
  \citenamefont {Liu}, \citenamefont {Li}, \citenamefont {Zhang},\ and\
  \citenamefont {Yin}}]{Yinarxiv2024}%
  \BibitemOpen
  \bibfield  {author} {\bibinfo {author} {\bibfnamefont {W.}~\bibnamefont
  {Wang}}, \bibinfo {author} {\bibfnamefont {S.}~\bibnamefont {Liu}}, \bibinfo
  {author} {\bibfnamefont {J.}~\bibnamefont {Li}}, \bibinfo {author}
  {\bibfnamefont {S.-X.}\ \bibnamefont {Zhang}}, \ and\ \bibinfo {author}
  {\bibfnamefont {S.}~\bibnamefont {Yin}},\ }\href {\doibase
  10.48550/arXiv.2411.06648} {\bibfield  {journal} {\bibinfo  {journal} {arXiv:
  2411.06648}\ } (\bibinfo {year} {2024}),\
  10.48550/arXiv.2411.06648}\BibitemShut {NoStop}%
\bibitem [{\citenamefont {Yin}\ \emph {et~al.}(2016)\citenamefont {Yin},
  \citenamefont {Lo},\ and\ \citenamefont {Chen}}]{Yin2016prb}%
  \BibitemOpen
  \bibfield  {author} {\bibinfo {author} {\bibfnamefont {S.}~\bibnamefont
  {Yin}}, \bibinfo {author} {\bibfnamefont {C.-Y.}\ \bibnamefont {Lo}}, \ and\
  \bibinfo {author} {\bibfnamefont {P.}~\bibnamefont {Chen}},\ }\bibfield
  {title} {\enquote {\bibinfo {title} {Scaling in driven dynamics starting in
  the vicinity of a quantum critical point},}\ }\href {\doibase
  10.1103/PhysRevB.94.064302} {\bibfield  {journal} {\bibinfo  {journal} {Phys.
  Rev. B}\ }\textbf {\bibinfo {volume} {94}},\ \bibinfo {pages} {064302}
  (\bibinfo {year} {2016})}\BibitemShut {NoStop}%
\bibitem [{\citenamefont {Zeng}\ \emph
  {et~al.}(2024{\natexlab{a}})\citenamefont {Zeng}, \citenamefont {Yu},
  \citenamefont {Li}, \citenamefont {Li},\ and\ \citenamefont
  {Yin}}]{Zeng2024arx_1}%
  \BibitemOpen
  \bibfield  {author} {\bibinfo {author} {\bibfnamefont {Z.}~\bibnamefont
  {Zeng}}, \bibinfo {author} {\bibfnamefont {Y.-K.}\ \bibnamefont {Yu}},
  \bibinfo {author} {\bibfnamefont {Z.-X.}\ \bibnamefont {Li}}, \bibinfo
  {author} {\bibfnamefont {Z.-X.}\ \bibnamefont {Li}}, \ and\ \bibinfo {author}
  {\bibfnamefont {S.}~\bibnamefont {Yin}},\ }\bibfield  {title} {\enquote
  {\bibinfo {title} {{Finite-time Scaling beyond the Kibble-Zurek Prerequisite:
  Driven Critical Dynamics in Strongly Interacting Dirac Systems}},}\ }\href
  {https://arxiv.org/abs/2403.19258} {\bibfield  {journal} {\bibinfo  {journal}
  {arXiv:2403.19258}\ } (\bibinfo {year} {2024}{\natexlab{a}})}\BibitemShut
  {NoStop}%
\bibitem [{\citenamefont {Zeng}\ \emph
  {et~al.}(2024{\natexlab{b}})\citenamefont {Zeng}, \citenamefont {Yu},
  \citenamefont {Li},\ and\ \citenamefont {Yin}}]{Zeng2024arx_2}%
  \BibitemOpen
  \bibfield  {author} {\bibinfo {author} {\bibfnamefont {Z.}~\bibnamefont
  {Zeng}}, \bibinfo {author} {\bibfnamefont {Y.-K.}\ \bibnamefont {Yu}},
  \bibinfo {author} {\bibfnamefont {Z.-X.}\ \bibnamefont {Li}}, \ and\ \bibinfo
  {author} {\bibfnamefont {S.}~\bibnamefont {Yin}},\ }\bibfield  {title}
  {\enquote {\bibinfo {title} {Nonequilibrium critical dynamics with emergent
  supersymmetry},}\ }\href {https://arxiv.org/abs/2408.06138} {\bibfield
  {journal} {\bibinfo  {journal} {arXiv: 2408.06138}\ } (\bibinfo {year}
  {2024}{\natexlab{b}})}\BibitemShut {NoStop}%
\bibitem [{\citenamefont {Deng}\ \emph {et~al.}(2009)\citenamefont {Deng},
  \citenamefont {Ortiz},\ and\ \citenamefont {Viola}}]{Deng2008}%
  \BibitemOpen
  \bibfield  {author} {\bibinfo {author} {\bibfnamefont {S.}~\bibnamefont
  {Deng}}, \bibinfo {author} {\bibfnamefont {G.}~\bibnamefont {Ortiz}}, \ and\
  \bibinfo {author} {\bibfnamefont {L.}~\bibnamefont {Viola}},\ }\bibfield
  {title} {\enquote {\bibinfo {title} {Dynamical non-ergodic scaling in
  continuous finite-order quantum phase transitions},}\ }\href {\doibase
  10.1209/0295-5075/84/67008} {\bibfield  {journal} {\bibinfo  {journal}
  {Europhysics Letters}\ }\textbf {\bibinfo {volume} {84}},\ \bibinfo {pages}
  {67008} (\bibinfo {year} {2009})}\BibitemShut {NoStop}%
\bibitem [{\citenamefont {Kolodrubetz}\ \emph {et~al.}(2012)\citenamefont
  {Kolodrubetz}, \citenamefont {Clark},\ and\ \citenamefont
  {Huse}}]{Huse2012prl}%
  \BibitemOpen
  \bibfield  {author} {\bibinfo {author} {\bibfnamefont {M.}~\bibnamefont
  {Kolodrubetz}}, \bibinfo {author} {\bibfnamefont {B.~K.}\ \bibnamefont
  {Clark}}, \ and\ \bibinfo {author} {\bibfnamefont {D.~A.}\ \bibnamefont
  {Huse}},\ }\bibfield  {title} {\enquote {\bibinfo {title} {Nonequilibrium
  dynamic critical scaling of the quantum ising chain},}\ }\href {\doibase
  10.1103/PhysRevLett.109.015701} {\bibfield  {journal} {\bibinfo  {journal}
  {Phys. Rev. Lett.}\ }\textbf {\bibinfo {volume} {109}},\ \bibinfo {pages}
  {015701} (\bibinfo {year} {2012})}\BibitemShut {NoStop}%
\bibitem [{\citenamefont {Liu}\ \emph {et~al.}(2014)\citenamefont {Liu},
  \citenamefont {Polkovnikov},\ and\ \citenamefont {Sandvik}}]{liuprb2014}%
  \BibitemOpen
  \bibfield  {author} {\bibinfo {author} {\bibfnamefont {C.-W.}\ \bibnamefont
  {Liu}}, \bibinfo {author} {\bibfnamefont {A.}~\bibnamefont {Polkovnikov}}, \
  and\ \bibinfo {author} {\bibfnamefont {A.~W.}\ \bibnamefont {Sandvik}},\
  }\bibfield  {title} {\enquote {\bibinfo {title} {Dynamic scaling at classical
  phase transitions approached through nonequilibrium quenching},}\ }\href
  {\doibase 10.1103/PhysRevB.89.054307} {\bibfield  {journal} {\bibinfo
  {journal} {Phys. Rev. B}\ }\textbf {\bibinfo {volume} {89}},\ \bibinfo
  {pages} {054307} (\bibinfo {year} {2014})}\BibitemShut {NoStop}%
\bibitem [{\citenamefont {Francuz}\ \emph {et~al.}(2016)\citenamefont
  {Francuz}, \citenamefont {Dziarmaga}, \citenamefont {Gardas},\ and\
  \citenamefont {Zurek}}]{Zurek2016prb}%
  \BibitemOpen
  \bibfield  {author} {\bibinfo {author} {\bibfnamefont {A.}~\bibnamefont
  {Francuz}}, \bibinfo {author} {\bibfnamefont {J.}~\bibnamefont {Dziarmaga}},
  \bibinfo {author} {\bibfnamefont {B.}~\bibnamefont {Gardas}}, \ and\ \bibinfo
  {author} {\bibfnamefont {W.~H.}\ \bibnamefont {Zurek}},\ }\bibfield  {title}
  {\enquote {\bibinfo {title} {Space and time renormalization in phase
  transition dynamics},}\ }\href {\doibase 10.1103/PhysRevB.93.075134}
  {\bibfield  {journal} {\bibinfo  {journal} {Phys. Rev. B}\ }\textbf {\bibinfo
  {volume} {93}},\ \bibinfo {pages} {075134} (\bibinfo {year}
  {2016})}\BibitemShut {NoStop}%
\bibitem [{\citenamefont {Hendry}\ \emph {et~al.}(1994)\citenamefont {Hendry},
  \citenamefont {Lawson}, \citenamefont {Lee}, \citenamefont {McClintock},\
  and\ \citenamefont {Williams}}]{Hendry1994nature}%
  \BibitemOpen
  \bibfield  {author} {\bibinfo {author} {\bibfnamefont {P.~C.}\ \bibnamefont
  {Hendry}}, \bibinfo {author} {\bibfnamefont {N.~S.}\ \bibnamefont {Lawson}},
  \bibinfo {author} {\bibfnamefont {R.~A.~M.}\ \bibnamefont {Lee}}, \bibinfo
  {author} {\bibfnamefont {P.~V.~E.}\ \bibnamefont {McClintock}}, \ and\
  \bibinfo {author} {\bibfnamefont {C.~D.~H.}\ \bibnamefont {Williams}},\
  }\bibfield  {title} {\enquote {\bibinfo {title} {Generation of defects in
  superfluid 4he as an analogue of the formation of cosmic strings},}\ }\href
  {\doibase 10.1038/368315a0} {\bibfield  {journal} {\bibinfo  {journal}
  {Nature}\ }\textbf {\bibinfo {volume} {368}},\ \bibinfo {pages} {315--317}
  (\bibinfo {year} {1994})}\BibitemShut {NoStop}%
\bibitem [{\citenamefont {Dodd}\ \emph {et~al.}(1998)\citenamefont {Dodd},
  \citenamefont {Hendry}, \citenamefont {Lawson}, \citenamefont {McClintock},\
  and\ \citenamefont {Williams}}]{Dodd1998prl}%
  \BibitemOpen
  \bibfield  {author} {\bibinfo {author} {\bibfnamefont {M.~E.}\ \bibnamefont
  {Dodd}}, \bibinfo {author} {\bibfnamefont {P.~C.}\ \bibnamefont {Hendry}},
  \bibinfo {author} {\bibfnamefont {N.~S.}\ \bibnamefont {Lawson}}, \bibinfo
  {author} {\bibfnamefont {P.~V.~E.}\ \bibnamefont {McClintock}}, \ and\
  \bibinfo {author} {\bibfnamefont {C.~D.~H.}\ \bibnamefont {Williams}},\
  }\bibfield  {title} {\enquote {\bibinfo {title} {Nonappearance of vortices in
  fast mechanical expansions of liquid ${}^{4}\mathrm{He}$ through the lambda
  transition},}\ }\href {\doibase 10.1103/PhysRevLett.81.3703} {\bibfield
  {journal} {\bibinfo  {journal} {Phys. Rev. Lett.}\ }\textbf {\bibinfo
  {volume} {81}},\ \bibinfo {pages} {3703--3706} (\bibinfo {year}
  {1998})}\BibitemShut {NoStop}%
\bibitem [{\citenamefont {Bäuerle}\ \emph {et~al.}(1996)\citenamefont
  {Bäuerle}, \citenamefont {Bunkov}, \citenamefont {Fisher}, \citenamefont
  {Godfrin},\ and\ \citenamefont {Pickett}}]{Bunkov1996nature}%
  \BibitemOpen
  \bibfield  {author} {\bibinfo {author} {\bibfnamefont {C.}~\bibnamefont
  {Bäuerle}}, \bibinfo {author} {\bibfnamefont {Y.~M.}\ \bibnamefont
  {Bunkov}}, \bibinfo {author} {\bibfnamefont {S.~N.}\ \bibnamefont {Fisher}},
  \bibinfo {author} {\bibfnamefont {H.}~\bibnamefont {Godfrin}}, \ and\
  \bibinfo {author} {\bibfnamefont {G.~R.}\ \bibnamefont {Pickett}},\
  }\bibfield  {title} {\enquote {\bibinfo {title} {Laboratory simulation of
  cosmic string formation in the early universe using superfluid 3he},}\ }\href
  {\doibase 10.1038/382332a0} {\bibfield  {journal} {\bibinfo  {journal}
  {Nature}\ }\textbf {\bibinfo {volume} {382}},\ \bibinfo {pages} {332--334}
  (\bibinfo {year} {1996})}\BibitemShut {NoStop}%
\bibitem [{\citenamefont {Ruutu}\ \emph {et~al.}(1996)\citenamefont {Ruutu},
  \citenamefont {Eltsov}, \citenamefont {Gill}, \citenamefont {Kibble},
  \citenamefont {Krusius}, \citenamefont {Makhlin}, \citenamefont {Plaçais},
  \citenamefont {Volovik},\ and\ \citenamefont {Xu}}]{Xuwen1996nature}%
  \BibitemOpen
  \bibfield  {author} {\bibinfo {author} {\bibfnamefont {V.~M.~H.}\
  \bibnamefont {Ruutu}}, \bibinfo {author} {\bibfnamefont {V.~B.}\ \bibnamefont
  {Eltsov}}, \bibinfo {author} {\bibfnamefont {A.~J.}\ \bibnamefont {Gill}},
  \bibinfo {author} {\bibfnamefont {T.~W.~B.}\ \bibnamefont {Kibble}}, \bibinfo
  {author} {\bibfnamefont {M.}~\bibnamefont {Krusius}}, \bibinfo {author}
  {\bibfnamefont {Y.~G.}\ \bibnamefont {Makhlin}}, \bibinfo {author}
  {\bibfnamefont {B.}~\bibnamefont {Plaçais}}, \bibinfo {author}
  {\bibfnamefont {G.~E.}\ \bibnamefont {Volovik}}, \ and\ \bibinfo {author}
  {\bibfnamefont {W.}~\bibnamefont {Xu}},\ }\bibfield  {title} {\enquote
  {\bibinfo {title} {Vortex formation in neutron-irradiated superfluid 3he as
  an analogue of cosmological defect formation},}\ }\href {\doibase
  10.1038/382334a0} {\bibfield  {journal} {\bibinfo  {journal} {Nature}\
  }\textbf {\bibinfo {volume} {382}},\ \bibinfo {pages} {334--336} (\bibinfo
  {year} {1996})}\BibitemShut {NoStop}%
\bibitem [{\citenamefont {Carmi}\ \emph {et~al.}(2000)\citenamefont {Carmi},
  \citenamefont {Polturak},\ and\ \citenamefont {Koren}}]{Carmi2000prl}%
  \BibitemOpen
  \bibfield  {author} {\bibinfo {author} {\bibfnamefont {R.}~\bibnamefont
  {Carmi}}, \bibinfo {author} {\bibfnamefont {E.}~\bibnamefont {Polturak}}, \
  and\ \bibinfo {author} {\bibfnamefont {G.}~\bibnamefont {Koren}},\ }\bibfield
   {title} {\enquote {\bibinfo {title} {Observation of spontaneous flux
  generation in a multi-josephson-junction loop},}\ }\href {\doibase
  10.1103/PhysRevLett.84.4966} {\bibfield  {journal} {\bibinfo  {journal}
  {Phys. Rev. Lett.}\ }\textbf {\bibinfo {volume} {84}},\ \bibinfo {pages}
  {4966--4969} (\bibinfo {year} {2000})}\BibitemShut {NoStop}%
\bibitem [{\citenamefont {Monaco}\ \emph {et~al.}(2002)\citenamefont {Monaco},
  \citenamefont {Mygind},\ and\ \citenamefont {Rivers}}]{Monaco2002prl}%
  \BibitemOpen
  \bibfield  {author} {\bibinfo {author} {\bibfnamefont {R.}~\bibnamefont
  {Monaco}}, \bibinfo {author} {\bibfnamefont {J.}~\bibnamefont {Mygind}}, \
  and\ \bibinfo {author} {\bibfnamefont {R.~J.}\ \bibnamefont {Rivers}},\
  }\bibfield  {title} {\enquote {\bibinfo {title} {Zurek-kibble domain
  structures: The dynamics of spontaneous vortex formation in annular josephson
  tunnel junctions},}\ }\href {\doibase 10.1103/PhysRevLett.89.080603}
  {\bibfield  {journal} {\bibinfo  {journal} {Phys. Rev. Lett.}\ }\textbf
  {\bibinfo {volume} {89}},\ \bibinfo {pages} {080603} (\bibinfo {year}
  {2002})}\BibitemShut {NoStop}%
\bibitem [{\citenamefont {Monaco}\ \emph {et~al.}(2009)\citenamefont {Monaco},
  \citenamefont {Mygind}, \citenamefont {Rivers},\ and\ \citenamefont
  {Koshelets}}]{Monacoprb2009}%
  \BibitemOpen
  \bibfield  {author} {\bibinfo {author} {\bibfnamefont {R.}~\bibnamefont
  {Monaco}}, \bibinfo {author} {\bibfnamefont {J.}~\bibnamefont {Mygind}},
  \bibinfo {author} {\bibfnamefont {R.~J.}\ \bibnamefont {Rivers}}, \ and\
  \bibinfo {author} {\bibfnamefont {V.~P.}\ \bibnamefont {Koshelets}},\
  }\bibfield  {title} {\enquote {\bibinfo {title} {Spontaneous fluxoid
  formation in superconducting loops},}\ }\href {\doibase
  10.1103/PhysRevB.80.180501} {\bibfield  {journal} {\bibinfo  {journal} {Phys.
  Rev. B}\ }\textbf {\bibinfo {volume} {80}},\ \bibinfo {pages} {180501}
  (\bibinfo {year} {2009})}\BibitemShut {NoStop}%
\bibitem [{\citenamefont {Su}\ \emph {et~al.}(2013)\citenamefont {Su},
  \citenamefont {Gou}, \citenamefont {Bradley}, \citenamefont {Fialko},\ and\
  \citenamefont {Brand}}]{Brand2013prl}%
  \BibitemOpen
  \bibfield  {author} {\bibinfo {author} {\bibfnamefont {S.-W.}\ \bibnamefont
  {Su}}, \bibinfo {author} {\bibfnamefont {S.-C.}\ \bibnamefont {Gou}},
  \bibinfo {author} {\bibfnamefont {A.}~\bibnamefont {Bradley}}, \bibinfo
  {author} {\bibfnamefont {O.}~\bibnamefont {Fialko}}, \ and\ \bibinfo {author}
  {\bibfnamefont {J.}~\bibnamefont {Brand}},\ }\bibfield  {title} {\enquote
  {\bibinfo {title} {Kibble-zurek scaling and its breakdown for spontaneous
  generation of josephson vortices in bose-einstein condensates},}\ }\href
  {\doibase 10.1103/PhysRevLett.110.215302} {\bibfield  {journal} {\bibinfo
  {journal} {Phys. Rev. Lett.}\ }\textbf {\bibinfo {volume} {110}},\ \bibinfo
  {pages} {215302} (\bibinfo {year} {2013})}\BibitemShut {NoStop}%
\bibitem [{\citenamefont {Labeyrie}\ and\ \citenamefont
  {Kaiser}(2016)}]{Labeyrie2016prl}%
  \BibitemOpen
  \bibfield  {author} {\bibinfo {author} {\bibfnamefont {G.}~\bibnamefont
  {Labeyrie}}\ and\ \bibinfo {author} {\bibfnamefont {R.}~\bibnamefont
  {Kaiser}},\ }\bibfield  {title} {\enquote {\bibinfo {title} {Kibble-zurek
  mechanism in the self-organization of a cold atomic cloud},}\ }\href
  {\doibase 10.1103/PhysRevLett.117.275701} {\bibfield  {journal} {\bibinfo
  {journal} {Phys. Rev. Lett.}\ }\textbf {\bibinfo {volume} {117}},\ \bibinfo
  {pages} {275701} (\bibinfo {year} {2016})}\BibitemShut {NoStop}%
\bibitem [{\citenamefont {Navon}\ \emph {et~al.}(2015)\citenamefont {Navon},
  \citenamefont {Gaunt}, \citenamefont {Smith},\ and\ \citenamefont
  {Hadzibabic}}]{Navon2015science}%
  \BibitemOpen
  \bibfield  {author} {\bibinfo {author} {\bibfnamefont {N.}~\bibnamefont
  {Navon}}, \bibinfo {author} {\bibfnamefont {A.~L.}\ \bibnamefont {Gaunt}},
  \bibinfo {author} {\bibfnamefont {R.~P.}\ \bibnamefont {Smith}}, \ and\
  \bibinfo {author} {\bibfnamefont {Z.}~\bibnamefont {Hadzibabic}},\ }\bibfield
   {title} {\enquote {\bibinfo {title} {Critical dynamics of spontaneous
  symmetry breaking in a homogeneous bose gas},}\ }\href {\doibase
  10.1126/science.1258676} {\bibfield  {journal} {\bibinfo  {journal}
  {Science}\ }\textbf {\bibinfo {volume} {347}},\ \bibinfo {pages} {167--170}
  (\bibinfo {year} {2015})}\BibitemShut {NoStop}%
\bibitem [{\citenamefont {Sunami}\ \emph {et~al.}(2023)\citenamefont {Sunami},
  \citenamefont {Singh}, \citenamefont {Garrick}, \citenamefont {Beregi},
  \citenamefont {Barker}, \citenamefont {Luksch}, \citenamefont {Bentine},
  \citenamefont {Mathey},\ and\ \citenamefont {Foot}}]{Shinichi2023science}%
  \BibitemOpen
  \bibfield  {author} {\bibinfo {author} {\bibfnamefont {S.}~\bibnamefont
  {Sunami}}, \bibinfo {author} {\bibfnamefont {V.~P.}\ \bibnamefont {Singh}},
  \bibinfo {author} {\bibfnamefont {D.}~\bibnamefont {Garrick}}, \bibinfo
  {author} {\bibfnamefont {A.}~\bibnamefont {Beregi}}, \bibinfo {author}
  {\bibfnamefont {A.~J.}\ \bibnamefont {Barker}}, \bibinfo {author}
  {\bibfnamefont {K.}~\bibnamefont {Luksch}}, \bibinfo {author} {\bibfnamefont
  {E.}~\bibnamefont {Bentine}}, \bibinfo {author} {\bibfnamefont
  {L.}~\bibnamefont {Mathey}}, \ and\ \bibinfo {author} {\bibfnamefont {C.~J.}\
  \bibnamefont {Foot}},\ }\bibfield  {title} {\enquote {\bibinfo {title}
  {Universal scaling of the dynamic bkt transition in quenched 2d bose
  gases},}\ }\href {\doibase 10.1126/science.abq6753} {\bibfield  {journal}
  {\bibinfo  {journal} {Science}\ }\textbf {\bibinfo {volume} {382}},\ \bibinfo
  {pages} {443--447} (\bibinfo {year} {2023})}\BibitemShut {NoStop}%
\bibitem [{\citenamefont {Chuang}\ \emph {et~al.}(1991)\citenamefont {Chuang},
  \citenamefont {Durrer}, \citenamefont {Turok},\ and\ \citenamefont
  {Yurke}}]{Chuang1991sci}%
  \BibitemOpen
  \bibfield  {author} {\bibinfo {author} {\bibfnamefont {I.}~\bibnamefont
  {Chuang}}, \bibinfo {author} {\bibfnamefont {R.}~\bibnamefont {Durrer}},
  \bibinfo {author} {\bibfnamefont {N.}~\bibnamefont {Turok}}, \ and\ \bibinfo
  {author} {\bibfnamefont {B.}~\bibnamefont {Yurke}},\ }\bibfield  {title}
  {\enquote {\bibinfo {title} {Cosmology in the laboratory: Defect dynamics in
  liquid crystals},}\ }\href {\doibase 10.1126/science.251.4999.1336}
  {\bibfield  {journal} {\bibinfo  {journal} {Science}\ }\textbf {\bibinfo
  {volume} {251}},\ \bibinfo {pages} {1336--1342} (\bibinfo {year}
  {1991})}\BibitemShut {NoStop}%
\bibitem [{\citenamefont {Ducci}\ \emph {et~al.}(1999)\citenamefont {Ducci},
  \citenamefont {Ramazza}, \citenamefont {Gonz\'alez-Vi\~nas},\ and\
  \citenamefont {Arecchi}}]{Ducci1999prl}%
  \BibitemOpen
  \bibfield  {author} {\bibinfo {author} {\bibfnamefont {S.}~\bibnamefont
  {Ducci}}, \bibinfo {author} {\bibfnamefont {P.~L.}\ \bibnamefont {Ramazza}},
  \bibinfo {author} {\bibfnamefont {W.}~\bibnamefont {Gonz\'alez-Vi\~nas}}, \
  and\ \bibinfo {author} {\bibfnamefont {F.~T.}\ \bibnamefont {Arecchi}},\
  }\bibfield  {title} {\enquote {\bibinfo {title} {Order parameter
  fragmentation after a symmetry-breaking transition},}\ }\href {\doibase
  10.1103/PhysRevLett.83.5210} {\bibfield  {journal} {\bibinfo  {journal}
  {Phys. Rev. Lett.}\ }\textbf {\bibinfo {volume} {83}},\ \bibinfo {pages}
  {5210--5213} (\bibinfo {year} {1999})}\BibitemShut {NoStop}%
\bibitem [{\citenamefont {Keesling}\ \emph {et~al.}(2019)\citenamefont
  {Keesling}, \citenamefont {Omran}, \citenamefont {Levine}, \citenamefont
  {Bernien}, \citenamefont {Pichler}, \citenamefont {Choi}, \citenamefont
  {Samajdar}, \citenamefont {Schwartz}, \citenamefont {Silvi}, \citenamefont
  {Sachdev}, \citenamefont {Zoller}, \citenamefont {Endres}, \citenamefont
  {Greiner}, \citenamefont {Vuleti{\'{c}}},\ and\ \citenamefont
  {Lukin}}]{Keesling2019}%
  \BibitemOpen
  \bibfield  {author} {\bibinfo {author} {\bibfnamefont {A.}~\bibnamefont
  {Keesling}}, \bibinfo {author} {\bibfnamefont {A.}~\bibnamefont {Omran}},
  \bibinfo {author} {\bibfnamefont {H.}~\bibnamefont {Levine}}, \bibinfo
  {author} {\bibfnamefont {H.}~\bibnamefont {Bernien}}, \bibinfo {author}
  {\bibfnamefont {H.}~\bibnamefont {Pichler}}, \bibinfo {author} {\bibfnamefont
  {S.}~\bibnamefont {Choi}}, \bibinfo {author} {\bibfnamefont {R.}~\bibnamefont
  {Samajdar}}, \bibinfo {author} {\bibfnamefont {S.}~\bibnamefont {Schwartz}},
  \bibinfo {author} {\bibfnamefont {P.}~\bibnamefont {Silvi}}, \bibinfo
  {author} {\bibfnamefont {S.}~\bibnamefont {Sachdev}}, \bibinfo {author}
  {\bibfnamefont {P.}~\bibnamefont {Zoller}}, \bibinfo {author} {\bibfnamefont
  {M.}~\bibnamefont {Endres}}, \bibinfo {author} {\bibfnamefont
  {M.}~\bibnamefont {Greiner}}, \bibinfo {author} {\bibfnamefont
  {V.}~\bibnamefont {Vuleti{\'{c}}}}, \ and\ \bibinfo {author} {\bibfnamefont
  {M.~D.}\ \bibnamefont {Lukin}},\ }\bibfield  {title} {\enquote {\bibinfo
  {title} {Quantum kibble--zurek mechanism and critical dynamics on a
  programmable rydberg simulator},}\ }\href {\doibase
  10.1038/s41586-019-1070-1} {\bibfield  {journal} {\bibinfo  {journal}
  {Nature}\ }\textbf {\bibinfo {volume} {568}},\ \bibinfo {pages} {207--211}
  (\bibinfo {year} {2019})}\BibitemShut {NoStop}%
\bibitem [{\citenamefont {Ebadi}\ \emph {et~al.}(2021)\citenamefont {Ebadi},
  \citenamefont {Wang}, \citenamefont {Levine}, \citenamefont {Keesling},
  \citenamefont {Semeghini}, \citenamefont {Omran}, \citenamefont {Bluvstein},
  \citenamefont {Samajdar}, \citenamefont {Pichler}, \citenamefont {Ho},
  \citenamefont {Choi}, \citenamefont {Sachdev}, \citenamefont {Greiner},
  \citenamefont {Vuleti{\'{c}}},\ and\ \citenamefont {Lukin}}]{Ebadi2021}%
  \BibitemOpen
  \bibfield  {author} {\bibinfo {author} {\bibfnamefont {S.}~\bibnamefont
  {Ebadi}}, \bibinfo {author} {\bibfnamefont {T.~T.}\ \bibnamefont {Wang}},
  \bibinfo {author} {\bibfnamefont {H.}~\bibnamefont {Levine}}, \bibinfo
  {author} {\bibfnamefont {A.}~\bibnamefont {Keesling}}, \bibinfo {author}
  {\bibfnamefont {G.}~\bibnamefont {Semeghini}}, \bibinfo {author}
  {\bibfnamefont {A.}~\bibnamefont {Omran}}, \bibinfo {author} {\bibfnamefont
  {D.}~\bibnamefont {Bluvstein}}, \bibinfo {author} {\bibfnamefont
  {R.}~\bibnamefont {Samajdar}}, \bibinfo {author} {\bibfnamefont
  {H.}~\bibnamefont {Pichler}}, \bibinfo {author} {\bibfnamefont {W.~W.}\
  \bibnamefont {Ho}}, \bibinfo {author} {\bibfnamefont {S.}~\bibnamefont
  {Choi}}, \bibinfo {author} {\bibfnamefont {S.}~\bibnamefont {Sachdev}},
  \bibinfo {author} {\bibfnamefont {M.}~\bibnamefont {Greiner}}, \bibinfo
  {author} {\bibfnamefont {V.}~\bibnamefont {Vuleti{\'{c}}}}, \ and\ \bibinfo
  {author} {\bibfnamefont {M.~D.}\ \bibnamefont {Lukin}},\ }\bibfield  {title}
  {\enquote {\bibinfo {title} {Quantum phases of matter on a 256-atom
  programmable quantum simulator},}\ }\href {\doibase
  10.1038/s41586-021-03582-4} {\bibfield  {journal} {\bibinfo  {journal}
  {Nature}\ }\textbf {\bibinfo {volume} {595}},\ \bibinfo {pages} {227--232}
  (\bibinfo {year} {2021})}\BibitemShut {NoStop}%
\bibitem [{\citenamefont {King}\ \emph {et~al.}(2022)\citenamefont {King},
  \citenamefont {Suzuki}, \citenamefont {Raymond}, \citenamefont {Zucca},
  \citenamefont {Lanting}, \citenamefont {Altomare}, \citenamefont {Berkley},
  \citenamefont {Ejtemaee}, \citenamefont {Hoskinson}, \citenamefont {Huang},
  \citenamefont {Ladizinsky}, \citenamefont {MacDonald}, \citenamefont
  {Marsden}, \citenamefont {Oh}, \citenamefont {Poulin-Lamarre}, \citenamefont
  {Reis}, \citenamefont {Rich}, \citenamefont {Sato}, \citenamefont
  {Whittaker}, \citenamefont {Yao}, \citenamefont {Harris}, \citenamefont
  {Lidar}, \citenamefont {Nishimori},\ and\ \citenamefont {Amin}}]{King2022}%
  \BibitemOpen
  \bibfield  {author} {\bibinfo {author} {\bibfnamefont {A.~D.}\ \bibnamefont
  {King}}, \bibinfo {author} {\bibfnamefont {S.}~\bibnamefont {Suzuki}},
  \bibinfo {author} {\bibfnamefont {J.}~\bibnamefont {Raymond}}, \bibinfo
  {author} {\bibfnamefont {A.}~\bibnamefont {Zucca}}, \bibinfo {author}
  {\bibfnamefont {T.}~\bibnamefont {Lanting}}, \bibinfo {author} {\bibfnamefont
  {F.}~\bibnamefont {Altomare}}, \bibinfo {author} {\bibfnamefont {A.~J.}\
  \bibnamefont {Berkley}}, \bibinfo {author} {\bibfnamefont {S.}~\bibnamefont
  {Ejtemaee}}, \bibinfo {author} {\bibfnamefont {E.}~\bibnamefont {Hoskinson}},
  \bibinfo {author} {\bibfnamefont {S.}~\bibnamefont {Huang}}, \bibinfo
  {author} {\bibfnamefont {E.}~\bibnamefont {Ladizinsky}}, \bibinfo {author}
  {\bibfnamefont {A.~J.~R.}\ \bibnamefont {MacDonald}}, \bibinfo {author}
  {\bibfnamefont {G.}~\bibnamefont {Marsden}}, \bibinfo {author} {\bibfnamefont
  {T.}~\bibnamefont {Oh}}, \bibinfo {author} {\bibfnamefont {G.}~\bibnamefont
  {Poulin-Lamarre}}, \bibinfo {author} {\bibfnamefont {M.}~\bibnamefont
  {Reis}}, \bibinfo {author} {\bibfnamefont {C.}~\bibnamefont {Rich}}, \bibinfo
  {author} {\bibfnamefont {Y.}~\bibnamefont {Sato}}, \bibinfo {author}
  {\bibfnamefont {J.~D.}\ \bibnamefont {Whittaker}}, \bibinfo {author}
  {\bibfnamefont {J.}~\bibnamefont {Yao}}, \bibinfo {author} {\bibfnamefont
  {R.}~\bibnamefont {Harris}}, \bibinfo {author} {\bibfnamefont {D.~A.}\
  \bibnamefont {Lidar}}, \bibinfo {author} {\bibfnamefont {H.}~\bibnamefont
  {Nishimori}}, \ and\ \bibinfo {author} {\bibfnamefont {M.~H.}\ \bibnamefont
  {Amin}},\ }\bibfield  {title} {\enquote {\bibinfo {title} {Coherent quantum
  annealing in a programmable 2,000{\thinspace}qubit ising chain},}\ }\href
  {\doibase 10.1038/s41567-022-01741-6} {\bibfield  {journal} {\bibinfo
  {journal} {Nature Physics}\ }\textbf {\bibinfo {volume} {18}},\ \bibinfo
  {pages} {1324--1328} (\bibinfo {year} {2022})}\BibitemShut {NoStop}%
\bibitem [{\citenamefont {Chae}\ \emph {et~al.}(2012)\citenamefont {Chae},
  \citenamefont {Lee}, \citenamefont {Horibe}, \citenamefont {Tanimura},
  \citenamefont {Mori}, \citenamefont {Gao}, \citenamefont {Carr},\ and\
  \citenamefont {Cheong}}]{Chae2012prl}%
  \BibitemOpen
  \bibfield  {author} {\bibinfo {author} {\bibfnamefont {S.~C.}\ \bibnamefont
  {Chae}}, \bibinfo {author} {\bibfnamefont {N.}~\bibnamefont {Lee}}, \bibinfo
  {author} {\bibfnamefont {Y.}~\bibnamefont {Horibe}}, \bibinfo {author}
  {\bibfnamefont {M.}~\bibnamefont {Tanimura}}, \bibinfo {author}
  {\bibfnamefont {S.}~\bibnamefont {Mori}}, \bibinfo {author} {\bibfnamefont
  {B.}~\bibnamefont {Gao}}, \bibinfo {author} {\bibfnamefont {S.}~\bibnamefont
  {Carr}}, \ and\ \bibinfo {author} {\bibfnamefont {S.-W.}\ \bibnamefont
  {Cheong}},\ }\bibfield  {title} {\enquote {\bibinfo {title} {{Direct
  Observation of the Proliferation of Ferroelectric Loop Domains and
  Vortex-Antivortex Pairs}},}\ }\href {\doibase 10.1103/PhysRevLett.108.167603}
  {\bibfield  {journal} {\bibinfo  {journal} {Phys. Rev. Lett.}\ }\textbf
  {\bibinfo {volume} {108}},\ \bibinfo {pages} {167603} (\bibinfo {year}
  {2012})}\BibitemShut {NoStop}%
\bibitem [{\citenamefont {Griffin}\ \emph {et~al.}(2012)\citenamefont
  {Griffin}, \citenamefont {Lilienblum}, \citenamefont {Delaney}, \citenamefont
  {Kumagai}, \citenamefont {Fiebig},\ and\ \citenamefont
  {Spaldin}}]{Griffin2012prx}%
  \BibitemOpen
  \bibfield  {author} {\bibinfo {author} {\bibfnamefont {S.~M.}\ \bibnamefont
  {Griffin}}, \bibinfo {author} {\bibfnamefont {M.}~\bibnamefont {Lilienblum}},
  \bibinfo {author} {\bibfnamefont {K.~T.}\ \bibnamefont {Delaney}}, \bibinfo
  {author} {\bibfnamefont {Y.}~\bibnamefont {Kumagai}}, \bibinfo {author}
  {\bibfnamefont {M.}~\bibnamefont {Fiebig}}, \ and\ \bibinfo {author}
  {\bibfnamefont {N.~A.}\ \bibnamefont {Spaldin}},\ }\bibfield  {title}
  {\enquote {\bibinfo {title} {{Scaling Behavior and Beyond Equilibrium in the
  Hexagonal Manganites}},}\ }\href {\doibase 10.1103/PhysRevX.2.041022}
  {\bibfield  {journal} {\bibinfo  {journal} {Phys. Rev. X}\ }\textbf {\bibinfo
  {volume} {2}},\ \bibinfo {pages} {041022} (\bibinfo {year}
  {2012})}\BibitemShut {NoStop}%
\bibitem [{\citenamefont {Lin}\ \emph {et~al.}(2014)\citenamefont {Lin},
  \citenamefont {Wang}, \citenamefont {Kamiya}, \citenamefont {Chern},
  \citenamefont {Fan}, \citenamefont {Fan}, \citenamefont {Casas},
  \citenamefont {Liu}, \citenamefont {Kiryukhin}, \citenamefont {Zurek},
  \citenamefont {Batista},\ and\ \citenamefont {Cheong}}]{Lin2014natphy}%
  \BibitemOpen
  \bibfield  {author} {\bibinfo {author} {\bibfnamefont {S.-Z.}\ \bibnamefont
  {Lin}}, \bibinfo {author} {\bibfnamefont {X.}~\bibnamefont {Wang}}, \bibinfo
  {author} {\bibfnamefont {Y.}~\bibnamefont {Kamiya}}, \bibinfo {author}
  {\bibfnamefont {G.-W.}\ \bibnamefont {Chern}}, \bibinfo {author}
  {\bibfnamefont {F.}~\bibnamefont {Fan}}, \bibinfo {author} {\bibfnamefont
  {D.}~\bibnamefont {Fan}}, \bibinfo {author} {\bibfnamefont {B.}~\bibnamefont
  {Casas}}, \bibinfo {author} {\bibfnamefont {Y.}~\bibnamefont {Liu}}, \bibinfo
  {author} {\bibfnamefont {V.}~\bibnamefont {Kiryukhin}}, \bibinfo {author}
  {\bibfnamefont {W.~H.}\ \bibnamefont {Zurek}}, \bibinfo {author}
  {\bibfnamefont {C.~D.}\ \bibnamefont {Batista}}, \ and\ \bibinfo {author}
  {\bibfnamefont {S.-W.}\ \bibnamefont {Cheong}},\ }\bibfield  {title}
  {\enquote {\bibinfo {title} {{Topological defects as relics of emergent
  continuous symmetry and Higgs condensation of disorder in
  ferroelectrics}},}\ }\href {\doibase 10.1038/nphys3142} {\bibfield  {journal}
  {\bibinfo  {journal} {Nature Physics}\ }\textbf {\bibinfo {volume} {10}},\
  \bibinfo {pages} {970--977} (\bibinfo {year} {2014})}\BibitemShut {NoStop}%
\bibitem [{\citenamefont {Meier}\ \emph {et~al.}(2017)\citenamefont {Meier},
  \citenamefont {Lilienblum}, \citenamefont {Griffin}, \citenamefont {Conder},
  \citenamefont {Pomjakushina}, \citenamefont {Yan}, \citenamefont {Bourret},
  \citenamefont {Meier}, \citenamefont {Lichtenberg}, \citenamefont {Salje},
  \citenamefont {Spaldin}, \citenamefont {Fiebig},\ and\ \citenamefont
  {Cano}}]{Meier2017prx}%
  \BibitemOpen
  \bibfield  {author} {\bibinfo {author} {\bibfnamefont {Q.~N.}\ \bibnamefont
  {Meier}}, \bibinfo {author} {\bibfnamefont {M.}~\bibnamefont {Lilienblum}},
  \bibinfo {author} {\bibfnamefont {S.~M.}\ \bibnamefont {Griffin}}, \bibinfo
  {author} {\bibfnamefont {K.}~\bibnamefont {Conder}}, \bibinfo {author}
  {\bibfnamefont {E.}~\bibnamefont {Pomjakushina}}, \bibinfo {author}
  {\bibfnamefont {Z.}~\bibnamefont {Yan}}, \bibinfo {author} {\bibfnamefont
  {E.}~\bibnamefont {Bourret}}, \bibinfo {author} {\bibfnamefont
  {D.}~\bibnamefont {Meier}}, \bibinfo {author} {\bibfnamefont
  {F.}~\bibnamefont {Lichtenberg}}, \bibinfo {author} {\bibfnamefont
  {E.~K.~H.}\ \bibnamefont {Salje}}, \bibinfo {author} {\bibfnamefont {N.~A.}\
  \bibnamefont {Spaldin}}, \bibinfo {author} {\bibfnamefont {M.}~\bibnamefont
  {Fiebig}}, \ and\ \bibinfo {author} {\bibfnamefont {A.}~\bibnamefont
  {Cano}},\ }\bibfield  {title} {\enquote {\bibinfo {title} {{Global Formation
  of Topological Defects in the Multiferroic Hexagonal Manganites}},}\ }\href
  {\doibase 10.1103/PhysRevX.7.041014} {\bibfield  {journal} {\bibinfo
  {journal} {Phys. Rev. X}\ }\textbf {\bibinfo {volume} {7}},\ \bibinfo {pages}
  {041014} (\bibinfo {year} {2017})}\BibitemShut {NoStop}%
\bibitem [{\citenamefont {Skj\ae{}rv\o{}}\ \emph {et~al.}(2019)\citenamefont
  {Skj\ae{}rv\o{}}, \citenamefont {Meier}, \citenamefont {Feygenson},
  \citenamefont {Spaldin}, \citenamefont {Billinge}, \citenamefont {Bozin},\
  and\ \citenamefont {Selbach}}]{Skjaervo2019prx}%
  \BibitemOpen
  \bibfield  {author} {\bibinfo {author} {\bibfnamefont {S.~H.}\ \bibnamefont
  {Skj\ae{}rv\o{}}}, \bibinfo {author} {\bibfnamefont {Q.~N.}\ \bibnamefont
  {Meier}}, \bibinfo {author} {\bibfnamefont {M.}~\bibnamefont {Feygenson}},
  \bibinfo {author} {\bibfnamefont {N.~A.}\ \bibnamefont {Spaldin}}, \bibinfo
  {author} {\bibfnamefont {S.~J.~L.}\ \bibnamefont {Billinge}}, \bibinfo
  {author} {\bibfnamefont {E.~S.}\ \bibnamefont {Bozin}}, \ and\ \bibinfo
  {author} {\bibfnamefont {S.~M.}\ \bibnamefont {Selbach}},\ }\bibfield
  {title} {\enquote {\bibinfo {title} {{Unconventional Continuous Structural
  Disorder at the Order-Disorder Phase Transition in the Hexagonal
  Manganites}},}\ }\href {\doibase 10.1103/PhysRevX.9.031001} {\bibfield
  {journal} {\bibinfo  {journal} {Phys. Rev. X}\ }\textbf {\bibinfo {volume}
  {9}},\ \bibinfo {pages} {031001} (\bibinfo {year} {2019})}\BibitemShut
  {NoStop}%
\bibitem [{\citenamefont {Meier}\ \emph {et~al.}(2020)\citenamefont {Meier},
  \citenamefont {Stucky}, \citenamefont {Teyssier}, \citenamefont {Griffin},
  \citenamefont {van~der Marel},\ and\ \citenamefont {Spaldin}}]{Meier2020prb}%
  \BibitemOpen
  \bibfield  {author} {\bibinfo {author} {\bibfnamefont {Q.~N.}\ \bibnamefont
  {Meier}}, \bibinfo {author} {\bibfnamefont {A.}~\bibnamefont {Stucky}},
  \bibinfo {author} {\bibfnamefont {J.}~\bibnamefont {Teyssier}}, \bibinfo
  {author} {\bibfnamefont {S.~M.}\ \bibnamefont {Griffin}}, \bibinfo {author}
  {\bibfnamefont {D.}~\bibnamefont {van~der Marel}}, \ and\ \bibinfo {author}
  {\bibfnamefont {N.~A.}\ \bibnamefont {Spaldin}},\ }\bibfield  {title}
  {\enquote {\bibinfo {title} {Manifestation of structural higgs and goldstone
  modes in the hexagonal manganites},}\ }\href {\doibase
  10.1103/PhysRevB.102.014102} {\bibfield  {journal} {\bibinfo  {journal}
  {Phys. Rev. B}\ }\textbf {\bibinfo {volume} {102}},\ \bibinfo {pages}
  {014102} (\bibinfo {year} {2020})}\BibitemShut {NoStop}%
\bibitem [{\citenamefont {Zhang}\ \emph {et~al.}(2021)\citenamefont {Zhang},
  \citenamefont {Ye},\ and\ \citenamefont {Li}}]{Zhang2021prb}%
  \BibitemOpen
  \bibfield  {author} {\bibinfo {author} {\bibfnamefont {X.}~\bibnamefont
  {Zhang}}, \bibinfo {author} {\bibfnamefont {Q.-J.}\ \bibnamefont {Ye}}, \
  and\ \bibinfo {author} {\bibfnamefont {X.-Z.}\ \bibnamefont {Li}},\
  }\bibfield  {title} {\enquote {\bibinfo {title} {{Structural phase transition
  and Goldstone-like mode in hexagonal ${\mathrm{BaMnO}}_{3}$}},}\ }\href
  {\doibase 10.1103/PhysRevB.103.024101} {\bibfield  {journal} {\bibinfo
  {journal} {Phys. Rev. B}\ }\textbf {\bibinfo {volume} {103}},\ \bibinfo
  {pages} {024101} (\bibinfo {year} {2021})}\BibitemShut {NoStop}%
\bibitem [{\citenamefont {Sandvik}\ \emph {et~al.}(2023)\citenamefont
  {Sandvik}, \citenamefont {M\"u{}ller}, \citenamefont {\AA{}nes},
  \citenamefont {Zahn}, \citenamefont {He}, \citenamefont {Fiebig},
  \citenamefont {Lottermoser}, \citenamefont {Rojac}, \citenamefont {Meier},\
  and\ \citenamefont {Schulthei\ss}}]{OWSandvik2023nl}%
  \BibitemOpen
  \bibfield  {author} {\bibinfo {author} {\bibfnamefont {O.~W.}\ \bibnamefont
  {Sandvik}}, \bibinfo {author} {\bibfnamefont {A.~M.}\ \bibnamefont
  {M\"u{}ller}}, \bibinfo {author} {\bibfnamefont {H.~W.}\ \bibnamefont
  {\AA{}nes}}, \bibinfo {author} {\bibfnamefont {M.}~\bibnamefont {Zahn}},
  \bibinfo {author} {\bibfnamefont {J.}~\bibnamefont {He}}, \bibinfo {author}
  {\bibfnamefont {M.}~\bibnamefont {Fiebig}}, \bibinfo {author} {\bibfnamefont
  {T.}~\bibnamefont {Lottermoser}}, \bibinfo {author} {\bibfnamefont
  {T.}~\bibnamefont {Rojac}}, \bibinfo {author} {\bibfnamefont
  {D.}~\bibnamefont {Meier}}, \ and\ \bibinfo {author} {\bibfnamefont
  {J.}~\bibnamefont {Schulthei\ss}},\ }\bibfield  {title} {\enquote {\bibinfo
  {title} {{Pressure Control of Nonferroelastic Ferroelectric Domains in
  ErMnO$_3$}},}\ }\href {\doibase 10.1021/acs.nanolett.3c01638} {\bibfield
  {journal} {\bibinfo  {journal} {Nano Letters}\ }\textbf {\bibinfo {volume}
  {23}},\ \bibinfo {pages} {6994--7000} (\bibinfo {year} {2023})}\BibitemShut
  {NoStop}%
\bibitem [{\citenamefont {Kang}\ \emph {et~al.}(2023)\citenamefont {Kang},
  \citenamefont {Gao}, \citenamefont {Guo}, \citenamefont {Zhu}, \citenamefont
  {Huang}, \citenamefont {Hong}, \citenamefont {Cheong},\ and\ \citenamefont
  {Wang}}]{Kang2023jap}%
  \BibitemOpen
  \bibfield  {author} {\bibinfo {author} {\bibfnamefont {J.}~\bibnamefont
  {Kang}}, \bibinfo {author} {\bibfnamefont {Z.}~\bibnamefont {Gao}}, \bibinfo
  {author} {\bibfnamefont {C.}~\bibnamefont {Guo}}, \bibinfo {author}
  {\bibfnamefont {W.}~\bibnamefont {Zhu}}, \bibinfo {author} {\bibfnamefont
  {H.}~\bibnamefont {Huang}}, \bibinfo {author} {\bibfnamefont
  {J.}~\bibnamefont {Hong}}, \bibinfo {author} {\bibfnamefont {S.-W.}\
  \bibnamefont {Cheong}}, \ and\ \bibinfo {author} {\bibfnamefont
  {X.}~\bibnamefont {Wang}},\ }\bibfield  {title} {\enquote {\bibinfo {title}
  {{A snapshot of domain evolution between topological vortex and stripe in
  ferroelectric hexagonal ErMnO3}},}\ }\href {\doibase 10.1063/5.0138096}
  {\bibfield  {journal} {\bibinfo  {journal} {Journal of Applied Physics}\
  }\textbf {\bibinfo {volume} {133}},\ \bibinfo {pages} {124102} (\bibinfo
  {year} {2023})}\BibitemShut {NoStop}%
\bibitem [{\citenamefont {Baghizadeh}\ \emph {et~al.}(2019)\citenamefont
  {Baghizadeh}, \citenamefont {Mirzadeh~Vaghefi}, \citenamefont {Alikin},
  \citenamefont {Amaral}, \citenamefont {Amaral},\ and\ \citenamefont
  {Vieira}}]{Baghizadeh2019jpcc}%
  \BibitemOpen
  \bibfield  {author} {\bibinfo {author} {\bibfnamefont {A.}~\bibnamefont
  {Baghizadeh}}, \bibinfo {author} {\bibfnamefont {P.}~\bibnamefont
  {Mirzadeh~Vaghefi}}, \bibinfo {author} {\bibfnamefont {D.~O.}\ \bibnamefont
  {Alikin}}, \bibinfo {author} {\bibfnamefont {J.~S.}\ \bibnamefont {Amaral}},
  \bibinfo {author} {\bibfnamefont {V.~S.}\ \bibnamefont {Amaral}}, \ and\
  \bibinfo {author} {\bibfnamefont {J.~M.}\ \bibnamefont {Vieira}},\ }\bibfield
   {title} {\enquote {\bibinfo {title} {{Link of Weak Ferromagnetism to
  Emergence of Topological Vortices in Bulk Ceramics of
  h-${\mathrm{LuMn}}_{x}{\mathrm{O}}_{3}$ Manganite}},}\ }\href {\doibase
  10.1021/acs.jpcc.8b11253} {\bibfield  {journal} {\bibinfo  {journal} {The
  Journal of Physical Chemistry C}\ }\textbf {\bibinfo {volume} {123}},\
  \bibinfo {pages} {6158--6166} (\bibinfo {year} {2019})}\BibitemShut {NoStop}%
\bibitem [{\citenamefont {Juraschek}\ \emph {et~al.}(2020)\citenamefont
  {Juraschek}, \citenamefont {Meier},\ and\ \citenamefont
  {Narang}}]{Juraschek2020prl}%
  \BibitemOpen
  \bibfield  {author} {\bibinfo {author} {\bibfnamefont {D.~M.}\ \bibnamefont
  {Juraschek}}, \bibinfo {author} {\bibfnamefont {Q.~N.}\ \bibnamefont
  {Meier}}, \ and\ \bibinfo {author} {\bibfnamefont {P.}~\bibnamefont
  {Narang}},\ }\bibfield  {title} {\enquote {\bibinfo {title} {Parametric
  excitation of an optically silent goldstone-like phonon mode},}\ }\href
  {\doibase 10.1103/PhysRevLett.124.117401} {\bibfield  {journal} {\bibinfo
  {journal} {Phys. Rev. Lett.}\ }\textbf {\bibinfo {volume} {124}},\ \bibinfo
  {pages} {117401} (\bibinfo {year} {2020})}\BibitemShut {NoStop}%
\bibitem [{\citenamefont {Oshikawa}(2000)}]{Oshikawa2000prb}%
  \BibitemOpen
  \bibfield  {author} {\bibinfo {author} {\bibfnamefont {M.}~\bibnamefont
  {Oshikawa}},\ }\bibfield  {title} {\enquote {\bibinfo {title} {{Ordered phase
  and scaling in ${Z}_{n}$ models and the three-state antiferromagnetic Potts
  model in three dimensions}},}\ }\href {\doibase 10.1103/PhysRevB.61.3430}
  {\bibfield  {journal} {\bibinfo  {journal} {Phys. Rev. B}\ }\textbf {\bibinfo
  {volume} {61}},\ \bibinfo {pages} {3430--3434} (\bibinfo {year}
  {2000})}\BibitemShut {NoStop}%
\bibitem [{\citenamefont {Lou}\ \emph {et~al.}(2007)\citenamefont {Lou},
  \citenamefont {Sandvik},\ and\ \citenamefont {Balents}}]{Lou2007prl}%
  \BibitemOpen
  \bibfield  {author} {\bibinfo {author} {\bibfnamefont {J.}~\bibnamefont
  {Lou}}, \bibinfo {author} {\bibfnamefont {A.~W.}\ \bibnamefont {Sandvik}}, \
  and\ \bibinfo {author} {\bibfnamefont {L.}~\bibnamefont {Balents}},\
  }\bibfield  {title} {\enquote {\bibinfo {title} {{Emergence of U(1) Symmetry
  in the 3D $XY$ Model with ${Z}_{q}$ Anisotropy}},}\ }\href {\doibase
  10.1103/PhysRevLett.99.207203} {\bibfield  {journal} {\bibinfo  {journal}
  {Phys. Rev. Lett.}\ }\textbf {\bibinfo {volume} {99}},\ \bibinfo {pages}
  {207203} (\bibinfo {year} {2007})}\BibitemShut {NoStop}%
\bibitem [{\citenamefont {Okubo}\ \emph {et~al.}(2015)\citenamefont {Okubo},
  \citenamefont {Oshikawa}, \citenamefont {Watanabe},\ and\ \citenamefont
  {Kawashima}}]{Okubo2015prb}%
  \BibitemOpen
  \bibfield  {author} {\bibinfo {author} {\bibfnamefont {T.}~\bibnamefont
  {Okubo}}, \bibinfo {author} {\bibfnamefont {K.}~\bibnamefont {Oshikawa}},
  \bibinfo {author} {\bibfnamefont {H.}~\bibnamefont {Watanabe}}, \ and\
  \bibinfo {author} {\bibfnamefont {N.}~\bibnamefont {Kawashima}},\ }\bibfield
  {title} {\enquote {\bibinfo {title} {{Scaling relation for dangerously
  irrelevant symmetry-breaking fields}},}\ }\href {\doibase
  10.1103/PhysRevB.91.174417} {\bibfield  {journal} {\bibinfo  {journal} {Phys.
  Rev. B}\ }\textbf {\bibinfo {volume} {91}},\ \bibinfo {pages} {174417}
  (\bibinfo {year} {2015})}\BibitemShut {NoStop}%
\bibitem [{\citenamefont {L\'eonard}\ and\ \citenamefont
  {Delamotte}(2015)}]{Leonard2015prl}%
  \BibitemOpen
  \bibfield  {author} {\bibinfo {author} {\bibfnamefont {F.}~\bibnamefont
  {L\'eonard}}\ and\ \bibinfo {author} {\bibfnamefont {B.}~\bibnamefont
  {Delamotte}},\ }\bibfield  {title} {\enquote {\bibinfo {title} {{Critical
  Exponents Can Be Different on the Two Sides of a Transition: A Generic
  Mechanism}},}\ }\href {\doibase 10.1103/PhysRevLett.115.200601} {\bibfield
  {journal} {\bibinfo  {journal} {Phys. Rev. Lett.}\ }\textbf {\bibinfo
  {volume} {115}},\ \bibinfo {pages} {200601} (\bibinfo {year}
  {2015})}\BibitemShut {NoStop}%
\bibitem [{\citenamefont {Pujari}\ \emph {et~al.}(2015)\citenamefont {Pujari},
  \citenamefont {Alet},\ and\ \citenamefont {Damle}}]{Pujari2015prb}%
  \BibitemOpen
  \bibfield  {author} {\bibinfo {author} {\bibfnamefont {S.}~\bibnamefont
  {Pujari}}, \bibinfo {author} {\bibfnamefont {F.}~\bibnamefont {Alet}}, \ and\
  \bibinfo {author} {\bibfnamefont {K.}~\bibnamefont {Damle}},\ }\bibfield
  {title} {\enquote {\bibinfo {title} {{Transitions to valence-bond solid order
  in a honeycomb lattice antiferromagnet}},}\ }\href {\doibase
  10.1103/PhysRevB.91.104411} {\bibfield  {journal} {\bibinfo  {journal} {Phys.
  Rev. B}\ }\textbf {\bibinfo {volume} {91}},\ \bibinfo {pages} {104411}
  (\bibinfo {year} {2015})}\BibitemShut {NoStop}%
\bibitem [{\citenamefont {Ding}\ \emph {et~al.}(2016)\citenamefont {Ding},
  \citenamefont {Bl\"ote},\ and\ \citenamefont {Deng}}]{Ding2016prb}%
  \BibitemOpen
  \bibfield  {author} {\bibinfo {author} {\bibfnamefont {C.}~\bibnamefont
  {Ding}}, \bibinfo {author} {\bibfnamefont {H.~W.~J.}\ \bibnamefont
  {Bl\"ote}}, \ and\ \bibinfo {author} {\bibfnamefont {Y.}~\bibnamefont
  {Deng}},\ }\bibfield  {title} {\enquote {\bibinfo {title} {{Emergent O($n$)
  symmetry in a series of three-dimensional Potts models}},}\ }\href {\doibase
  10.1103/PhysRevB.94.104402} {\bibfield  {journal} {\bibinfo  {journal} {Phys.
  Rev. B}\ }\textbf {\bibinfo {volume} {94}},\ \bibinfo {pages} {104402}
  (\bibinfo {year} {2016})}\BibitemShut {NoStop}%
\bibitem [{\citenamefont {Hasenbusch}\ and\ \citenamefont
  {Vicari}(2011)}]{Hasenbusch2011prb}%
  \BibitemOpen
  \bibfield  {author} {\bibinfo {author} {\bibfnamefont {M.}~\bibnamefont
  {Hasenbusch}}\ and\ \bibinfo {author} {\bibfnamefont {E.}~\bibnamefont
  {Vicari}},\ }\bibfield  {title} {\enquote {\bibinfo {title} {{Anisotropic
  perturbations in three-dimensional O($N$)-symmetric vector models}},}\ }\href
  {\doibase 10.1103/PhysRevB.84.125136} {\bibfield  {journal} {\bibinfo
  {journal} {Phys. Rev. B}\ }\textbf {\bibinfo {volume} {84}},\ \bibinfo
  {pages} {125136} (\bibinfo {year} {2011})}\BibitemShut {NoStop}%
\bibitem [{\citenamefont {Shao}\ \emph {et~al.}(2020)\citenamefont {Shao},
  \citenamefont {Guo},\ and\ \citenamefont {Sandvik}}]{Shao2020prl}%
  \BibitemOpen
  \bibfield  {author} {\bibinfo {author} {\bibfnamefont {H.}~\bibnamefont
  {Shao}}, \bibinfo {author} {\bibfnamefont {W.}~\bibnamefont {Guo}}, \ and\
  \bibinfo {author} {\bibfnamefont {A.~W.}\ \bibnamefont {Sandvik}},\
  }\bibfield  {title} {\enquote {\bibinfo {title} {{Monte Carlo Renormalization
  Flows in the Space of Relevant and Irrelevant Operators: Application to
  Three-Dimensional Clock Models}},}\ }\href {\doibase
  10.1103/PhysRevLett.124.080602} {\bibfield  {journal} {\bibinfo  {journal}
  {Phys. Rev. Lett.}\ }\textbf {\bibinfo {volume} {124}},\ \bibinfo {pages}
  {080602} (\bibinfo {year} {2020})}\BibitemShut {NoStop}%
\bibitem [{\citenamefont {Patil}\ \emph {et~al.}(2021)\citenamefont {Patil},
  \citenamefont {Shao},\ and\ \citenamefont {Sandvik}}]{Patil2021prb}%
  \BibitemOpen
  \bibfield  {author} {\bibinfo {author} {\bibfnamefont {P.}~\bibnamefont
  {Patil}}, \bibinfo {author} {\bibfnamefont {H.}~\bibnamefont {Shao}}, \ and\
  \bibinfo {author} {\bibfnamefont {A.~W.}\ \bibnamefont {Sandvik}},\
  }\bibfield  {title} {\enquote {\bibinfo {title} {{Unconventional U(1) to
  ${Z}_{q}$ crossover in quantum and classical $q$-state clock models}},}\
  }\href {\doibase 10.1103/PhysRevB.103.054418} {\bibfield  {journal} {\bibinfo
   {journal} {Phys. Rev. B}\ }\textbf {\bibinfo {volume} {103}},\ \bibinfo
  {pages} {054418} (\bibinfo {year} {2021})}\BibitemShut {NoStop}%
\bibitem [{\citenamefont {Li}\ \emph {et~al.}(2017)\citenamefont {Li},
  \citenamefont {Jiang}, \citenamefont {Jian},\ and\ \citenamefont
  {Yao}}]{Li2017nc}%
  \BibitemOpen
  \bibfield  {author} {\bibinfo {author} {\bibfnamefont {Z.-X.}\ \bibnamefont
  {Li}}, \bibinfo {author} {\bibfnamefont {Y.-F.}\ \bibnamefont {Jiang}},
  \bibinfo {author} {\bibfnamefont {S.-K.}\ \bibnamefont {Jian}}, \ and\
  \bibinfo {author} {\bibfnamefont {H.}~\bibnamefont {Yao}},\ }\bibfield
  {title} {\enquote {\bibinfo {title} {{Fermion-induced quantum critical
  points}},}\ }\href {\doibase 10.1038/s41467-017-00167-6} {\bibfield
  {journal} {\bibinfo  {journal} {Nature Communications}\ }\textbf {\bibinfo
  {volume} {8}},\ \bibinfo {pages} {314} (\bibinfo {year} {2017})}\BibitemShut
  {NoStop}%
\bibitem [{\citenamefont {Jian}\ and\ \citenamefont
  {Yao}(2017{\natexlab{a}})}]{Jian2017prb}%
  \BibitemOpen
  \bibfield  {author} {\bibinfo {author} {\bibfnamefont {S.-K.}\ \bibnamefont
  {Jian}}\ and\ \bibinfo {author} {\bibfnamefont {H.}~\bibnamefont {Yao}},\
  }\bibfield  {title} {\enquote {\bibinfo {title} {Fermion-induced quantum
  critical points in three-dimensional weyl semimetals},}\ }\href {\doibase
  10.1103/PhysRevB.96.155112} {\bibfield  {journal} {\bibinfo  {journal} {Phys.
  Rev. B}\ }\textbf {\bibinfo {volume} {96}},\ \bibinfo {pages} {155112}
  (\bibinfo {year} {2017}{\natexlab{a}})}\BibitemShut {NoStop}%
\bibitem [{\citenamefont {Jian}\ and\ \citenamefont
  {Yao}(2017{\natexlab{b}})}]{Jian2018prb}%
  \BibitemOpen
  \bibfield  {author} {\bibinfo {author} {\bibfnamefont {S.-K.}\ \bibnamefont
  {Jian}}\ and\ \bibinfo {author} {\bibfnamefont {H.}~\bibnamefont {Yao}},\
  }\bibfield  {title} {\enquote {\bibinfo {title} {Fermion-induced quantum
  critical points in two-dimensional dirac semimetals},}\ }\href {\doibase
  10.1103/PhysRevB.96.195162} {\bibfield  {journal} {\bibinfo  {journal} {Phys.
  Rev. B}\ }\textbf {\bibinfo {volume} {96}},\ \bibinfo {pages} {195162}
  (\bibinfo {year} {2017}{\natexlab{b}})}\BibitemShut {NoStop}%
\bibitem [{\citenamefont {Senthil}\ \emph
  {et~al.}(2004{\natexlab{a}})\citenamefont {Senthil}, \citenamefont
  {Vishwanath}, \citenamefont {Balents}, \citenamefont {Sachdev},\ and\
  \citenamefont {Fisher}}]{Senthil2004sci}%
  \BibitemOpen
  \bibfield  {author} {\bibinfo {author} {\bibfnamefont {T.}~\bibnamefont
  {Senthil}}, \bibinfo {author} {\bibfnamefont {A.}~\bibnamefont {Vishwanath}},
  \bibinfo {author} {\bibfnamefont {L.}~\bibnamefont {Balents}}, \bibinfo
  {author} {\bibfnamefont {S.}~\bibnamefont {Sachdev}}, \ and\ \bibinfo
  {author} {\bibfnamefont {M.~P.~A.}\ \bibnamefont {Fisher}},\ }\bibfield
  {title} {\enquote {\bibinfo {title} {Deconfined quantum critical points},}\
  }\href {\doibase 10.1126/science.1091806} {\bibfield  {journal} {\bibinfo
  {journal} {Science}\ }\textbf {\bibinfo {volume} {303}},\ \bibinfo {pages}
  {1490--1494} (\bibinfo {year} {2004}{\natexlab{a}})}\BibitemShut {NoStop}%
\bibitem [{\citenamefont {Senthil}\ \emph
  {et~al.}(2004{\natexlab{b}})\citenamefont {Senthil}, \citenamefont {Balents},
  \citenamefont {Sachdev}, \citenamefont {Vishwanath},\ and\ \citenamefont
  {Fisher}}]{Senthil2004prb}%
  \BibitemOpen
  \bibfield  {author} {\bibinfo {author} {\bibfnamefont {T.}~\bibnamefont
  {Senthil}}, \bibinfo {author} {\bibfnamefont {L.}~\bibnamefont {Balents}},
  \bibinfo {author} {\bibfnamefont {S.}~\bibnamefont {Sachdev}}, \bibinfo
  {author} {\bibfnamefont {A.}~\bibnamefont {Vishwanath}}, \ and\ \bibinfo
  {author} {\bibfnamefont {M.~P.~A.}\ \bibnamefont {Fisher}},\ }\bibfield
  {title} {\enquote {\bibinfo {title} {{Quantum criticality beyond the
  Landau-Ginzburg-Wilson paradigm}},}\ }\href {\doibase
  10.1103/PhysRevB.70.144407} {\bibfield  {journal} {\bibinfo  {journal} {Phys.
  Rev. B}\ }\textbf {\bibinfo {volume} {70}},\ \bibinfo {pages} {144407}
  (\bibinfo {year} {2004}{\natexlab{b}})}\BibitemShut {NoStop}%
\bibitem [{\citenamefont {Nahum}\ \emph {et~al.}(2015)\citenamefont {Nahum},
  \citenamefont {Serna}, \citenamefont {Chalker}, \citenamefont {Ortu\~no},\
  and\ \citenamefont {Somoza}}]{Nahum2015prl}%
  \BibitemOpen
  \bibfield  {author} {\bibinfo {author} {\bibfnamefont {A.}~\bibnamefont
  {Nahum}}, \bibinfo {author} {\bibfnamefont {P.}~\bibnamefont {Serna}},
  \bibinfo {author} {\bibfnamefont {J.~T.}\ \bibnamefont {Chalker}}, \bibinfo
  {author} {\bibfnamefont {M.}~\bibnamefont {Ortu\~no}}, \ and\ \bibinfo
  {author} {\bibfnamefont {A.~M.}\ \bibnamefont {Somoza}},\ }\bibfield  {title}
  {\enquote {\bibinfo {title} {{Emergent SO($5$) Symmetry at the N\'eel to
  Valence-Bond-Solid Transition}},}\ }\href {\doibase
  10.1103/PhysRevLett.115.267203} {\bibfield  {journal} {\bibinfo  {journal}
  {Phys. Rev. Lett.}\ }\textbf {\bibinfo {volume} {115}},\ \bibinfo {pages}
  {267203} (\bibinfo {year} {2015})}\BibitemShut {NoStop}%
\bibitem [{\citenamefont {Wang}\ \emph {et~al.}(2017)\citenamefont {Wang},
  \citenamefont {Nahum}, \citenamefont {Metlitski}, \citenamefont {Xu},\ and\
  \citenamefont {Senthil}}]{Wang2017prx}%
  \BibitemOpen
  \bibfield  {author} {\bibinfo {author} {\bibfnamefont {C.}~\bibnamefont
  {Wang}}, \bibinfo {author} {\bibfnamefont {A.}~\bibnamefont {Nahum}},
  \bibinfo {author} {\bibfnamefont {M.~A.}\ \bibnamefont {Metlitski}}, \bibinfo
  {author} {\bibfnamefont {C.}~\bibnamefont {Xu}}, \ and\ \bibinfo {author}
  {\bibfnamefont {T.}~\bibnamefont {Senthil}},\ }\bibfield  {title} {\enquote
  {\bibinfo {title} {{Deconfined Quantum Critical Points: Symmetries and
  Dualities}},}\ }\href {\doibase 10.1103/PhysRevX.7.031051} {\bibfield
  {journal} {\bibinfo  {journal} {Phys. Rev. X}\ }\textbf {\bibinfo {volume}
  {7}},\ \bibinfo {pages} {031051} (\bibinfo {year} {2017})}\BibitemShut
  {NoStop}%
\bibitem [{\citenamefont {Takahashi}\ and\ \citenamefont
  {Sandvik}(2020)}]{Takahashi2020prr}%
  \BibitemOpen
  \bibfield  {author} {\bibinfo {author} {\bibfnamefont {J.}~\bibnamefont
  {Takahashi}}\ and\ \bibinfo {author} {\bibfnamefont {A.~W.}\ \bibnamefont
  {Sandvik}},\ }\bibfield  {title} {\enquote {\bibinfo {title} {{Valence-bond
  solids, vestigial order, and emergent SO(5) symmetry in a two-dimensional
  quantum magnet}},}\ }\href {\doibase 10.1103/PhysRevResearch.2.033459}
  {\bibfield  {journal} {\bibinfo  {journal} {Phys. Rev. Res.}\ }\textbf
  {\bibinfo {volume} {2}},\ \bibinfo {pages} {033459} (\bibinfo {year}
  {2020})}\BibitemShut {NoStop}%
\bibitem [{\citenamefont {Ma}\ \emph {et~al.}(2019)\citenamefont {Ma},
  \citenamefont {You},\ and\ \citenamefont {Meng}}]{Ma2019prl}%
  \BibitemOpen
  \bibfield  {author} {\bibinfo {author} {\bibfnamefont {N.}~\bibnamefont
  {Ma}}, \bibinfo {author} {\bibfnamefont {Y.-Z.}\ \bibnamefont {You}}, \ and\
  \bibinfo {author} {\bibfnamefont {Z.~Y.}\ \bibnamefont {Meng}},\ }\bibfield
  {title} {\enquote {\bibinfo {title} {Role of noether's theorem at the
  deconfined quantum critical point},}\ }\href {\doibase
  10.1103/PhysRevLett.122.175701} {\bibfield  {journal} {\bibinfo  {journal}
  {Phys. Rev. Lett.}\ }\textbf {\bibinfo {volume} {122}},\ \bibinfo {pages}
  {175701} (\bibinfo {year} {2019})}\BibitemShut {NoStop}%
\bibitem [{\citenamefont {Torres}\ \emph {et~al.}(2018)\citenamefont {Torres},
  \citenamefont {Classen}, \citenamefont {Herbut},\ and\ \citenamefont
  {Scherer}}]{Torres2018prb}%
  \BibitemOpen
  \bibfield  {author} {\bibinfo {author} {\bibfnamefont {E.}~\bibnamefont
  {Torres}}, \bibinfo {author} {\bibfnamefont {L.}~\bibnamefont {Classen}},
  \bibinfo {author} {\bibfnamefont {I.~F.}\ \bibnamefont {Herbut}}, \ and\
  \bibinfo {author} {\bibfnamefont {M.~M.}\ \bibnamefont {Scherer}},\
  }\bibfield  {title} {\enquote {\bibinfo {title} {Fermion-induced quantum
  criticality with two length scales in dirac systems},}\ }\href {\doibase
  10.1103/PhysRevB.97.125137} {\bibfield  {journal} {\bibinfo  {journal} {Phys.
  Rev. B}\ }\textbf {\bibinfo {volume} {97}},\ \bibinfo {pages} {125137}
  (\bibinfo {year} {2018})}\BibitemShut {NoStop}%
\bibitem [{\citenamefont {Classen}\ \emph {et~al.}(2017)\citenamefont
  {Classen}, \citenamefont {Herbut},\ and\ \citenamefont
  {Scherer}}]{Classen2017prb}%
  \BibitemOpen
  \bibfield  {author} {\bibinfo {author} {\bibfnamefont {L.}~\bibnamefont
  {Classen}}, \bibinfo {author} {\bibfnamefont {I.~F.}\ \bibnamefont {Herbut}},
  \ and\ \bibinfo {author} {\bibfnamefont {M.~M.}\ \bibnamefont {Scherer}},\
  }\bibfield  {title} {\enquote {\bibinfo {title} {Fluctuation-induced
  continuous transition and quantum criticality in dirac semimetals},}\ }\href
  {\doibase 10.1103/PhysRevB.96.115132} {\bibfield  {journal} {\bibinfo
  {journal} {Phys. Rev. B}\ }\textbf {\bibinfo {volume} {96}},\ \bibinfo
  {pages} {115132} (\bibinfo {year} {2017})}\BibitemShut {NoStop}%
\bibitem [{\citenamefont {Nelson}(1976)}]{Nelson1976prb}%
  \BibitemOpen
  \bibfield  {author} {\bibinfo {author} {\bibfnamefont {D.~R.}\ \bibnamefont
  {Nelson}},\ }\bibfield  {title} {\enquote {\bibinfo {title}
  {{Coexistence-curve singularities in isotropic ferromagnets}},}\ }\href
  {\doibase 10.1103/PhysRevB.13.2222} {\bibfield  {journal} {\bibinfo
  {journal} {Phys. Rev. B}\ }\textbf {\bibinfo {volume} {13}},\ \bibinfo
  {pages} {2222--2230} (\bibinfo {year} {1976})}\BibitemShut {NoStop}%
\bibitem [{\citenamefont {Miyashita}(1997)}]{Miyashita1997jpsj}%
  \BibitemOpen
  \bibfield  {author} {\bibinfo {author} {\bibfnamefont {S.}~\bibnamefont
  {Miyashita}},\ }\bibfield  {title} {\enquote {\bibinfo {title} {{Nature of
  the Ordered Phase and the Critical Properties of the Three Dimensional
  Six-State Clock Model}},}\ }\href {\doibase 10.1143/JPSJ.66.3411} {\bibfield
  {journal} {\bibinfo  {journal} {Journal of the Physical Society of Japan}\
  }\textbf {\bibinfo {volume} {66}},\ \bibinfo {pages} {3411--3420} (\bibinfo
  {year} {1997})}\BibitemShut {NoStop}%
\bibitem [{\citenamefont {Shao}\ \emph {et~al.}(2016)\citenamefont {Shao},
  \citenamefont {Guo},\ and\ \citenamefont {Sandvik}}]{Shao2016Sci}%
  \BibitemOpen
  \bibfield  {author} {\bibinfo {author} {\bibfnamefont {H.}~\bibnamefont
  {Shao}}, \bibinfo {author} {\bibfnamefont {W.}~\bibnamefont {Guo}}, \ and\
  \bibinfo {author} {\bibfnamefont {A.~W.}\ \bibnamefont {Sandvik}},\
  }\bibfield  {title} {\enquote {\bibinfo {title} {Quantum criticality with two
  length scales},}\ }\href {\doibase 10.1126/science.aad5007} {\bibfield
  {journal} {\bibinfo  {journal} {Science}\ }\textbf {\bibinfo {volume}
  {352}},\ \bibinfo {pages} {213--216} (\bibinfo {year} {2016})}\BibitemShut
  {NoStop}%
\bibitem [{\citenamefont {Amit}\ and\ \citenamefont
  {Peliti}(1982)}]{Amit1982annph}%
  \BibitemOpen
  \bibfield  {author} {\bibinfo {author} {\bibfnamefont {D.~J.}\ \bibnamefont
  {Amit}}\ and\ \bibinfo {author} {\bibfnamefont {L.}~\bibnamefont {Peliti}},\
  }\bibfield  {title} {\enquote {\bibinfo {title} {{On dangerous irrelevant
  operators}},}\ }\href {\doibase https://doi.org/10.1016/0003-4916(82)90159-2}
  {\bibfield  {journal} {\bibinfo  {journal} {Annals of Physics}\ }\textbf
  {\bibinfo {volume} {140}},\ \bibinfo {pages} {207--231} (\bibinfo {year}
  {1982})}\BibitemShut {NoStop}%
\bibitem [{\citenamefont {Shu}\ \emph {et~al.}(2023)\citenamefont {Shu},
  \citenamefont {Jian}, \citenamefont {Sandvik},\ and\ \citenamefont
  {Yin}}]{Shu2023kz}%
  \BibitemOpen
  \bibfield  {author} {\bibinfo {author} {\bibfnamefont {Y.-R.}\ \bibnamefont
  {Shu}}, \bibinfo {author} {\bibfnamefont {S.-K.}\ \bibnamefont {Jian}},
  \bibinfo {author} {\bibfnamefont {A.~W.}\ \bibnamefont {Sandvik}}, \ and\
  \bibinfo {author} {\bibfnamefont {S.}~\bibnamefont {Yin}},\ }\bibfield
  {title} {\enquote {\bibinfo {title} {{Equilibration of Topological Defects at
  the Deconfined Quantum Critical Point}},}\ }\href
  {https://arxiv.org/abs/2305.04771} {\bibfield  {journal} {\bibinfo  {journal}
  {arXiv:2305.04771}\ } (\bibinfo {year} {2023})}\BibitemShut {NoStop}%
\bibitem [{\citenamefont {Xiang}\ \emph {et~al.}(2024)\citenamefont {Xiang},
  \citenamefont {Zhang}, \citenamefont {Gao}, \citenamefont {Schmidt},
  \citenamefont {Schmalzl}, \citenamefont {Wang}, \citenamefont {Li},
  \citenamefont {Xi}, \citenamefont {Liu}, \citenamefont {Jin}, \citenamefont
  {Li}, \citenamefont {Shen}, \citenamefont {Chen}, \citenamefont {Qi},
  \citenamefont {Wan}, \citenamefont {Jin}, \citenamefont {Li}, \citenamefont
  {Sun},\ and\ \citenamefont {Su}}]{Xiang2024nat}%
  \BibitemOpen
  \bibfield  {author} {\bibinfo {author} {\bibfnamefont {J.}~\bibnamefont
  {Xiang}}, \bibinfo {author} {\bibfnamefont {C.}~\bibnamefont {Zhang}},
  \bibinfo {author} {\bibfnamefont {Y.}~\bibnamefont {Gao}}, \bibinfo {author}
  {\bibfnamefont {W.}~\bibnamefont {Schmidt}}, \bibinfo {author} {\bibfnamefont
  {K.}~\bibnamefont {Schmalzl}}, \bibinfo {author} {\bibfnamefont {C.-W.}\
  \bibnamefont {Wang}}, \bibinfo {author} {\bibfnamefont {B.}~\bibnamefont
  {Li}}, \bibinfo {author} {\bibfnamefont {N.}~\bibnamefont {Xi}}, \bibinfo
  {author} {\bibfnamefont {X.-Y.}\ \bibnamefont {Liu}}, \bibinfo {author}
  {\bibfnamefont {H.}~\bibnamefont {Jin}}, \bibinfo {author} {\bibfnamefont
  {G.}~\bibnamefont {Li}}, \bibinfo {author} {\bibfnamefont {J.}~\bibnamefont
  {Shen}}, \bibinfo {author} {\bibfnamefont {Z.}~\bibnamefont {Chen}}, \bibinfo
  {author} {\bibfnamefont {Y.}~\bibnamefont {Qi}}, \bibinfo {author}
  {\bibfnamefont {Y.}~\bibnamefont {Wan}}, \bibinfo {author} {\bibfnamefont
  {W.}~\bibnamefont {Jin}}, \bibinfo {author} {\bibfnamefont {W.}~\bibnamefont
  {Li}}, \bibinfo {author} {\bibfnamefont {P.}~\bibnamefont {Sun}}, \ and\
  \bibinfo {author} {\bibfnamefont {G.}~\bibnamefont {Su}},\ }\bibfield
  {title} {\enquote {\bibinfo {title} {{Giant magnetocaloric effect in spin
  supersolid candidate ${\mathrm{Na}_{2}}{\mathrm{BaCo(PO}}_{4})_{2}$}},}\
  }\href {\doibase 10.1038/s41586-023-06885-w} {\bibfield  {journal} {\bibinfo
  {journal} {Nature}\ }\textbf {\bibinfo {volume} {625}},\ \bibinfo {pages}
  {270--275} (\bibinfo {year} {2024})}\BibitemShut {NoStop}%
\bibitem [{\citenamefont {Chi}\ \emph {et~al.}(2024)\citenamefont {Chi},
  \citenamefont {Hu}, \citenamefont {Liao},\ and\ \citenamefont
  {Xiang}}]{Chi2024}%
  \BibitemOpen
  \bibfield  {author} {\bibinfo {author} {\bibfnamefont {R.}~\bibnamefont
  {Chi}}, \bibinfo {author} {\bibfnamefont {J.}~\bibnamefont {Hu}}, \bibinfo
  {author} {\bibfnamefont {H.-J.}\ \bibnamefont {Liao}}, \ and\ \bibinfo
  {author} {\bibfnamefont {T.}~\bibnamefont {Xiang}},\ }\bibfield  {title}
  {\enquote {\bibinfo {title} {{Dynamical Spectra of Spin Supersolid States in
  Triangular Antiferromagnets}},}\ }\href {https://arxiv.org/abs/2404.14163}
  {\bibfield  {journal} {\bibinfo  {journal} {arXiv:2024.14163}\ } (\bibinfo
  {year} {2024})}\BibitemShut {NoStop}%
\bibitem [{\citenamefont {Gao}\ \emph {et~al.}()\citenamefont {Gao},
  \citenamefont {Zhang}, \citenamefont {Xiang}, \citenamefont {Yu},
  \citenamefont {Lu}, \citenamefont {Sun}, \citenamefont {Jin}, \citenamefont
  {Su},\ and\ \citenamefont {Li}}]{Gao2024}%
  \BibitemOpen
  \bibfield  {author} {\bibinfo {author} {\bibfnamefont {Y.}~\bibnamefont
  {Gao}}, \bibinfo {author} {\bibfnamefont {C.}~\bibnamefont {Zhang}}, \bibinfo
  {author} {\bibfnamefont {J.}~\bibnamefont {Xiang}}, \bibinfo {author}
  {\bibfnamefont {D.}~\bibnamefont {Yu}}, \bibinfo {author} {\bibfnamefont
  {X.}~\bibnamefont {Lu}}, \bibinfo {author} {\bibfnamefont {P.}~\bibnamefont
  {Sun}}, \bibinfo {author} {\bibfnamefont {W.}~\bibnamefont {Jin}}, \bibinfo
  {author} {\bibfnamefont {G.}~\bibnamefont {Su}}, \ and\ \bibinfo {author}
  {\bibfnamefont {W.}~\bibnamefont {Li}},\ }\bibfield  {title} {\enquote
  {\bibinfo {title} {Spin supersolid phase and double magnon-roton excitations
  in a cobalt-based triangular lattice},}\ }\href@noop {} {\ }\BibitemShut
  {NoStop}%
\bibitem [{\citenamefont {Domb}\ and\ \citenamefont
  {Lebowitz}(1983)}]{Nelson1983book}%
  \BibitemOpen
  \bibfield  {author} {\bibinfo {author} {\bibfnamefont {C.}~\bibnamefont
  {Domb}}\ and\ \bibinfo {author} {\bibfnamefont {J.}~\bibnamefont
  {Lebowitz}},\ }\href@noop {} {\emph {\bibinfo {title} {{Phase Transitions and
  Critical Phenomena, Vol. 7.}}}}\ (\bibinfo  {publisher} {Academic Press},\
  \bibinfo {year} {1983})\BibitemShut {NoStop}%
\bibitem [{\citenamefont {Li}\ \emph {et~al.}(2020)\citenamefont {Li},
  \citenamefont {Yang}, \citenamefont {Xie}, \citenamefont {Tu}, \citenamefont
  {Liao},\ and\ \citenamefont {Xiang}}]{xiang2020pre}%
  \BibitemOpen
  \bibfield  {author} {\bibinfo {author} {\bibfnamefont {Z.-Q.}\ \bibnamefont
  {Li}}, \bibinfo {author} {\bibfnamefont {L.-P.}\ \bibnamefont {Yang}},
  \bibinfo {author} {\bibfnamefont {Z.~Y.}\ \bibnamefont {Xie}}, \bibinfo
  {author} {\bibfnamefont {H.-H.}\ \bibnamefont {Tu}}, \bibinfo {author}
  {\bibfnamefont {H.-J.}\ \bibnamefont {Liao}}, \ and\ \bibinfo {author}
  {\bibfnamefont {T.}~\bibnamefont {Xiang}},\ }\bibfield  {title} {\enquote
  {\bibinfo {title} {Critical properties of the two-dimensional $q$-state clock
  model},}\ }\href {\doibase 10.1103/PhysRevE.101.060105} {\bibfield  {journal}
  {\bibinfo  {journal} {Phys. Rev. E}\ }\textbf {\bibinfo {volume} {101}},\
  \bibinfo {pages} {060105} (\bibinfo {year} {2020})}\BibitemShut {NoStop}%
\bibitem [{\citenamefont {Ueno}\ and\ \citenamefont
  {Mitsubo}(1991)}]{Ueno1991prb}%
  \BibitemOpen
  \bibfield  {author} {\bibinfo {author} {\bibfnamefont {Y.}~\bibnamefont
  {Ueno}}\ and\ \bibinfo {author} {\bibfnamefont {K.}~\bibnamefont {Mitsubo}},\
  }\bibfield  {title} {\enquote {\bibinfo {title} {{Incompletely ordered phase
  in the three-dimensional six-state clock model: Evidence for an absence of
  ordered phases of XY character}},}\ }\href {\doibase
  10.1103/PhysRevB.43.8654} {\bibfield  {journal} {\bibinfo  {journal} {Phys.
  Rev. B}\ }\textbf {\bibinfo {volume} {43}},\ \bibinfo {pages} {8654--8657}
  (\bibinfo {year} {1991})}\BibitemShut {NoStop}%
\bibitem [{\citenamefont {Chubukov}\ \emph {et~al.}(1994)\citenamefont
  {Chubukov}, \citenamefont {Sachdev},\ and\ \citenamefont
  {Ye}}]{Chubukov1994prb}%
  \BibitemOpen
  \bibfield  {author} {\bibinfo {author} {\bibfnamefont {A.~V.}\ \bibnamefont
  {Chubukov}}, \bibinfo {author} {\bibfnamefont {S.}~\bibnamefont {Sachdev}}, \
  and\ \bibinfo {author} {\bibfnamefont {J.}~\bibnamefont {Ye}},\ }\bibfield
  {title} {\enquote {\bibinfo {title} {{Theory of two-dimensional quantum
  Heisenberg antiferromagnets with a nearly critical ground state}},}\ }\href
  {\doibase 10.1103/PhysRevB.49.11919} {\bibfield  {journal} {\bibinfo
  {journal} {Phys. Rev. B}\ }\textbf {\bibinfo {volume} {49}},\ \bibinfo
  {pages} {11919--11961} (\bibinfo {year} {1994})}\BibitemShut {NoStop}%
\bibitem [{\citenamefont {Banerjee}\ \emph {et~al.}(2018)\citenamefont
  {Banerjee}, \citenamefont {Chandrasekharan},\ and\ \citenamefont
  {Orlando}}]{Banerjee2018prl}%
  \BibitemOpen
  \bibfield  {author} {\bibinfo {author} {\bibfnamefont {D.}~\bibnamefont
  {Banerjee}}, \bibinfo {author} {\bibfnamefont {S.}~\bibnamefont
  {Chandrasekharan}}, \ and\ \bibinfo {author} {\bibfnamefont {D.}~\bibnamefont
  {Orlando}},\ }\bibfield  {title} {\enquote {\bibinfo {title} {{Conformal
  Dimensions via Large Charge Expansion}},}\ }\href {\doibase
  10.1103/PhysRevLett.120.061603} {\bibfield  {journal} {\bibinfo  {journal}
  {Phys. Rev. Lett.}\ }\textbf {\bibinfo {volume} {120}},\ \bibinfo {pages}
  {061603} (\bibinfo {year} {2018})}\BibitemShut {NoStop}%
\bibitem [{\citenamefont {Binder}\ and\ \citenamefont
  {Heermann}(2010)}]{binderbook}%
  \BibitemOpen
  \bibfield  {author} {\bibinfo {author} {\bibfnamefont {K.}~\bibnamefont
  {Binder}}\ and\ \bibinfo {author} {\bibfnamefont {D.~W.}\ \bibnamefont
  {Heermann}},\ }\href@noop {} {\emph {\bibinfo {title} {{Monte Carlo
  Simulation in Statistical Physics}}}}\ (\bibinfo  {publisher} {Springer
  Berlin, Heidelberg},\ \bibinfo {year} {2010})\BibitemShut {NoStop}%
\bibitem [{\citenamefont {Campostrini}\ \emph {et~al.}(2006)\citenamefont
  {Campostrini}, \citenamefont {Hasenbusch}, \citenamefont {Pelissetto},\ and\
  \citenamefont {Vicari}}]{Campostrini2006prb}%
  \BibitemOpen
  \bibfield  {author} {\bibinfo {author} {\bibfnamefont {M.}~\bibnamefont
  {Campostrini}}, \bibinfo {author} {\bibfnamefont {M.}~\bibnamefont
  {Hasenbusch}}, \bibinfo {author} {\bibfnamefont {A.}~\bibnamefont
  {Pelissetto}}, \ and\ \bibinfo {author} {\bibfnamefont {E.}~\bibnamefont
  {Vicari}},\ }\bibfield  {title} {\enquote {\bibinfo {title} {{Theoretical
  estimates of the critical exponents of the superfluid transition in
  $^{4}\mathrm{He}$ by lattice methods}},}\ }\href {\doibase
  10.1103/PhysRevB.74.144506} {\bibfield  {journal} {\bibinfo  {journal} {Phys.
  Rev. B}\ }\textbf {\bibinfo {volume} {74}},\ \bibinfo {pages} {144506}
  (\bibinfo {year} {2006})}\BibitemShut {NoStop}%
\bibitem [{\citenamefont {Campostrini}\ \emph {et~al.}(2001)\citenamefont
  {Campostrini}, \citenamefont {Hasenbusch}, \citenamefont {Pelissetto},
  \citenamefont {Rossi},\ and\ \citenamefont {Vicari}}]{Campostrini2001prb}%
  \BibitemOpen
  \bibfield  {author} {\bibinfo {author} {\bibfnamefont {M.}~\bibnamefont
  {Campostrini}}, \bibinfo {author} {\bibfnamefont {M.}~\bibnamefont
  {Hasenbusch}}, \bibinfo {author} {\bibfnamefont {A.}~\bibnamefont
  {Pelissetto}}, \bibinfo {author} {\bibfnamefont {P.}~\bibnamefont {Rossi}}, \
  and\ \bibinfo {author} {\bibfnamefont {E.}~\bibnamefont {Vicari}},\
  }\bibfield  {title} {\enquote {\bibinfo {title} {{Critical behavior of the
  three-dimensional $\mathrm{XY}$ universality class}},}\ }\href {\doibase
  10.1103/PhysRevB.63.214503} {\bibfield  {journal} {\bibinfo  {journal} {Phys.
  Rev. B}\ }\textbf {\bibinfo {volume} {63}},\ \bibinfo {pages} {214503}
  (\bibinfo {year} {2001})}\BibitemShut {NoStop}%
\bibitem [{\citenamefont {Chester}\ \emph {et~al.}(2020)\citenamefont
  {Chester}, \citenamefont {Landry}, \citenamefont {Liu}, \citenamefont
  {Poland}, \citenamefont {Simmons-Duffin}, \citenamefont {Su},\ and\
  \citenamefont {Vichi}}]{Chester2020jhep}%
  \BibitemOpen
  \bibfield  {author} {\bibinfo {author} {\bibfnamefont {S.~M.}\ \bibnamefont
  {Chester}}, \bibinfo {author} {\bibfnamefont {W.}~\bibnamefont {Landry}},
  \bibinfo {author} {\bibfnamefont {J.}~\bibnamefont {Liu}}, \bibinfo {author}
  {\bibfnamefont {D.}~\bibnamefont {Poland}}, \bibinfo {author} {\bibfnamefont
  {D.}~\bibnamefont {Simmons-Duffin}}, \bibinfo {author} {\bibfnamefont
  {N.}~\bibnamefont {Su}}, \ and\ \bibinfo {author} {\bibfnamefont
  {A.}~\bibnamefont {Vichi}},\ }\bibfield  {title} {\enquote {\bibinfo {title}
  {{Carving out OPE space and precise $O(2)$ model critical exponents}},}\
  }\href {\doibase 10.1007/JHEP06(2020)142} {\bibfield  {journal} {\bibinfo
  {journal} {Journal of High Energy Physics}\ }\textbf {\bibinfo {volume}
  {2020}},\ \bibinfo {pages} {142} (\bibinfo {year} {2020})}\BibitemShut
  {NoStop}%
\bibitem [{\citenamefont {Adzhemyan}\ \emph {et~al.}(2022)\citenamefont
  {Adzhemyan}, \citenamefont {Evdokimov}, \citenamefont {Hnatič},
  \citenamefont {Ivanova}, \citenamefont {Kompaniets}, \citenamefont {Kudlis},\
  and\ \citenamefont {Zakharov}}]{Adzhemyan2022pa}%
  \BibitemOpen
  \bibfield  {author} {\bibinfo {author} {\bibfnamefont {L.}~\bibnamefont
  {Adzhemyan}}, \bibinfo {author} {\bibfnamefont {D.}~\bibnamefont
  {Evdokimov}}, \bibinfo {author} {\bibfnamefont {M.}~\bibnamefont {Hnatič}},
  \bibinfo {author} {\bibfnamefont {E.}~\bibnamefont {Ivanova}}, \bibinfo
  {author} {\bibfnamefont {M.}~\bibnamefont {Kompaniets}}, \bibinfo {author}
  {\bibfnamefont {A.}~\bibnamefont {Kudlis}}, \ and\ \bibinfo {author}
  {\bibfnamefont {D.}~\bibnamefont {Zakharov}},\ }\bibfield  {title} {\enquote
  {\bibinfo {title} {Model a of critical dynamics: 5-loop $\varepsilon$
  expansion study},}\ }\href {\doibase
  https://doi.org/10.1016/j.physa.2022.127530} {\bibfield  {journal} {\bibinfo
  {journal} {Physica A: Statistical Mechanics and its Applications}\ }\textbf
  {\bibinfo {volume} {600}},\ \bibinfo {pages} {127530} (\bibinfo {year}
  {2022})}\BibitemShut {NoStop}%
\bibitem [{\citenamefont {Shu}\ \emph {et~al.}(2024)\citenamefont {Shu},
  \citenamefont {Liao},\ and\ \citenamefont {Yin}}]{Shu2024prb}%
  \BibitemOpen
  \bibfield  {author} {\bibinfo {author} {\bibfnamefont {Y.-R.}\ \bibnamefont
  {Shu}}, \bibinfo {author} {\bibfnamefont {T.}~\bibnamefont {Liao}}, \ and\
  \bibinfo {author} {\bibfnamefont {S.}~\bibnamefont {Yin}},\ }\bibfield
  {title} {\enquote {\bibinfo {title} {Relaxation critical dynamics with
  emergent symmetry},}\ }\href {\doibase 10.1103/PhysRevB.110.134306}
  {\bibfield  {journal} {\bibinfo  {journal} {Phys. Rev. B}\ }\textbf {\bibinfo
  {volume} {110}},\ \bibinfo {pages} {134306} (\bibinfo {year}
  {2024})}\BibitemShut {NoStop}%
\bibitem [{\citenamefont {Ali}\ \emph {et~al.}(2024)\citenamefont {Ali},
  \citenamefont {Xu}, \citenamefont {Bernoudy}, \citenamefont {Nocera},
  \citenamefont {King},\ and\ \citenamefont {Banerjee}}]{Ali2024}%
  \BibitemOpen
  \bibfield  {author} {\bibinfo {author} {\bibfnamefont {A.}~\bibnamefont
  {Ali}}, \bibinfo {author} {\bibfnamefont {H.}~\bibnamefont {Xu}}, \bibinfo
  {author} {\bibfnamefont {W.}~\bibnamefont {Bernoudy}}, \bibinfo {author}
  {\bibfnamefont {A.}~\bibnamefont {Nocera}}, \bibinfo {author} {\bibfnamefont
  {A.~D.}\ \bibnamefont {King}}, \ and\ \bibinfo {author} {\bibfnamefont
  {A.}~\bibnamefont {Banerjee}},\ }\bibfield  {title} {\enquote {\bibinfo
  {title} {{Quantum Quench Dynamics of Geometrically Frustrated Ising
  Models}},}\ }\href {https://arxiv.org/abs/2403.00091} {\bibfield  {journal}
  {\bibinfo  {journal} {arXiv:2023.00091}\ } (\bibinfo {year}
  {2024})}\BibitemShut {NoStop}%
\end{thebibliography}%

\end{document}